%% file: main.tex
\documentclass[preprint,12pt]{elsarticle}

\input{preamble}
\usepackage{etoolbox} 

\journal{}

\begin{document}

\begin{frontmatter}



\title{CLIP: A CUDA-Accelerated Lattice Boltzmann Framework for Interfacial Phenomena with Application to Liquid Jet Simulations}





\author[label1]{Mehdi Shadkhah\fnref{fn1}}
\author[label1]{Mohammad Taeibi Rahni}
\author[label1]{Azadeh Kebriaee\corref{cor1}}
\author[label1]{Mohammad Reza Salimi}

\affiliation[label1]{organization={Sharif University of Technology}, state={Tehran}, country={Iran}}

\fntext[fn1]{Currently Ph.D. student at Iowa State University, Ames, IA, USA}
\cortext[cor1]{Corresponding author. Email: kebriaee@sharif.edu}


\begin{abstract}
This work introduces \href{https://github.com/mshadkhah/CLIP}{\textcolor{green!40!black}{CLIP}}, a CUDA-accelerated phase-field lattice Boltzmann framework for simulating immiscible two-phase flows with high density and viscosity ratios in both two- and three-dimensional domains. By leveraging GPU parallelism, the framework delivers substantial computational speedups, enabling large-scale simulations to be performed efficiently on standard desktop hardware without the need for high-performance computing clusters. It employs the Weighted Multi-Relaxation Time (WMRT) collision operator to enhance numerical stability and improve interface tracking under challenging multiphase conditions. The model is validated through a series of benchmark cases, including capillary wave dynamics, stationary drop tests, two-phase Poiseuille flow, shear-driven interface deformation, and Rayleigh–Taylor instability. It is further applied to simulate liquid jet breakup, capturing the transition from dripping to jetting regimes and identifying a critical Weber number of approximately 2.2. The results closely match experimental observations, offering detailed insights into breakup length, drop size distributions, and flow regime transitions. With its efficiency, accuracy, and scalability, the proposed framework serves as a powerful and accessible tool for investigating complex interfacial phenomena in multiphase flow physics.
\end{abstract}



\begin{keyword}
Lattice Boltzmann method \sep phase-field model \sep two-phase flow \sep high-density ratio \sep GPU acceleration \sep liquid jet breakup \sep dripping and jetting regimes
\end{keyword}

\end{frontmatter}



\input{Sections/Introduction}

\input{Sections/Literature}
\input{Sections/Methodology}

\input{Sections/Implementation}

\input{Sections/Validation}
\input{Sections/Results}
\input{Sections/Conclusion}

\input{Sections/Appendix}











\clearpage
\bibliographystyle{elsarticle-num-names}
\bibliography{bibliography}




\end{document}

%% file: preamble.tex
\usepackage{amsmath} 
\usepackage{amssymb} 
\usepackage{paralist}
\usepackage[misc]{ifsym}
\usepackage{epsfig} 
\usepackage{epstopdf} 
\usepackage[colorlinks=true]{hyperref}
\allowdisplaybreaks

\usepackage{tikz}
\usetikzlibrary{3d}

\usepackage{booktabs}
\usepackage{graphicx}
\usepackage{pgfplots}
\usepackage{subcaption}
\usepackage{natbib}
\usepackage{amsmath,amssymb,amsfonts}
\usepackage{graphicx}
\usepackage{textcomp}
\usepackage{xcolor}
\usepackage{algorithm}
\usepackage{algpseudocode}
\usepackage{hyperref}
\usepackage{multirow}
 \usepackage{array}
 \usepackage{geometry}
\usepackage{tabularx}
\usepackage{tabularray}
\usepackage{nicefrac,xfrac}
\usepackage{threeparttable}

\def\BibTeX{{\rm B\kern-.05em{\sc i\kern-.025em b}\kern-.08em
    T\kern-.1667em\lower.7ex\hbox{E}\kern-.125emX}}

\pgfplotsset{
compat=1.8,
legend image code/.code={
\draw[mark repeat=2,mark phase=2]
plot coordinates {
(0cm,0cm)
(0.15cm,0cm)        
(0.3cm,0cm)         
};%
}
}

\usepackage{amsthm}

\newcommand{\cref}[2]{\hyperref[#2]{#1~\ref*{#2}}}
\newcommand{\figref}[1]{\hyperref[#1]{Fig.~\ref*{#1}}}
\newcommand{\secref}[1]{\hyperref[#1]{Sec.~\ref*{#1}}}
\newcommand{\tabref}[1]{\hyperref[#1]{Tab.~\ref*{#1}}}
\newcommand{\eqnref}[1]{\hyperref[#1]{Eq.~(\ref*{#1})}}
\hypersetup{
	colorlinks,
	linkcolor={blue!50!black},
	citecolor={blue!50!black},
	urlcolor={blue!80!black},
	anchorcolor = {blue!80!black},
	filecolor = {blue!80!black},
	menucolor = {blue!80!black},
	runcolor = {blue!80!black}
}

\definecolor{cpu3}{HTML}{F44336}
\definecolor{cpu4}{HTML}{2196F3}
\definecolor{cpu1}{HTML}{4CAF50}
\definecolor{cpu2}{HTML}{FFC107}
\definecolor{gpu3}{HTML}{EF9A9A}
\definecolor{gpu4}{HTML}{90CAF9}
\definecolor{gpu1}{HTML}{A5D6A7}
\definecolor{gpu2}{HTML}{FFE082}

\definecolor{cpu5}{HTML}{9932CC}

\definecolor{sq_b1}{RGB}{37,52,148}
\definecolor{sq_b2}{RGB}{44,127,184}
\definecolor{sq_b3}{RGB}{65,182,196}
\definecolor{sq_b4}{RGB}{127,205,187}
\definecolor{sq_b5}{RGB}{199,233,180}
\definecolor{sq_b6}{RGB}{255,255,204}

\definecolor{sq_r1}{RGB}{189,0,38}
\definecolor{sq_r2}{RGB}{240,59,32}
\definecolor{sq_r3}{RGB}{253,141,60}
\definecolor{sq_r4}{RGB}{254,178,76}
\definecolor{sq_r5}{RGB}{254,217,118}
\definecolor{sq_r6}{RGB}{255,255,178}

\definecolor{sq_g1}{RGB}{0,104,55}
\definecolor{sq_g2}{RGB}{49,163,84}
\definecolor{sq_g3}{RGB}{120,198,121}
\definecolor{sq_g4}{RGB}{173,221,142}
\definecolor{sq_g5}{RGB}{217,240,163}
\definecolor{sq_g6}{RGB}{255,255,204}

\definecolor{div_c1}{RGB}{230,171,2}
\definecolor{div_c2}{RGB}{102,166,30}
\definecolor{div_c3}{RGB}{231,41,138}
\definecolor{div_c4}{RGB}{117,112,179}
\definecolor{div_c5}{RGB}{217,95,2}
\definecolor{div_c6}{RGB}{27,158,119}
\definecolor{div_c7}{RGB}{215,48,39}

\definecolor{div_d1}{RGB}{215,25,28}
\definecolor{div_d2}{RGB}{253,174,97}
\definecolor{div_d3}{RGB}{255,255,191}
\definecolor{div_d4}{RGB}{171,217,233}
\definecolor{div_d5}{RGB}{44,123,182}
\definecolor{ao}{RGB}{0.0, 128, 0.0}

\definecolor{ActiveElement}{RGB}{147,194,74}
\definecolor{InterceptedElement}{RGB}{255,208,48}
\definecolor{FalseInterceptedElement}{RGB}{92,91,255}
\definecolor{lightprofgreen}{RGB}{60,179,113} 


%% file: Sections/Introduction.tex
\section{Introduction}
\label{sec:introduction}

Multiphase flows with complex interfaces are ubiquitous in both natural and industrial processes. A notable example of such flows is liquid injection into a gaseous environment, which is widely utilized in diverse applications such as drug delivery, dosage devices, electronic cooling systems, and surface coating. These flows have been extensively studied through theoretical, experimental, and numerical approaches \cite{Batchelor1962,Michalke1984,Morris1988} since Rayleigh's seminal work on jet instability \cite{Rayleigh1878}. Understanding the behavior of liquid jets in this study focuses on three key areas: liquid jet systems, the lattice Boltzmann method, and GPU parallelization, all of which are extensively reviewed and analyzed.

In this paper, a three-dimensional phase-field model is developed to investigate fluid dynamics, with a particular focus on the breakup of immiscible liquid jets. The solver utilizes a D3Q19 lattice to accurately simulate two-phase flows without restrictions on density or viscosity contrasts. Interface tracking is achieved using a phase-field approach based on the Allen-Cahn equation, ensuring numerical stability and accuracy. Additionally, the D3Q19 version of the Weighted Multi-Relaxation Time (WMRT) collision operator is employed to study high-momentum physics effectively. 

To optimize performance, a parallel algorithm is implemented on the CUDA platform, enabling the simulation of large lattice domains with reduced computational costs. The proposed model's accuracy is validated through several benchmark tests, including the stationary drop test, two-phase Poiseuille flow, circular interface in shear flow, and Rayleigh-Taylor instability. Finally, the model's capabilities are demonstrated by simulating liquid-jet breakup with a high-density ratio, with results compared to experimental data. The paper's organization is as follows: \secref{sec:literature} reviews the literature; \secref{sec:methodology} outlines the development of the three-dimensional phase-field LB model; \secref{sec:implementation} describes the parallel algorithm and GPU implementation; \secref{sec:validation} details the benchmark tests for accuracy validation; \secref{sec:results} presents the application to liquid-jet breakup; and \secref{sec:conclusion} concludes with a summary of findings.

%% file: Sections/Literature.tex
\section{Literature Review}
\label{sec:literature}

This section provides a comprehensive overview of previous studies related to liquid jet dynamics, numerical modeling techniques, and interface capturing methods. Emphasis is placed on understanding breakup mechanisms at low Weber numbers and recent advances in computational frameworks for simulating multiphase flows. The review is organized into three main parts: (i) numerical methods used in the simulation of liquid jets, focusing on interfacial instabilities and breakup regimes; (ii) developments in the Lattice Boltzmann Method (LBM) for multiphase and high-density ratio flows; and (iii) GPU-based parallelization strategies to enable efficient large-scale simulations.

\subsection{Review of numerical studies on liquid jets}
\label{sec:review_jet}

Understanding the behavior of liquid jets at low Weber numbers has long been a subject of significant interest due to its relevance in natural and industrial processes. At low injection velocities, jets interact with the nozzle geometry and ambient gas, producing fragmentation into droplets with distinct breakup patterns. The breakup length—defined as the distance from the nozzle exit to the end of the continuous liquid column—depends on the flow regime \cite{Mayinger2006}. Ohnesorge \cite{Ohnesorge1936} introduced a classical framework to categorize these regimes into (0) dripping, (I) Rayleigh breakup, (II) wind-induced breakup, and (III) atomization.

The dependence of jet breakup on governing dimensionless numbers has been illustrated using the Ohnesorge diagram, shown in \figref{fig:ohnesorge_diagram}, which delineates transitions between regimes based on Reynolds and Ohnesorge numbers \cite{Bonhoeffer2017}. For Weber numbers below 100 ($We < 100$), the breakup behavior is notably distinct. In such regimes, the breakup length is primarily influenced by the Weber number and, to a lesser extent, the nozzle exit diameter. Viscous, gravitational, inertial, and surface tension forces jointly govern the breakup dynamics at these conditions \cite{Rajendran2012}.

\input{Figures/ohnesorge}

To gain deeper insights into the physics governing jet breakup and atomization, numerical simulations have played a vital role. High-fidelity methods such as adaptive volume-of-fluid (VOF) have proven effective in resolving the interface dynamics during primary breakup. These simulations capture interfacial instabilities, vorticity-dominated structures, and the subsequent formation of small droplets \cite{Fuster2009}. The accuracy of such simulations is highly sensitive to mesh resolution, particularly in regions with strong droplet-vorticity coupling.

In addition to conventional VOF approaches, the original formulation of the Volume of Fluid method introduced by Hirt and Nichols \cite{Hirt1981} remains foundational for capturing sharp fluid interfaces in free boundary problems. This method tracks the fractional volume of fluid in each computational cell, allowing robust treatment of highly deformable interfaces while maintaining the conservation of mass. Meanwhile, the phase field method (or diffuse interface method) has emerged as a powerful alternative for modeling multiphase flows by treating the interface as a smooth transition zone governed by thermodynamic principles. It leverages the Cahn–Hilliard framework to model interfacial dynamics with Korteweg stresses and has been particularly effective in simulating phenomena involving topological changes or near-critical behavior \cite{Lamorgese2011}.

Further improvements have been realized through hybrid methods such as the Level Set/Volume of Fluid/Geometric Front (LS/VOF/GF) approach, which combines the advantages of accurate interface representation, mass conservation, and geometric precision. This method enhances the quantitative prediction of spray atomization, particularly in regimes where experimental data are scarce. Numerical studies have shown that surface instabilities arising during intermittent liquid injection initially emerge as two-dimensional perturbations, which progressively evolve into complex three-dimensional wave structures as the instabilities grow \cite{Lebas2009, Shinjo2011}.

\subsection{Review of Lattice Boltzmann Method}
\label{sec:review_lbm}

The lattice Boltzmann method (LBM) has undergone significant advancements over recent decades, solidifying its position as a versatile tool for modeling multiphase flows with complex interfacial dynamics. LBM models for interfacial multiphase flows are typically categorized into five major types:
\begin{enumerate}
    \item Color-fluid model \cite{Gunstensen1991},
    \item Pseudopotential model \cite{Shan1993, Shan1994},
    \item Free-energy model \cite{Swift1996},
    \item Mean-field model \cite{He1999},
    \item Phase-field model \cite{Zu2013}.
\end{enumerate}

Each of these models addresses specific challenges in fluid dynamics but also presents limitations related to mass conservation, interfacial tension, spurious velocities, and handling high-density contrasts \cite{Falcucci2011}. Advances in LBM technology have mitigated many of these issues, extending the range of dimensionless number constraints and enhancing the applicability of these methods, as summarized in \tabref{tab:lbm_models}. Among these models, phase-field approaches have emerged as a robust framework for tracking fluid interfaces and ensuring mass conservation across fluid phases \cite{Zu2013}. Their reliance on the locality of collision processes enables efficient computation of normal vectors and high-performance parallel implementations.
\input{Tables/lbmModels}

The development of phase-field models has evolved significantly, with continuous efforts to enhance interface tracking and computational efficiency. Early advancements introduced pressure-based distribution functions to simulate incompressible multiphase flows, successfully validated through three-dimensional Rayleigh-Taylor instability simulations \cite{He1999_Zhang}. To further improve accuracy, conservative approaches leveraging the Allen-Cahn equation were incorporated, ensuring mass conservation via dual-compact upwind advection schemes \cite{Chiu2011}. At the same time, projection strategies utilizing continuity viscosity fluxes enhanced the robustness of these models in handling high-density and viscosity contrasts \cite{Zu2013}.

Building on these foundations, researchers turned their focus to refining interface representation and improving numerical methods. The integration of modified equilibrium distribution functions with the Allen-Cahn equation enabled sharper interface representations and efficient solutions through Chapman-Enskog analysis \cite{Ren2016}. Further enhancements in mass conservation were achieved by employing central moments for interface tracking, with Allen-Cahn dynamics offering greater numerical stability and accuracy \cite{Fakhari2019}. Recognizing the need for improved spatiotemporal accuracy, new methods incorporating central differencing schemes and high-order time integration techniques, such as Runge-Kutta methods, were later introduced to solve the Cahn-Hilliard equation \cite{Zhang2019}.

The most recent strides in phase-field modeling have extended the LBM framework to multi-component immiscible incompressible flows. By introducing generalized conservative phase-field models, researchers have successfully solved the generalized eikonal equation, ensuring adherence to volume fraction constraints while maintaining physical accuracy \cite{Hu2020}. Collectively, these advancements showcase the adaptability of LBM in addressing multiphase flow challenges, demonstrating how continuous refinements in phase-field models have led to improved stability, efficiency, and precision in simulating complex interfacial dynamics.

\subsection{Review of GPU parallelization}
\label{sec:review_gpu}
The simulation of complex three-dimensional domains using lattice Boltzmann multiphase models requires extensive computational resources. To address this challenge, parallel algorithms have become essential for reducing computational costs, especially with advancements in graphics processing units (GPUs) and parallel processing techniques. Leveraging these technologies, researchers have developed sophisticated methods to accelerate simulations, making them more efficient and scalable. In this study, the compute unified device architecture (CUDA) is employed to efficiently manage the extensive calculations required for three-dimensional domains, significantly reducing computational overhead.

The implementation of lattice Boltzmann methods (LBM) on modern GPUs has been a subject of significant interest. Early contributions include work by Januszewski and Kostura, who utilized high-level programming languages, such as Python, to develop efficient GPU-accelerated LBM simulations. Their study systematically compared various LBM models, highlighting the potential for GPUs to enhance computational performance across different configurations \cite{Januszewski2014}. 

Building on this foundation, Latt et al. introduced Palabos, an open-source parallel lattice Boltzmann library, designed for flexibility and high computational performance \cite{Latt2021}. Palabos supports a wide range of physical phenomena, including fluid-structure interactions, supersonic compressible flows, and complex flow regimes, making it a versatile tool for both research and practical applications.

%% file: Figures/ohnesorge.tex
\begin{figure}[htbp]
    \centering
    \begin{subfigure}[t]{0.48\textwidth} 
        \centering
        \includegraphics[width=\textwidth]{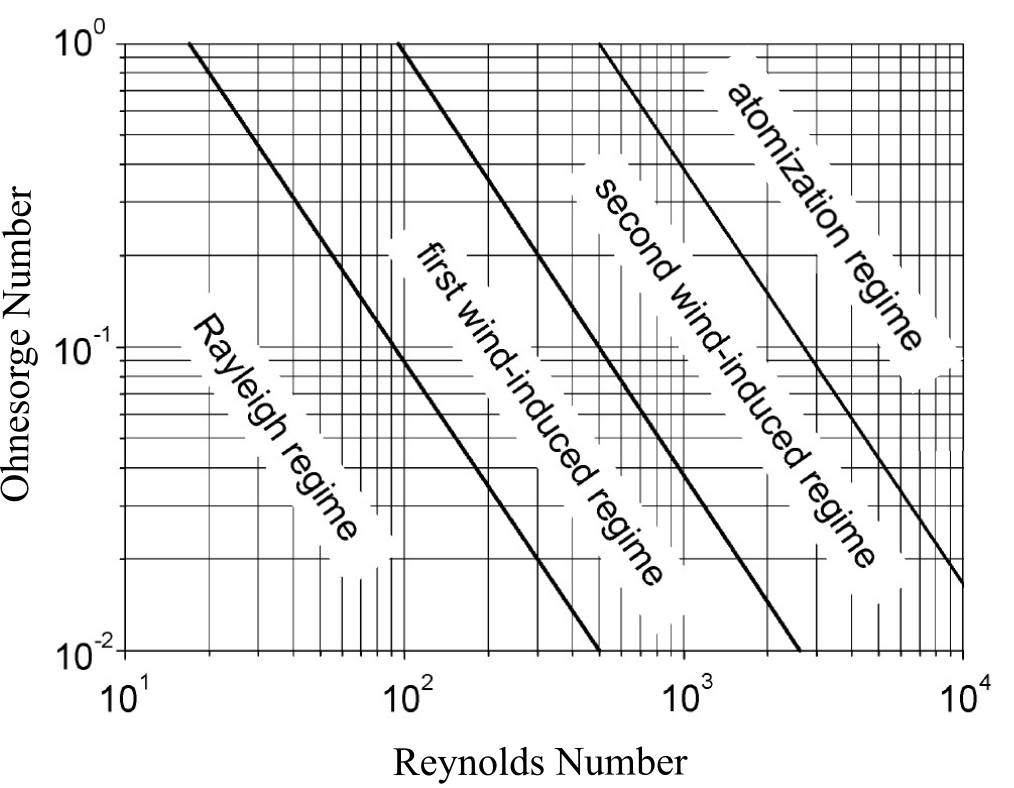} 
        \caption{Ohnesorge diagram illustrating jet breakup regimes as a function of Reynolds and Ohnesorge numbers. The identified regimes are: I. Dripping, II. Rayleigh breakup, III. First wind-induced breakup, IV. Second wind-induced breakup, and V. Atomization.}
        \label{fig:ohnesorge_diagram}
    \end{subfigure}
    \hspace{0.01\textwidth} 
    \begin{subfigure}[t]{0.48\textwidth} 
        \centering
        \includegraphics[width=\textwidth]{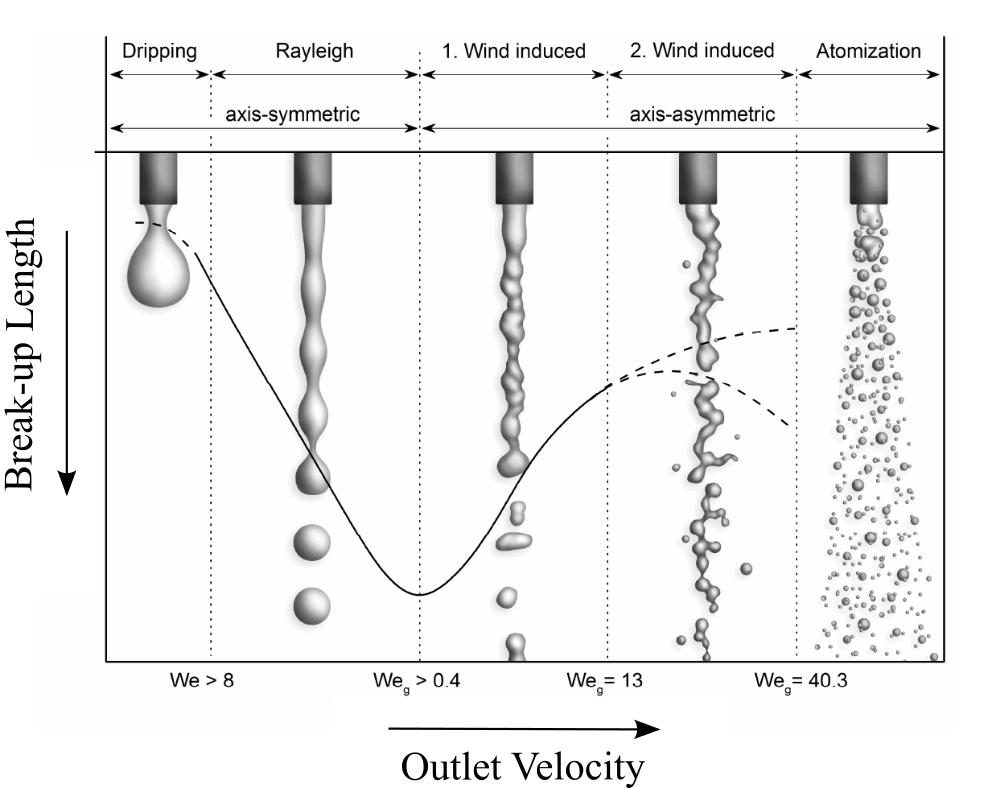} 
        \caption{Schematic representation of liquid jet breakup from a circular nozzle in quiescent air. The breakup regimes—dripping, Rayleigh, wind-induced, and atomization—are categorized by increasing outlet velocity and surface wave dynamics.}
        \label{fig:jet_breakup}
    \end{subfigure}
    \caption{(a) Ohnesorge diagram highlighting the transition between jet breakup regimes. (b) Visualization of jet breakup patterns as a function of Weber and outlet velocity, showcasing axisymmetric and non-axisymmetric behaviors.}
    \label{fig:jet_breakup_overview}
\end{figure}

%% file: Tables/lbmModels.tex
\begin{table}[htbp]
\caption{Application domains for multiphase LB models as a function of dimensionless numbers.}
\label{tab:lbm_models} 
\centering
\renewcommand{\arraystretch}{1.5} 
\resizebox{\textwidth}{!}{ 
\begin{tabular}{llccccl}
\hline
\textbf{Model} &  $\boldsymbol{\rho_L/\rho_V}$ & $\boldsymbol{{Re}_L}$ & $\boldsymbol{{We}_L}$ & $\boldsymbol{{Fr}}$ & \textbf{Applications} \\
\hline
Color-fluid           & $10^3$  & $>10^4$  & $>10^4$  & $>1$    & Miscible fluids \\
Pseudopotential       & $10^2$  & $<10^3$  & $<10^3$  & $>1$    & Micro-flows, ferrofluids, capillarity effects \\
Free-energy           & $10^3$  & $<10^4$  & $>10^3$  & $>1$    & Free jets, cavitation \\
Mean-field / Phase-field & $-$   & $>10^4$  & $>10^6$  & $<10$   & Large-scale sprays \\
\hline
\end{tabular}
}
\end{table}

%% file: Sections/Methodology.tex
\section{Methodology}
\label{sec:methodology}

This section outlines the methodology for simulating multiphase flows using a phase-field-based lattice Boltzmann framework. The approach integrates phase-field theory for capturing interfacial dynamics with the lattice Boltzmann equation (LBE) for solving hydrodynamic and interface tracking problems. Together, these methods enable accurate modeling of complex multiphase systems with high-density ratios and intricate interfacial phenomena.

\subsection{Macroscopic Equations}
\label{sec:macroscopic_eqns}

\textbf{Phase-Field Theory:} The phase-field theory provides a robust macroscopic framework for modeling interfacial dynamics in multiphase flows. This approach describes fluid interfaces with a finite thickness using the phase-field variable $\phi$, which smoothly transitions between 0 and 1. In this representation, $\phi = 1$ corresponds to the heavy fluid, $\phi = 0$ corresponds to the light fluid, and intermediate values define the diffuse interface. The governing equations of the phase-field model capture the evolution of $\phi$ and its interaction with the velocity field, providing a foundation for simulating complex multiphase phenomena.

The interface velocity can be decomposed into a normal interface speed and a velocity component due to external advection. The normal interface speed is proportional to the interface curvature. The evolution of the phase-field variable $\phi$ is described by the following equation \cite{SUN2007626}:
\begin{equation}
\frac{\partial \phi}{\partial t} + (\mathbf{u} \cdot \boldsymbol{\nabla}) \phi = \gamma |\boldsymbol{\nabla} \phi| \kappa, 
\label{eq:phase_field_start}
\end{equation}
where $\mathbf{u}$ is the velocity field (advection), $\gamma$ is a positive constant, and $\kappa$ is the curvature of the interface. 

The curvature $\kappa$ is defined in terms of the unit normal vector $\mathbf{n}$ as:
\begin{equation}
\kappa = \boldsymbol{\nabla} \cdot \mathbf{n}, \quad \mathbf{n} = \frac{\boldsymbol{\nabla} \phi}{|\boldsymbol{\nabla} \phi|}, 
\label{eq:curvature_def}
\end{equation}
where $\mathbf{n}$ represents the normal vector to the interface. By expanding the divergence term, the curvature can be expressed as:
\begin{equation}
\kappa = \boldsymbol{\nabla} \cdot \left( \frac{\boldsymbol{\nabla} \phi}{|\boldsymbol{\nabla} \phi|} \right) = \frac{\nabla^2 \phi}{|\boldsymbol{\nabla} \phi|} - \frac{\boldsymbol{\nabla} \phi \cdot \boldsymbol{\nabla} |\boldsymbol{\nabla} \phi|}{|\boldsymbol{\nabla} \phi|^2}. 
\label{eq:curvature_expanded}
\end{equation}

Substituting Eq.~\eqref{eq:curvature_expanded} into Eq.~\eqref{eq:phase_field_start} yields:
\begin{equation}
\frac{\partial \phi}{\partial t} + (\mathbf{u} \cdot \boldsymbol{\nabla}) \phi = \gamma \nabla^2 \phi - \gamma \mathbf{n} \cdot \boldsymbol{\nabla} |\boldsymbol{\nabla} \phi|. 
\label{eq:phase_field_with_curvature}
\end{equation}

Under equilibrium conditions, the kernel function for $\phi$ across a planar interface is given by:
\begin{equation}
\phi(z) = \frac{\phi_H + \phi_L}{2} + \frac{\phi_H - \phi_L}{2} \tanh\left(\frac{2z}{\xi}\right), 
\label{eq:kernel_function}
\end{equation}
where $z$ is the coordinate normal to the interface, $\xi$ is the interface thickness, and $\phi_H$ and $\phi_L$ are the values of $\phi$ in the heavy and light fluids, respectively. The gradient magnitude is derived as:
\begin{equation}
|\boldsymbol{\nabla} \phi| = \frac{\partial \phi}{\partial z} = \frac{-4(\phi - \phi_H)(\phi - \phi_L)}{\xi (\phi_H - \phi_L)}. 
\label{eq:gradient_magnitude}
\end{equation}

Using $\phi_0 = (\phi_H + \phi_L)/2$ and substituting Eq.~\eqref{eq:gradient_magnitude} into Eq.~\eqref{eq:phase_field_with_curvature}, the phase-field equation becomes:
\begin{equation}
\frac{\partial \phi}{\partial t} + (\mathbf{u} \cdot \boldsymbol{\nabla}) \phi = \gamma \nabla^2 \phi - \gamma \frac{1 - 4(\phi - \phi_0)^2}{\xi} \mathbf{n}.
\label{eq:phase_field_substituted}
\end{equation}

Here, $\gamma$, which controls the mobility of the interface, is replaced by $M$ to emphasize its role as a mobility parameter. By enforcing the incompressibility condition, $\boldsymbol{\nabla} \cdot \mathbf{u} = 0$, the conservative form of the Allen-Cahn equation is obtained as:
\begin{equation}
\frac{\partial \phi}{\partial t} + \boldsymbol{\nabla} \cdot (\phi \mathbf{u}) = \boldsymbol{\nabla} \cdot 
\left[ M \left( \boldsymbol{\nabla} \phi - \frac{1 - 4(\phi - \phi_0)^2}{\xi} \mathbf{n} \right) \right].
\label{eq:conservative_form}
\end{equation}

This formulation ensures mass conservation and maintains a sharp interface, making it particularly suitable for phase-field simulations in multiphase flows.


\textbf{Navier-Stokes Equations:} According to the phase-field model, the governing equations for incompressible multiphase flows are expressed as \cite{Ding2007, Li2012}:
\begin{subequations}
\begin{equation}
\frac{\partial \rho}{\partial t} + \boldsymbol{\nabla} \cdot (\rho \mathbf{u}) = 0,
\label{eq:continuity}
\end{equation}
\begin{eqnarray}
\rho \left( \frac{\partial \mathbf{u}}{\partial t} + (\mathbf{u} \cdot \boldsymbol{\nabla}) \mathbf{u} \right) &=& -\boldsymbol{\nabla} p + \boldsymbol{\nabla} \cdot 
\left[ \mu \left( \boldsymbol{\nabla} \mathbf{u} + (\boldsymbol{\nabla} \mathbf{u})^T \right) \right] + \mathbf{F}_s + \mathbf{F}_b,
\label{eq:momentum}
\end{eqnarray}
\end{subequations}
where $\rho$ is the fluid density, $\mathbf{u}$ is the velocity vector, $p$ is the macroscopic pressure, $\mu$ is the dynamic viscosity, $\mathbf{F}_s$ is the surface tension force, and $\mathbf{F}_b$ represents the body force.

The surface tension force $\mathbf{F}_s$ is modeled using the chemical potential $\mu_\phi$, and it is expressed as \cite{Jamet2002}:
\begin{equation}
\mathbf{F}_s = \mu_\phi \boldsymbol{\nabla} \phi,
\label{eq:surface_tension_force}
\end{equation}
where $\phi$ is the phase-field variable, and $\mu_\phi$ is the chemical potential given by \cite{JACQMIN2000}:
\begin{equation}
\mu_\phi = 4 \beta \phi \left(1 - \phi\right) \left(\phi - \frac{1}{2} \right) - \kappa \nabla^2 \phi.
\label{eq:chemical_potential}
\end{equation}

The coefficients $\beta$ and $\kappa$ are related to the interfacial thickness $\xi$ and the surface tension $\sigma$ as:
\begin{equation}
\beta = \frac{12 \sigma}{\xi}, \quad \kappa = \frac{3 \sigma \xi}{2}.
\label{eq:beta_kappa_relation}
\end{equation}

Equation~\eqref{eq:continuity} describes the conservation of mass, ensuring that the divergence of the mass flux is zero. Equation~\eqref{eq:momentum} represents the conservation of momentum, accounting for pressure gradients, viscous stresses, surface tension, and body forces. Together, these equations provide a macroscopic representation of multiphase flow dynamics, while the phase-field framework enables the accurate modeling of interface curvature and interfacial forces.

\subsection{Lattice Boltzmann Equations}
\label{sec:lbm_eqns}

\textbf{LBE for hydrodynamics:} The lattice Boltzmann method (LBM) has emerged as an effective computational tool for simulating multiphase flows, offering a kinetic-based approach to capture the complex dynamics of fluid interfaces. The consistency between the LBM and the Navier-Stokes equations, as demonstrated by Chapman-Enskog analysis, ensures its reliability for hydrodynamic simulations \cite{Sukop2006, Huang2015, Kruger2017}. In this study, we utilize the standard form of the lattice Boltzmann equation (LBE) to model hydrodynamic properties, such as velocity and pressure, which is expressed as \cite{Guo2002}:
\begin{equation}
f_a\left(\mathbf{x} + \mathbf{e}_a \delta t,\ t + \delta t\right) = f_a\left(\mathbf{x}, t\right) + \Omega_a(\mathbf{x}, t) + F_a(\mathbf{x}, t),
\label{eq:lbe_standard}
\end{equation}
where $f_a$ represents the hydrodynamic distribution function for incompressible fluids, $\Omega_a$ is the collision operator, and $F_a$ denotes the force term.

For this study, the three-dimensional nineteen-velocity (D3Q19) lattice structure is employed. The discrete lattice velocities $\mathbf{e}_i$ for the D3Q19 model are defined as:
\begin{equation}
\mathbf{e}_i = 
\begin{cases}
(0, 0, 0)c & i = 0, \\
(\pm1, 0, 0)c, (0, \pm1, 0)c, (0, 0, \pm1)c & i = 1-6, \\
(\pm1, \pm1, 0)c, (\pm1, 0, \pm1)c, (0, \pm1, \pm1)c & i = 7-18,
\end{cases}
\label{eq:d3q19_velocities}
\end{equation}
where $c_i = \delta_x / \delta_t$ is the lattice speed, with $\delta_x$ and $\delta_t$ representing the lattice spacing and time step, respectively. For simplicity, $\delta_x$ and $\delta_t$ are typically normalized to unity in lattice units.

The weight coefficients $w_i$ associated with the D3Q19 lattice are:
\begin{equation}
w_i = 
\begin{cases}
12/36 & i = 0, \\
2/36 & i = 1-6, \\
1/36 & i = 7-18.
\end{cases}
\label{eq:weights}
\end{equation}

\figref{fig:d3q19_lattice} illustrates the D3Q19 lattice structure, including its rest velocity and discrete velocity directions.

The collision step in the LBM is critical for simulating fluid interactions. To improve accuracy and stability, Fakhari et al. introduced a modified Crank-Nicholson discretization of the equilibrium distribution function \cite{Fakhari2017}. The modified equilibrium distribution function is expressed as:
\begin{equation}
\bar{f}_a^{eq} = f_a^{eq} - \frac{1}{2} F_a,
\label{eq:equilibrium_modified}
\end{equation}
where $f_a^{eq}$ is the equilibrium distribution function:
\begin{equation}
f_a^{eq} = p^\ast w_a + (\Gamma_a - w_a),
\label{eq:equilibrium}
\end{equation}
and $p^\ast = p / \rho c_s^2$ is the normalized pressure. Also, $\Gamma_a$ is the dimensionless distribution function, given by:
\begin{equation}
\Gamma_a = \omega_a \left[1 + \frac{\mathbf{e}_a \cdot \mathbf{u}}{c_s^2} + \frac{{(\mathbf{e}_a \cdot \mathbf{u})}^2}{2c_s^4} - \frac{\mathbf{u} \cdot \mathbf{u}}{2c_s^2}\right].
\label{eq:dimensionless_distribution}
\end{equation}
Here, $\phi$ is the phase-field variable, $\mathbf{u}$ is the velocity field, and $c_s = c / \sqrt{3}$ is the speed of sound in the lattice unit.
The force term $F_a$ is calculated as:
\begin{equation}
F_a(\mathbf{x}, t) = \delta t \, w_a \, \frac{\mathbf{e}_a \cdot \mathbf{F}}{\rho c_s^2},
\label{eq:force_term}
\end{equation}
where the total force $\mathbf{F}$ is given by:
\begin{equation}
\mathbf{F} = \mathbf{F}_s + \mathbf{F}_b + \mathbf{F}_p + \mathbf{F}_\mu.
\label{eq:total_force}
\end{equation}

Here, $\mathbf{F}_s$ is the surface tension force, $\mathbf{F}_b$ is the body force, $\mathbf{F}_p$ is the pressure force, and $\mathbf{F}_\mu$ represents the viscous force. The pressure and viscous forces are expressed as:
\begin{equation}
\mathbf{F}_p = -p^\ast c_s^2 \boldsymbol{\nabla} \rho,
\label{eq:pressure_force}
\end{equation}
\begin{equation}
\mathbf{F}_\mu = \nu \left[\boldsymbol{\nabla} \mathbf{u} + (\boldsymbol{\nabla} \mathbf{u})^T \right] \cdot \boldsymbol{\nabla} \rho,
\label{eq:viscous_force}
\end{equation}
where $\nu$ is the kinematic viscosity, related to the relaxation time $\tau$ as:
\begin{equation}
\nu = \tau c_s^2 \delta t.
\label{eq:kinematic_viscosity}
\end{equation}

To handle density and viscosity variations in multiphase flows, the relaxation time \(\tau\) is interpolated based on the phase-field variable \(\phi\) as:
\begin{equation}
\tau = \tau_L + (\phi - \phi_L)(\tau_H - \tau_L),
\label{eq:relaxation_time}
\end{equation}
where \(\tau_L\) and \(\tau_H\) are the relaxation times corresponding to the light and heavy fluids, respectively. The relaxation time \(\tau\) is directly related to the kinematic viscosity, which governs the diffusion of momentum within the fluid. Larger relaxation times result in higher viscosities, while smaller relaxation times correspond to lower viscosities, reflecting the properties of the two phases.

Similarly, the dynamic viscosity \(\mu\) is interpolated as:
\begin{equation}
\mu = \mu_L + (\phi - \phi_L)(\mu_H - \mu_L),
\label{eq:dynamic_viscosity}
\end{equation}
where \(\mu_L\) and \(\mu_H\) represent the dynamic viscosities of the light and heavy fluids, respectively. The dynamic viscosity \(\mu\) quantifies the fluid's resistance to deformation under shear stress and is crucial for accurately capturing the momentum transfer across the interface.

The collision operator \(\Omega_a\) plays a critical role in the lattice Boltzmann method and can take various forms depending on the desired balance between simplicity and stability. The simplest form is the single-relaxation-time (SRT) model, also known as the Bhatnagar-Gross-Krook (BGK) model \cite{Bhatnagar1954}, which is expressed as:
\begin{equation}
\Omega_a^{\text{BGK}} = -\frac{f_a - \bar{f}_a^{eq}}{\tau + 1/2},
\label{eq:bgk_collision}
\end{equation}
where \(\tau\) is the relaxation time, and \(\bar{f}_a^{eq}\) is the equilibrium distribution function. While the SRT model is computationally efficient and straightforward, its stability is limited in cases involving high Reynolds numbers, large density or viscosity contrasts, and complex geometries.

To address these limitations, the multi-relaxation-time (MRT) model was developed as an extension of the SRT approach, providing enhanced stability and accuracy by independently relaxing different moments of the distribution function \cite{Lallemand2000}. The MRT collision operator is given by:
\begin{equation}
\Omega_a^{\text{MRT}} = -M^{-1} \hat{S} M (f_a - \bar{f}_a^{eq}),
\label{eq:mrt_collision}
\end{equation}
where \(M\) is the transformation matrix that maps the distribution function from velocity space to moment space, and \(\hat{S}\) denotes a diagonal relaxation matrix for the D3Q19 lattice of the following form:
\begin{eqnarray}
\hat{S}=diag(1,\ 1,\ 1,\ 1 , \ s_\nu,\ s_\nu,\ s_\nu,\ s_\nu,\ s_\nu, 1,1,1,1,1,1,1,1,1,1).
\label{eq:23}
\end{eqnarray} 
Each diagonal element of \(\hat{S}\) corresponds to a specific relaxation rate, allowing precise control over the dissipation of different modes. By separating hydrodynamic moments, such as density and momentum, from non-hydrodynamic moments, the MRT model reduces unwanted couplings that can lead to numerical artifacts, particularly near interfaces or in flows with high Reynolds numbers. Furthermore, the independent tuning of relaxation rates enables the MRT model to effectively handle large density and viscosity contrasts, while mitigating issues related to insufficient dissipation of higher-order moments, a common drawback of the SRT approach.

Building on the MRT framework, the weighted multi-relaxation-time (WMRT) model introduces a non-uniform weight function to orthogonalize the moments, further improving numerical stability and reducing spurious couplings \cite{Dellar2002, Geier2015}. This enhancement makes the WMRT model particularly effective for simulating complex multiphase flows, such as those involving sharp interfaces, high-density ratios, and intricate geometries. Applications of the WMRT model include liquid jet breakup and interfacial instability problems, where maintaining stability and accuracy under challenging conditions is crucial.

LBM offers numerous advantages, particularly in its ability to compute gradients using the moments of the distribution function rather than traditional finite difference schemes. This enables efficient and local computation of the deviatoric stress tensor, making the method highly suitable for implementation on GPUs and other parallel architectures. In the context of multiphase flow simulations, the viscous force is a critical component that can be calculated locally. For the BGK model, the viscous force is computed as:
\begin{equation}
F_{\mu,i}^{\text{BGK}} = -\frac{\nu}{(\tau+1/2)c_s^2\delta t} 
\left[\sum_a e_{ai} e_{aj} (f_a - f_a^{eq})\right] 
\frac{\partial \rho}{\partial x_j},
\label{eq:viscous_force_bgk}
\end{equation}
where \(\nu\) is the kinematic viscosity, \(\tau\) is the relaxation time, \(c_s\) is the lattice speed of sound, and \(f_a^{eq}\) is the equilibrium distribution function. For the WMRT model, the viscous force calculation is further enhanced by leveraging the multi-relaxation-time framework:
\begin{eqnarray}
F_{\mu,i}^{\text{WMRT}} &=& -\frac{\nu}{c_s^2 \delta t} 
\left[\sum_\beta e_{\beta i} e_{\beta j} 
\sum_a (M^{-1} \hat{S} M)_{\beta \alpha} (f_a - f_a^{eq}) \right] 
\frac{\partial \rho}{\partial x_j}.
\label{eq:viscous_force_wmrt}
\end{eqnarray}
Here, \(M\) is the transformation matrix that maps the distribution function to moment space, and \(\hat{S}\) is the diagonal relaxation matrix containing relaxation rates for individual moments.

The LBM framework employs the standard streaming-collision process to compute fluid properties. After the collision step, the normalized pressure is updated as:
\begin{subequations}
\begin{eqnarray}
p^\ast &=& \sum_a f_a, \label{eq:normalized_pressure} \\
\mathbf{u} &=& \sum_a f_a \mathbf{e}_a + \frac{\mathbf{F}}{2 \rho} \delta t, 
\label{eq:velocity_update}
\end{eqnarray}
\end{subequations}
where \(p^\ast\) is the normalized pressure, \(\mathbf{u}\) is the velocity vector, and \(\mathbf{F}\) represents the total force acting on the fluid.

It is essential to note that the velocity update must be performed after the pressure calculation to ensure consistency and numerical stability. Additionally, the density gradient can be expressed in terms of the phase-field variable gradient as:
\begin{equation}
\boldsymbol{\nabla} \rho = (\rho_H - \rho_L) \boldsymbol{\nabla} \phi,
\label{eq:density_gradient_phase_field}
\end{equation}
where \(\rho_H\) and \(\rho_L\) are the densities of the heavy and light fluids, respectively, and \(\phi\) is the phase-field variable.

This formulation ensures accurate representation of multiphase dynamics, facilitating efficient and stable simulations even in scenarios with sharp interfaces and high-density contrasts.

\vspace{1em}

\textbf{LBE for interface tracking:} We employed the following lattice Boltzmann equation (LBE) to determine the interface between fluid phases \cite{Geier2015F}:
\begin{eqnarray}
g_a\left(\mathbf{x} + \mathbf{e}_a \delta t, t + \delta t\right) = g_a\left(\mathbf{x}, t\right) - \frac{g_a\left(\mathbf{x}, t\right) - \bar{g}_a^{eq}\left(\mathbf{x}, t\right)}{\tau_\phi + 1/2} + F_a^\phi(\mathbf{x}, t),
\label{eq:lbe_interface}
\end{eqnarray}
where $g_a$ represents the phase-field distribution function, $\tau_\phi$ is the dimensionless phase-field relaxation time, and $F_a^\phi$ is the forcing term. The forcing term is expressed as:
\begin{equation}
F_a^\phi\left(\mathbf{x}, t\right) = \delta t \frac{\left[1 - 4{(\phi - \phi_0)}^2\right]}{\xi} \omega_a \mathbf{e}_a \cdot \frac{\boldsymbol{\nabla} \phi}{|\boldsymbol{\nabla} \phi|},
\label{eq:forcing_term}
\end{equation}
where $\mathbf{e}_a$ and $\omega_a$ are the weight coefficients and the mesoscopic velocity set, respectively, as defined in Eqs.~\eqref{eq:d3q19_velocities} and~\eqref{eq:weights}. 

The equilibrium distribution function for the phase-field is defined as:
\begin{equation}
\bar{g}_a^{eq} = g_a^{eq} - \frac{1}{2} F_a^\phi,
\label{eq:equilibrium_with_force}
\end{equation}
where
\begin{equation}
g_a^{eq} = \phi \Gamma_a,
\label{eq:equilibrium_function}
\end{equation}
and $\Gamma_a$ is the dimensionless distribution function, given in Eq. \eqref{eq:dimensionless_distribution}.

The phase-field relaxation time $\tau_\phi$ is related to the mobility parameter $M$ by:
\begin{equation}
M = \tau_\phi c_s^2 \delta t.
\label{eq:mobility_relation}
\end{equation}
After completing the streaming-collision process, the phase-field variable $\phi$ is obtained by taking the zeroth-order moment of the distribution function:
\begin{equation}
\phi = \sum_a g_a.
\label{eq:phase_field_variable}
\end{equation}

Using the phase-field variable, the fluid density $\rho$ can be interpolated linearly as:
\begin{equation}
\rho = \rho_L + (\phi - \phi_L)(\rho_H - \rho_L),
\label{eq:density_interpolation}
\end{equation}
where $\rho_L$ and $\rho_H$ are the densities of the light and heavy fluids, respectively, and $\phi_L$ and $\phi_H$ correspond to their phase-field values.

The phase-field variable $\phi$ is the only nonlocal macroscopic variable in this model, which enhances computational parallelism. Gradients and Laplacians of $\phi$ are computed using second-order isotropic central differences \cite{Kumar2004, Mattila2014}:
\begin{subequations}
\begin{eqnarray}
\boldsymbol{\nabla} \phi &=& \frac{c}{c_s^2 \delta x} \sum_a \mathbf{e}_a \omega_a \phi\left(\mathbf{x} + \mathbf{e}_a \delta t, t\right),
\label{eq:gradient_phi}\\
\boldsymbol{\nabla}^2 \phi &=& \frac{2 c^2}{c_s^2 (\delta x)^2} \sum_a \omega_a \left[\phi\left(\mathbf{x} + \mathbf{e}_a \delta t, t\right) - \phi\left(\mathbf{x}, t\right)\right]. \qquad
\label{eq:laplacian_phi}
\end{eqnarray}
\end{subequations}

This framework ensures accurate modeling of the interface between fluid phases while maintaining computational efficiency through the LBE formulation.

\input{Figures/lattice}


%% file: Figures/lattice.tex
\begin{figure}[htbp]
    \centering
    \includegraphics[width=0.5\textwidth]{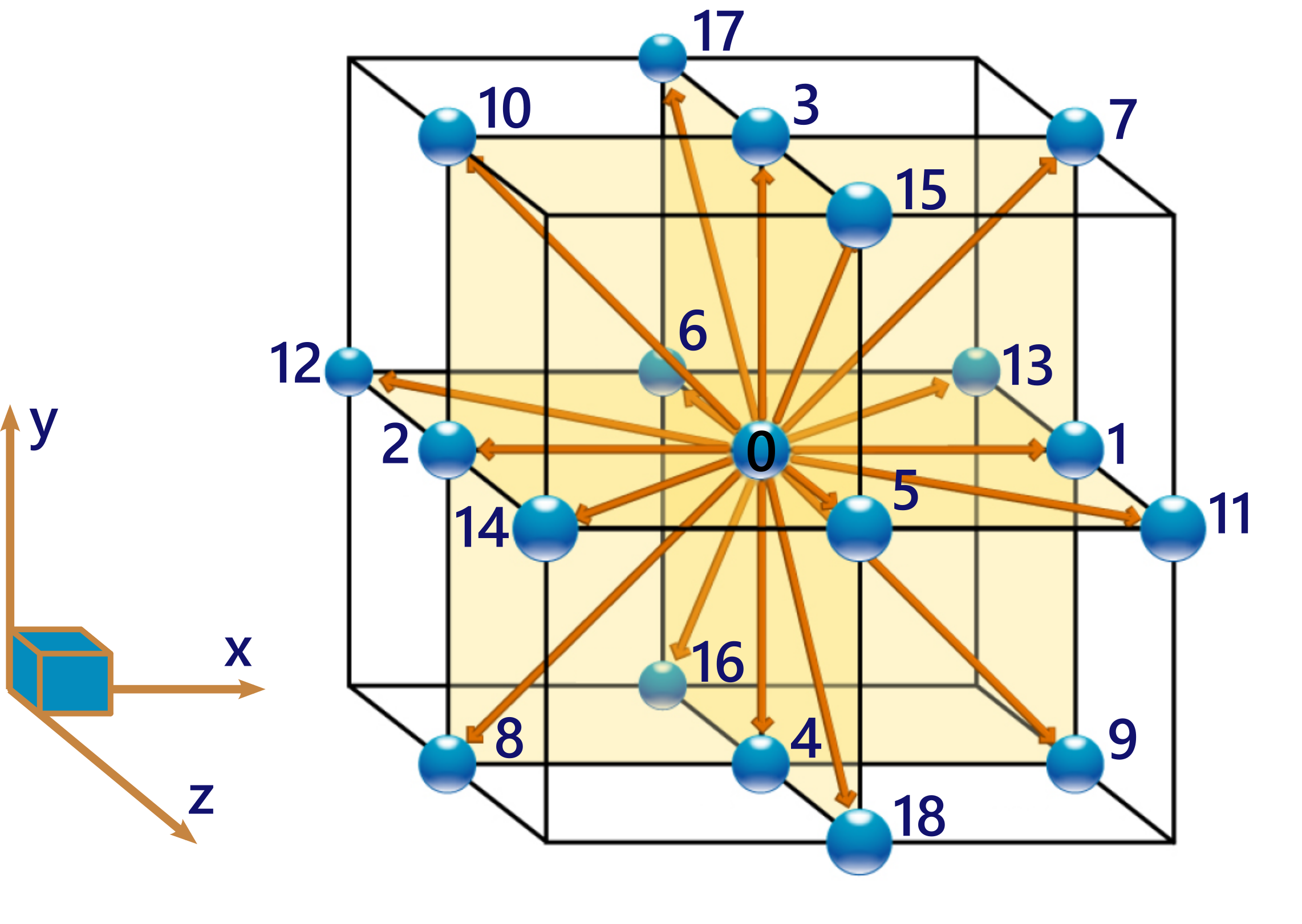} 
    \caption{Three-dimensional 19-velocity (D3Q19) lattice. The diagram illustrates the discrete velocity vectors in the D3Q19 model, including the central rest velocity and velocities along the axes, face diagonals, and space diagonals.}
    \label{fig:d3q19_lattice}
\end{figure}

%% file: Sections/Implementation.tex
\section{Implementation}
\label{sec:implementation}
The CUDA platform was utilized in this study by implementing an NVIDIA GPU to accelerate computations. A key distinction between traditional central processing units (CPUs) and graphics processing units (GPUs) lies in their architectural design. While a CPU typically has four to eight arithmetic logic units (ALUs), a GPU is equipped with thousands of ALUs, enabling it to efficiently handle massive parallel computations. This architectural advantage makes GPUs particularly suitable for computationally intensive tasks, such as fluid dynamics simulations.

As illustrated in \figref{fig:cuda_memory_model}, GPU threads can access a variety of memory types, including global memory, shared memory, registers, local memory, and constant memory. Global memory, which is relatively large in size, is accessible to all threads on the GPU as well as the host (CPU). However, global memory access is slower compared to other memory types.

Threads within a block can access shared memory, which is limited in size but offers significantly faster access speeds than global memory. Each thread also has access to a small number of discrete registers and a minimal amount of local memory for temporary data storage. Additionally, constant memory, located in device memory, allows the host to write and GPU threads to read. Due to its read-only nature and high access efficiency, constant memory is ideal for storing frequently accessed parameters.

These memory hierarchies are critical in achieving optimal performance in GPU-based simulations. The flexibility and programmability of GPUs have led to growing interest in their application to fluid dynamics simulations, enabling researchers to achieve substantial improvements in computational efficiency \cite{Li2012GPU}.

\input{Figures/gpu}

\subsection{Algorithm}

The implementation of the lattice Boltzmann method (LBM) involves three main computational procedures: collision, streaming, and updating macroscopic parameters. The collision operators (Algorithm~\ref{alg:collision}) handle the phase-field and hydrodynamic distribution functions by applying equilibrium and forcing terms to ensure accurate modeling of interfacial dynamics and flow properties. The streaming operator (Algorithm~\ref{alg:streaming}) facilitates data propagation across the lattice by shifting distribution functions to neighboring nodes based on their respective velocities. Finally, macroscopic parameters such as density, velocity, and pressure are updated using the distribution functions (Algorithm~\ref{alg:macroscopic}). These procedures collectively enable efficient and accurate simulation of multiphase flow phenomena while leveraging the parallel computational capabilities of GPU architectures.

\figref{fig:cpu_gpu_memory} illustrates the procedural differences between sequential program execution on the CPU and the parallelized implementation on the GPU. \figref{fig:cpu_gpu_comparison} highlights how GPU-based parallel computing enables simultaneous processing of computational nodes for collision, streaming, boundary conditions, and macroscopic variables, significantly improving efficiency compared to the sequential CPU-only approach. \figref{fig:memory_layout} demonstrates the memory layout transformation, where 2D array elements are placed into a 1D linear addressed system memory, optimizing memory access for GPU threads. This unified approach of parallelization and optimized memory handling further enhances the performance of the LBM framework for large-scale multiphase flow simulations.

\input{Figures/gpuVScpu}

\input{Figures/collisionAlg}

\input{Figures/streamingAlg}

\input{Figures/macroscopicAlg}

\subsection{Performance Analysis}

The performance of the lattice Boltzmann method (LBM) implementations was evaluated on different hardware configurations, including CPUs and GPUs, using the same algorithm across all platforms. For OpenMP implementations, the thread-based parallelism of the GPU was replaced with parallel \texttt{for} loops to distribute computations among CPU cores. \figref{fig:execution-breakdown} illustrates the average execution time breakdown for one iteration of a 3D D3Q19 lattice arrangement across mesh sizes ranging from 2 million to 64 million lattice nodes, performed on an NVIDIA A100 GPU. The figure highlights the contribution of individual computational components, such as collision, streaming, boundary conditions, and macroscopic variable updates, with the total execution time annotated on top of each bar. These results emphasize the scalability of the GPU implementation as the mesh size increases.

\figref{fig:gpu-execution-breakdown} further compares the average execution times for a fixed domain of 16 million lattice nodes across different NVIDIA GPUs, including the RTX6000ADA, V100, and A100, showcasing the significant performance improvements achieved with newer GPU architectures. \tabref{tab:cpu_vs_gpu_comparison} complements this analysis by providing a detailed comparison of execution times on an Intel i7-6700K CPU and an NVIDIA GTX 1060 GPU. It includes both sequential CPU computations and parallel OpenMP implementations, highlighting the speedup factors achieved through GPU acceleration and CPU parallelization. The reported times represent the average duration for a single LBM iteration, offering a fair basis for comparison across platforms. This table is particularly helpful for users who wish to run the framework on a standard desktop setup, providing practical insights into expected performance and optimization opportunities without requiring access to high-end computing resources.

\input{Figures/meshTime}

\input{Figures/deviceTime}

\input{Tables/performace}

%% file: Figures/gpu.tex
\begin{figure}[htbp]
    \centering
    \includegraphics[width=0.5\textwidth]{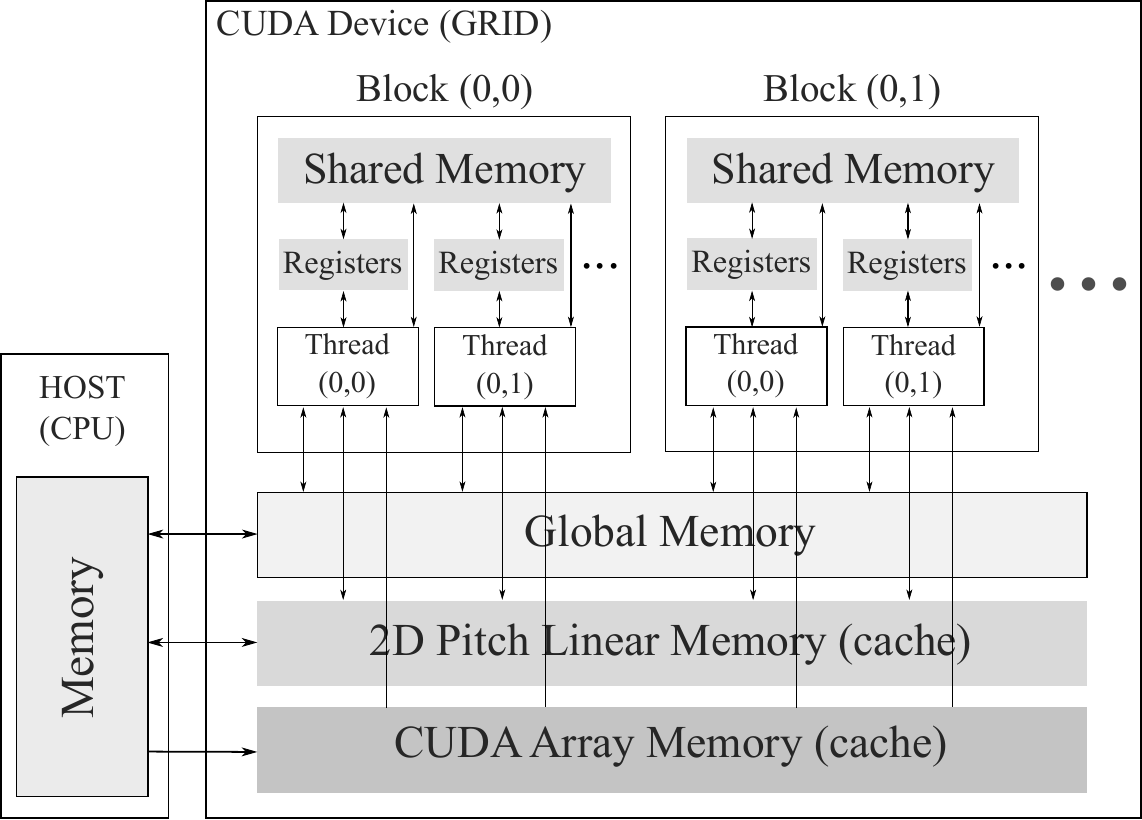} 
    \caption{CUDA device memory model (source from NVIDIA).}
    \label{fig:cuda_memory_model}
\end{figure}

%% file: Figures/gpuVScpu.tex
\begin{figure}[htbp]
    \centering
    \begin{subfigure}[t]{0.48\textwidth} 
        \centering
        \includegraphics[width=\textwidth]{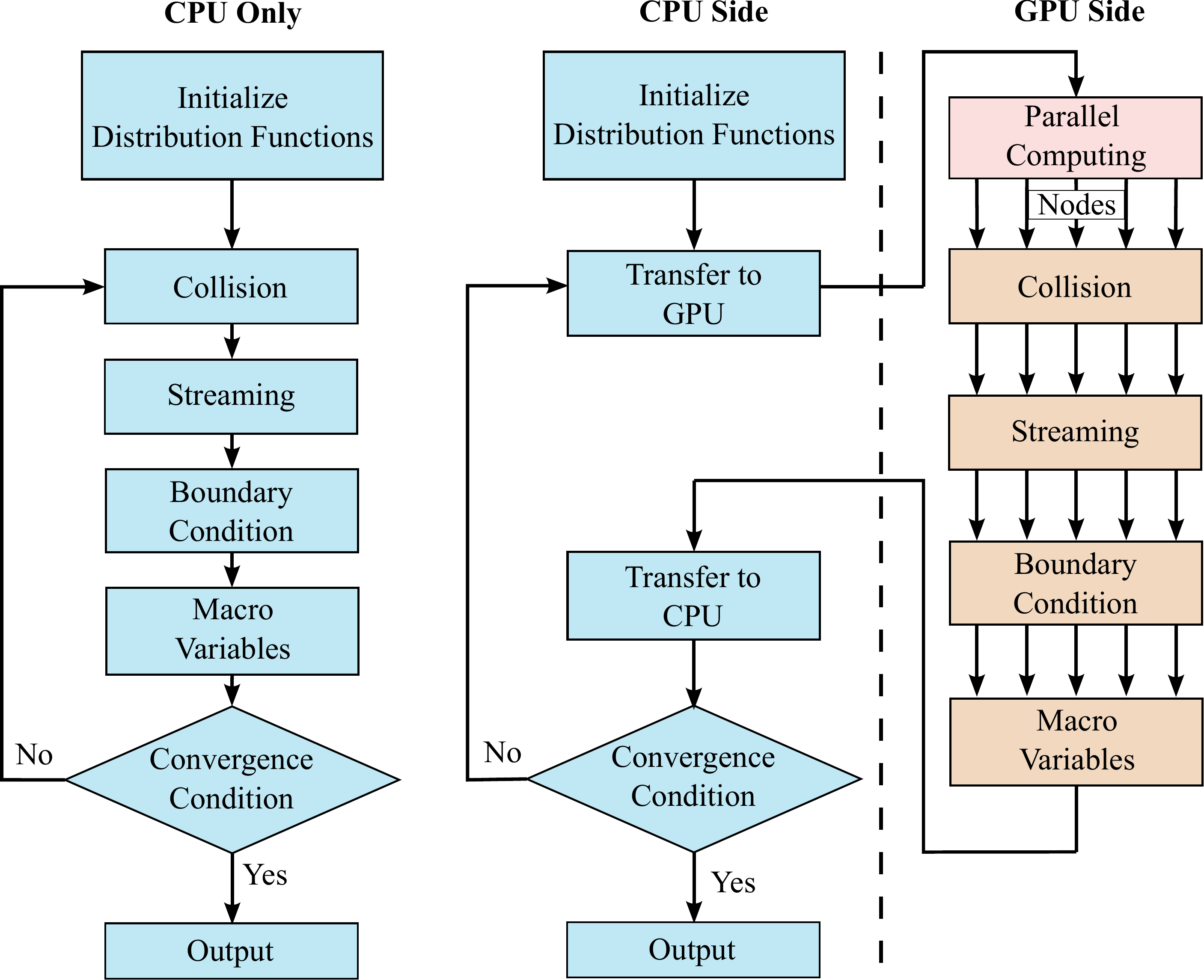} 
        \caption{Flowchart comparison of sequential program execution on CPU and hybrid execution on GPU, showing initialization, collision, streaming, boundary condition application, and macro variable computation. The GPU implementation includes parallelized computing nodes.}
        \label{fig:cpu_gpu_comparison}
    \end{subfigure}
    \hspace{0.02\textwidth} 
    \begin{subfigure}[t]{0.48\textwidth} 
        \centering
        \includegraphics[width=\textwidth]{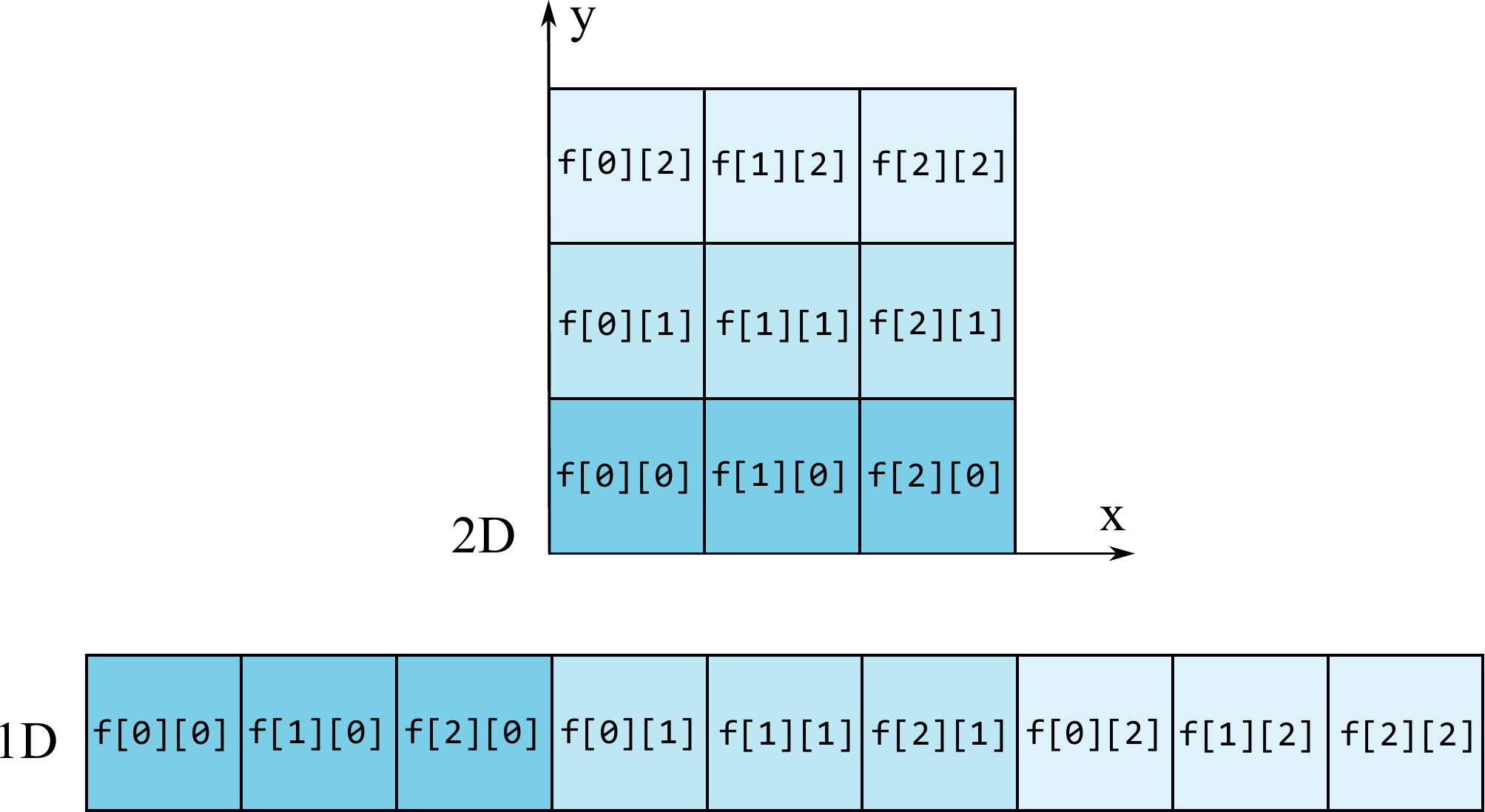} 
        \caption{Mapping of a 2D array into a 1D linear memory structure for GPU processing. Each 2D array element is accessed sequentially in a 1D address space to optimize memory access patterns.}
        \label{fig:memory_layout}
    \end{subfigure}
    \caption{(a) Comparative flowchart illustrating program execution steps on CPU versus GPU, emphasizing the advantages of GPU parallelization. (b) Conversion of a 2D array into a 1D memory layout for efficient GPU memory addressing.}
    \label{fig:cpu_gpu_memory}
\end{figure}

%% file: Figures/collisionAlg.tex
\begin{algorithm}[htbp]
\scriptsize
\caption{Collision Operators for Phase-Field and Hydrodynamic Distribution Functions}
\label{alg:collision}
\begin{algorithmic}[1]
\Require Phase-field distribution function $\textbf{g}$, hydrodynamic distribution function $\textbf{f}$, and relevant macroscopic variables
\Procedure{Phase-Field Collision}{} \Comment{Interface Tracking}
    \State \texttt{x, y, z} $\gets$ ThreadIndex.x, ThreadIndex.y, ThreadIndex.z
    \State \texttt{index} $\gets$ \Call{getIndex}{x, y, z}
    \If{\Call{IsActive}{index}}
        \For{each velocity direction $a$ in the lattice}
            \State \texttt{gEq} $\gets$ \texttt{$\phi$(index)} $\times$ \Call{ComputeEquilibrium}{$u$, $v$, $w$, $a$} 
            \Comment{Eqs.~\eqref{eq:dimensionless_distribution},~\eqref{eq:equilibrium_function}}
            \State \texttt{forcingTerm} $\gets$ \Call{ComputeForceTerm}{$\phi$, index, $a$} 
            \Comment{Eq.~\eqref{eq:forcing_term}}
            \State \texttt{gEqBar} $\gets$ \texttt{gEq} $- 0.5 \times$ \texttt{forcingTerm}
            \Comment{Eq.~\eqref{eq:equilibrium_with_force}}
            \State Update \texttt{g[index][a]} using Eq.~\eqref{eq:lbe_interface}
        \EndFor
    \EndIf
\EndProcedure

\Procedure{Hydrodynamic Collision}{} \Comment{Momentum and Mass Conservation}
    \State \texttt{x, y, z} $\gets$ ThreadIndex.x, ThreadIndex.y, ThreadIndex.z
    \State \texttt{index} $\gets$ \Call{getIndex}{x, y, z}
    \If{\Call{IsActive}{index}}
        \For{each velocity direction $a$ in the lattice}
            \State \texttt{fEqNoDim} $\gets$ \Call{ComputeEquilibrium}{$u$, $v$, $w$, $a$} 
            \Comment{Eq.~\eqref{eq:dimensionless_distribution}}
            \State \texttt{fEq[$a$]} $\gets$ \texttt{$p$(index)} $\times$ \texttt{wa[$a$]} $+$ \texttt{(fEqNoDim $-$ wa[$a$])}
            \Comment{Eq.~\eqref{eq:equilibrium}}
        \EndFor
        \State \texttt{relaxationTime} $\gets$ \Call{ComputeRelaxationTime}{$\phi$, $\tau_L$, $\tau_H$, index} 
        \Comment{Eq.~\eqref{eq:relaxation_time}}
        \State \texttt{viscousForce} $\gets$ \Call{ComputeViscousForce}{$\textbf{f}$, $\textbf{fEq}$, relaxationTime, index}
        \Comment{Eq.~\eqref{eq:viscous_force_wmrt}}
        \State \texttt{pressureForce} $\gets$ \Call{ComputePressureForce}{$p$, $\rho$, index}
        \Comment{Eq.~\eqref{eq:pressure_force}}
        \State \texttt{totalForce} $\gets$ \texttt{viscousForce $+$ surfaceForce $+$ pressureForce $+$ bodyForce}
        \Comment{Eq.~\eqref{eq:total_force}}
        \For{each velocity direction $a$ in the lattice}
            \State \texttt{effectiveForce} $\gets$ \texttt{totalForce.x $\times$ ex[$a$] $+$ totalForce.y $\times$ ey[$a$] $+$ totalForce.z $\times$ ez[$a$]}
            \Comment{Eq.~\eqref{eq:d3q19_velocities}}
            \State \texttt{forcingTerm[$a$]} $\gets$ \Call{ComputeForcingTerm}{effectiveForce}
            \Comment{Eq.~\eqref{eq:force_term}}
            \State \texttt{fEqBar[$a$]} $\gets$ \texttt{fEq[$a$] $- 0.5 \times$ forcingTerm[$a$]}
            \Comment{Eq.~\eqref{eq:equilibrium_modified}}
        \EndFor
        \State \texttt{collisionResult} $\gets$ \Call{ApplyCollision}{$\textbf{f}$, $\textbf{fEqBar}$}
        \Comment{Eq.~\eqref{eq:mrt_collision}}
        \State Update \texttt{f} using Eq.~\eqref{eq:lbe_standard}
    \EndIf
\EndProcedure

\end{algorithmic}
\end{algorithm}

%% file: Figures/streamingAlg.tex


\begin{algorithm}[htbp]
\scriptsize
\caption{Streaming Operators for Distribution Functions}
\label{alg:streaming}
\begin{algorithmic}[1]
\Require This algorithm is applicable to both phase-field ($\textbf{g}$) and hydrodynamic ($\textbf{f}$) distribution functions.

\Procedure{Streaming for Distribution Functions}{} \Comment{\texttt{distFunc} refers to the distribution function (e.g., $\textbf{g}$ or $\textbf{f}$)}
    \State \texttt{x, y, z} $\gets$ \texttt{ThreadIndex.x}, \texttt{ThreadIndex.y}, \texttt{ThreadIndex.z}
    \State \texttt{index} $\gets$ \Call{getIndex}{x, y, z}
    \If{\Call{IsActive}{index}}
        \For{each velocity direction $a$ in the lattice}
            \State \texttt{neighborX} $\gets$ \texttt{x} $-$ \texttt{ex[$a$]}
            \State \texttt{neighborY} $\gets$ \texttt{y} $-$ \texttt{ey[$a$]}
            \State \texttt{neighborZ} $\gets$ \texttt{z} $-$ \texttt{ez[$a$]}
            \State \texttt{distFunc[x, y, z, a]} $\gets$ \texttt{distFunc[neighborX, neighborY, neighborZ, a]}
            \Comment{Stream data from neighbor node}
        \EndFor
    \EndIf
\EndProcedure
\Ensure Updated $\textbf{g}$ and $\textbf{f}$ after streaming step
\end{algorithmic}
\end{algorithm}

%% file: Figures/macroscopicAlg.tex
\begin{algorithm}[t]
\scriptsize
\caption{Updating Macroscopic Parameters}
\label{alg:macroscopic}
\begin{algorithmic}[1]
\Require Phase-field ($\textbf{g}$) and hydrodynamic ($\textbf{f}$) distribution functions

\Procedure{Updating the Phase-Field Variable}{} \Comment{Interface Tracking}
    \State \texttt{x, y, z} $\gets$ \texttt{ThreadIndex.x}, \texttt{ThreadIndex.y}, \texttt{ThreadIndex.z}
    \State \texttt{index} $\gets$ \Call{getIndex}{x, y, z}
    \If{\Call{IsActive}{index}}
        \For{each velocity direction $a$ in the lattice}
            \State \texttt{$\phi$[index]} $\gets$ \texttt{$\phi$[index] + g[index][a]}
            \Comment{Eq. \eqref{eq:phase_field_variable}}
        \EndFor
        \State \texttt{$\rho$[index]} $\gets$ \texttt{$\rho_L$ + $\phi$[index] * ($\rho_H$ - $\rho_L$)}
        \Comment{Eq. \eqref{eq:density_interpolation}}
    \EndIf
\EndProcedure

\Procedure{Updating Hydrodynamic Parameters}{} \Comment{Hydrodynamic Calculations}
    \State \texttt{x, y, z} $\gets$ \texttt{ThreadIndex.x}, \texttt{ThreadIndex.y}, \texttt{ThreadIndex.z}
    \State \texttt{index} $\gets$ \Call{getIndex}{x, y, z}
    \If{\Call{IsActive}{index}}
        \For{each velocity direction $a$ in the lattice}
            \State \texttt{$p$[index]} $\gets$ \texttt{$p$[index] + f[index][a]}
            \Comment{Eq. \eqref{eq:normalized_pressure}}
            \State \texttt{fEqNoDim} $\gets$ \Call{EquilibriumNew}{u, v, w, a} 
            \Comment{Eq. \eqref{eq:dimensionless_distribution}}
            \State \texttt{fEq[a]} $\gets$ \texttt{$p$(index) * wa[a] + (fEqNoDim - wa[a])}
            \Comment{Eq. \eqref{eq:equilibrium}}
        \EndFor
        \State \texttt{relaxationTime} $\gets$ \Call{ComputeRelaxationTime}{$\phi$, $\tau_L$, $\tau_H$, index}
        \Comment{Eq. \eqref{eq:relaxation_time}}
        \State \texttt{viscousForce} $\gets$ \Call{ComputeViscousForce}{\textbf{f}, \textbf{fEq}, relaxationTime, index} 
        \Comment{Eq. \eqref{eq:viscous_force_wmrt}}
        \State \texttt{pressureForce} $\gets$ \Call{ComputePressureForce}{$p$, $\rho$, index} 
        \Comment{Eq. \eqref{eq:pressure_force}}
        \State \texttt{totalForce} $\gets$ \texttt{viscousForce + surfaceForce + pressureForce + bodyForce}
        \Comment{Eq. \eqref{eq:total_force}}
        
        \State \texttt{$u_x$, $u_y$, $u_z$} $\gets$ 0
        \For{each velocity direction $a$ in the lattice}
            \State \texttt{$u_x$} $\gets$ \texttt{f[index][ex[i]]}
            \State \texttt{$u_y$} $\gets$ \texttt{f[index][ey[i]]}
            \State \texttt{$u_z$} $\gets$ \texttt{f[index][ez[i]]}    
            \Comment{Eq. \eqref{eq:d3q19_velocities}}
        \EndFor
        \State \texttt{$u_x$} $\gets$ \texttt{$u_x$ + 0.5 * totalForce\_x / $\rho$[index]}
        \State \texttt{$u_y$} $\gets$ \texttt{$u_y$ + 0.5 * totalForce\_y / $\rho$[index]}
        \State \texttt{$u_z$} $\gets$ \texttt{$u_z$ + 0.5 * totalForce\_z / $\rho$[index]}
        \Comment{Eq. \eqref{eq:velocity_update}}
    \EndIf
\EndProcedure

\Ensure Updated $u_x, u_y, u_z, p, \phi$
\end{algorithmic}
\end{algorithm}

%% file: Figures/meshTime.tex
\begin{figure}[htbp]
  \centering
  \begin{tikzpicture}[scale=0.9]
  \begin{axis}[
  ybar stacked, bar width=10pt,    
  xlabel={Mesh size (Million lattice nodes) $\rightarrow$},
  ylabel={\small{Execution time (s) $\rightarrow$}},
  symbolic x coords={2, 4, 8, 16, 32, 64},
  xtick = data, 
  ymin=0,
ymax=0.25,
  width=10cm, height=8cm, 
  x tick label style={font=\small},
  legend pos=north west,
  legend style={at={(0.5,1.1)}, anchor=south, legend columns=4, row sep=0.5em,column sep=0.5em},
  grid=major
  ]

  \addplot [fill=div_d1]  table[x=Mesh, y=Collision, col sep=comma] {Images/meshTime.txt};
  \addplot [fill=div_d2]  table[x=Mesh, y=Streaming, col sep=comma] {Images/meshTime.txt};
  \addplot [fill=div_d3]  table[x=Mesh, y=BoundaryCondition, col sep=comma] {Images/meshTime.txt};
  \addplot [fill=div_d4]  table[x=Mesh, y=Macroscopic, col sep=comma] {Images/meshTime.txt};

\node[above] at (axis cs:2, 0.0056) {\small 0.0053};
  \node[above] at (axis cs:4, 0.012) {\small 0.0105};
    \node[above] at (axis cs:8, 0.023) {\small 0.0211};
  \node[above] at (axis cs:16, 0.048) {\small 0.0455};
    \node[above] at (axis cs:32, 0.095) {\small 0.0927};
        \node[above] at (axis cs:64, 0.19) {\small 0.1840};

  \legend{\small{Collision}, \small{Streaming}, \small{Boundary Condition}, \small{Macroscopic Variables}}

  \end{axis}
  \end{tikzpicture}
  \caption{\textit{Average execution time breakdown for one iteration on a 3D D3Q19 lattice arrangement}: Execution times for lattice Boltzmann simulation procedures (Collision, Streaming, Boundary Condition, and Macroscopic Variables) across mesh sizes ranging from 2M to 64M lattice nodes. The simulations were conducted on an NVIDIA A100 GPU. The total execution time for each GPU is annotated on top of the respective bar.}
  \label{fig:execution-breakdown}
\end{figure}
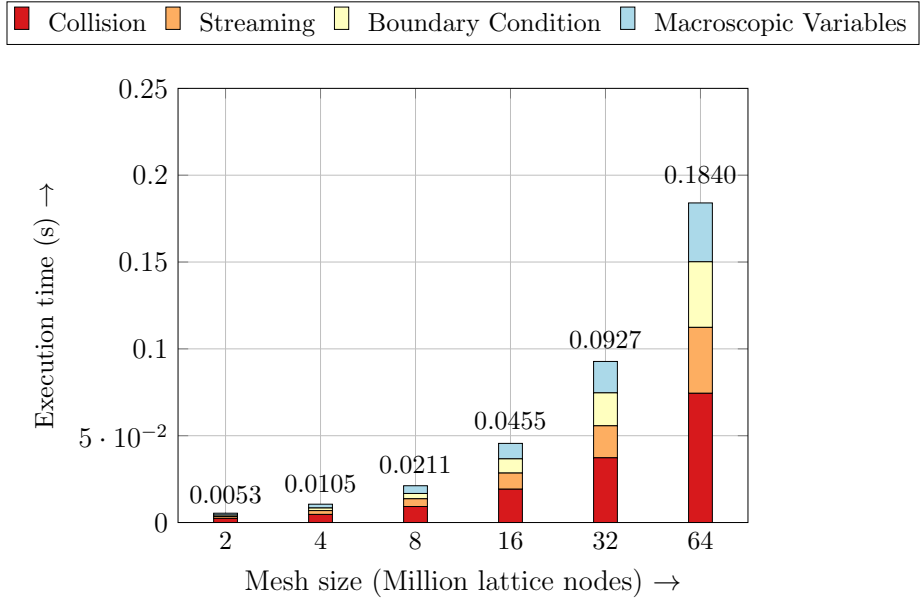

%% file: Figures/deviceTime.tex
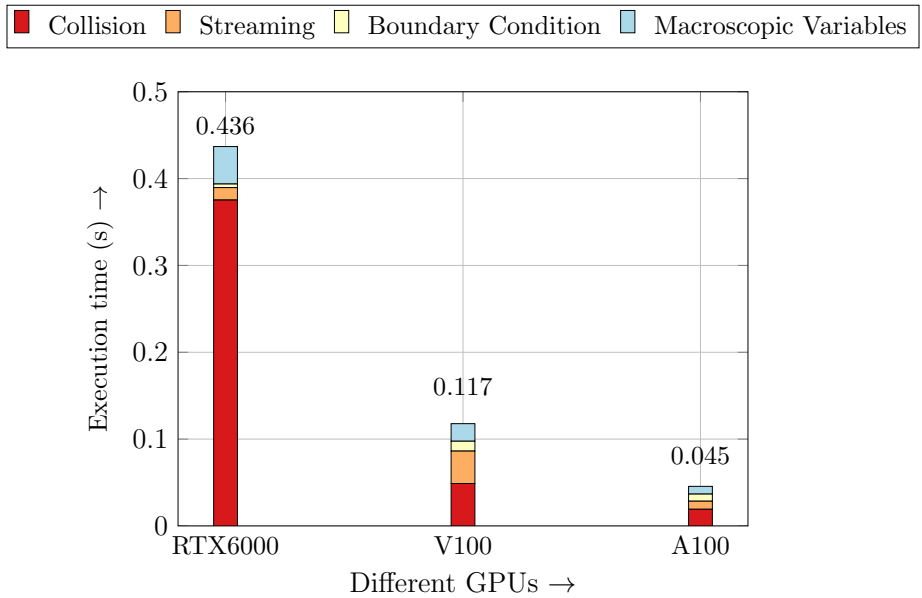
\begin{figure}[htbp]
 \centering
  \begin{tikzpicture}[scale=0.9]
    \begin{axis}[
  ybar stacked, bar width=10pt,    
  xlabel={Different GPUs $\rightarrow$},
  ylabel={\small{Execution time (s) $\rightarrow$}},
  symbolic x coords={RTX6000, V100, A100},
  xtick = data, 
  ymin=0,
  ymax=0.5,
  width=10cm, height=8cm, 
  x tick label style={font=\small},
  legend pos=north west,
  legend style={at={(0.5,1.1)}, anchor=south, legend columns=4, row sep=0.5em,column sep=0.5em},
  grid=major
  ]

  \addplot [fill=div_d1]  table[x=Device, y=Collision, col sep=comma] {Images/deviceTime.txt};
  \addplot [fill=div_d2]   table[x=Device, y=Streaming, col sep=comma] {Images/deviceTime.txt};
  \addplot [fill=div_d3]   table[x=Device, y=BoundaryCondition, col sep=comma] {Images/deviceTime.txt};
  \addplot [fill=div_d4]   table[x=Device, y=Macroscopic, col sep=comma] {Images/deviceTime.txt};

  \node[above] at (axis cs:RTX6000, 0.44) {\small 0.436};
  \node[above] at (axis cs:V100, 0.14) {\small 0.117};
  \node[above] at (axis cs:A100, 0.06) {\small 0.045};

  \legend{\small{Collision}, \small{Streaming}, \small{Boundary Condition}, \small{Macroscopic Variables}}

  \end{axis}
  \end{tikzpicture}
\caption{\textit{Average execution time breakdown for one iteration on a 3D D3Q19 lattice arrangement}: The figure illustrates the average execution times for various lattice Boltzmann simulation procedures (Collision, Streaming, Boundary Condition, and Macroscopic Variables) for a domain of 16 million lattice nodes on different NVIDIA GPUs, including RTX6000ADA, V100, and A100. The total execution time for each GPU is annotated on top of the respective bar.}

  \label{fig:gpu-execution-breakdown}
\end{figure}

%% file: Tables/performace.tex
\begin{table}[ht!]
\centering
\scriptsize
\renewcommand{\arraystretch}{1.5} 
\setlength{\tabcolsep}{5pt} 
\caption{Performance comparison of LBM implementations for different domain sizes on an Intel i7-6700K CPU and an NVIDIA GTX 1060 GPU. The table reports the average execution time per iteration for both sequential CPU computations and parallel OpenMP implementations, highlighting the speedup factors achieved through GPU acceleration and CPU parallelization.}

\label{tab:cpu_vs_gpu_comparison}
\begin{tabular}{l c c c}
\hline
\multirow{2}{*}{\textbf{Domain Size}} & \textbf{CPU Time (s)} & \textbf{GPU Time (s)} & \textbf{Speedup Factor - GPU} \\
                                      & \textbf{CPU Time (s) - OpenMP }               &       &     \textbf{Speedup Factor - OpenMP}                    \\
\hline
\multirow{2}{*}{$256 \times 64$ (D2Q9)} & {$0.4453$} & \multirow{2}{*}{$0.0094$} & {$48\times$} \\
                                        &     $0.0685$                     &                          &   {$7\times$}                       \\
\hline
\multirow{2}{*}{$256 \times 64 \times 64$ (D3Q19)} & {$102.1256$} & \multirow{2}{*}{$1.1875$} & {$86\times$} \\
                                                  &      $15.7115$                    &                          &      {$7\times$}                    \\
\hline
\end{tabular}
\end{table}

%% file: Sections/Validation.tex
\section{Validation}
\label{sec:validation}

The accuracy and stability of the proposed Lattice Boltzmann Method (LBM) model are assessed in this section using several benchmark cases. For 2D problems, the D2Q9 lattice model is utilized, while the D3Q19 lattice model is employed for 3D simulations. Gradients and Laplacians are approximated using second-order, isotropic centered differences, as described in Eqs.~\eqref{eq:gradient_phi} and \eqref{eq:laplacian_phi}. Additionally, \ref{sec:appendixB} provides a detailed discussion on the treatment of common boundary conditions, ensuring a comprehensive evaluation of the model's performance.

\subsection{Capillary Wave}
To validate the LBM simulations of two-phase flow, the dynamic behavior of capillary waves at the interface between two immiscible fluids is analyzed. In this study, a sinusoidal perturbation with a small amplitude $\eta_0$ and wave number $k$ is introduced to the initially quiescent interface. This setup serves as a rigorous test for the LBM framework, as it has a well-established analytical solution for cases where the two fluids have identical kinematic viscosities $\nu$ but differing densities. The temporal evolution of the interface amplitude $\eta(t)$ is used as a benchmark for the simulations. The analytical expression for the decay of the wave amplitude, $\eta(t)$, is given by~\cite{Prosperetti19811217}:

\begin{align}
\frac{\eta(t)}{\eta_0} &= \frac{4(1-4\gamma)\nu^2 k^4}{8(1-4\gamma)\nu^2 k^4 + \omega_0} \, \text{erfc}(\sqrt{\nu k^2 t}) + \sum_{i=1}^{4} \frac{z_i}{Z_i} \frac{\omega_0^2}{z_i^2 - \nu k^2} e^{(z_i^2 - \nu k^2)t} \, \text{erfc}(z_i\sqrt{\nu t}),
\label{eq:decay_wave_amplitude}
\end{align}
where $\omega_0 = \sqrt{\frac{\sigma k^3}{\rho_H + \rho_L}}$ is the angular frequency, $\gamma = \frac{\rho_H \rho_L}{(\rho_H + \rho_L)^2}$, and $Z_i = \prod_{\substack{1 \le j \le 4 \\ j \ne i}} (z_j - z_i)$. The evaluation of the complementary error function $\text{erfc}(z_i)$ is carried out by solving the following algebraic equation:

\begin{equation}
z^4 - 4\gamma \sqrt{\nu k^2} z^3 + 2 (1 - 6\gamma) \nu k^2 z^2 + 4 (1 - 3\gamma) (\nu k^2)^{3/2} z + (1 - 4\gamma) \nu k^2 + \omega_0^2 = 0.
\label{eq:algebraic_equation}
\end{equation}

\figref{fig:capillary_setup} illustrates the physical setup, where a lighter fluid with density $\rho_L$ overlays a heavier fluid with density $\rho_H$. The initial interface is described as $y = L + \eta_0 \cos(2 \pi x)$, where $\eta_0$ is the initial perturbation amplitude. The decay of this sinusoidal profile to a flat interface, influenced by viscosity and surface tension in the absence of external forces like gravity, provides a critical benchmark for validating the LBM framework.
\input{Figures/capillarySetup}

The computational domain is discretized into a $256 \times 512$ grid of lattice nodes. Free-slip boundary conditions are applied along the wave propagation direction, while no-slip conditions are imposed at the top and bottom boundaries. The parameters for the simulation are set as follows: $\eta_0 = 0.02$, $\sigma = 10^{-4}$, $\xi = 4$, and $M_\phi = 0.02$. Since the interface may not align perfectly with the grid points, the interface displacement $\eta(t)$ is interpolated from the phase-field variable $\phi$ using the relationship:

\begin{equation}
\eta(t) = y - \frac{\phi(x_{L0/2}, y)}{\phi(x_{L0/2}, y) - \phi(x_{L0/2}, y - 1)}, \quad \phi(x_{L0/2}, y)\phi(x_{L0/2}, y - 1) < 0.
\end{equation}

Both the length ($\eta$) and time scales ($t$) are normalized using the initial amplitude $\eta_0$ and the angular frequency $\omega_0$, respectively, defined as $\eta^* = \eta/\eta_0$ and $t^* = t\omega_0$. The angular frequency plays a crucial role in any wave system and depends on surface tension, viscosity, wave number, and density values. In this study, the angular frequency is derived under the assumption that both fluids have identical viscosities, with $\nu$ set to $0.005$ and $0.0005$. The wavelength is consistent with the grid size, $L_0 = 256$.

\input{Figures/capillary}

\figref{fig:capillary} compares the normalized interface amplitude $\eta^*$ as a function of normalized time $t^*$ between the LBM simulation and the analytical solution~\cite{Prosperetti19811217}. \figref{fig:capillary_a} corresponds to $\nu = 0.0005$, and \figref{fig:capillary_b} corresponds to $\nu = 0.005$. In both cases, the LBM results (blue circles) closely align with the analytical results (red line), demonstrating the model's accuracy in capturing the transient dynamics of capillary waves. This validation highlights the robustness of the proposed LBM framework in accurately capturing interfacial wave dynamics while efficiently handling varying viscosity, surface tension, and density ratios. The method’s inherent stability and parallel scalability make it well-suited for large-scale simulations of multiphase flows with complex interfacial interactions.

\subsection{Stationary Drop Test }

The validation of the proposed LBM model was conducted using the well-known static drop test. According to Laplace's law, the surface tension of a drop at equilibrium is proportional to the pressure difference across its interface, expressed as:
\begin{equation}
\Delta p = p_{in} - p_{out} = \frac{\sigma}{R},
\label{eq:laplace_law}
\end{equation}
where $R$ represents the radius of the drop in its equilibrium state and $\sigma$ is the surface tension coefficient. To verify this relationship, a three-dimensional computational domain of $100 \times 100 \times 100$ lattice points was generated, and stationary drops of varying radii were simulated within a periodic boundary domain. ~\tabref{tab:laplace_errors} summarizes the results, confirming the validity of Laplace's law.

The pressure difference $\Delta p$ is plotted against $1/R$ in \figref{fig:laplace_pressure}, along with analytical solutions for three different values of the surface tension coefficient $\sigma$. The results demonstrate an excellent agreement between the numerical and theoretical values, indicating the accuracy of the LBM simulations. Furthermore, \figref{fig:density_profile} presents the density profile and the drop interface after 50,000 iterations for a drop radius of $R=20$. This figure illustrates that the interface width remains confined to approximately 4 to 5 lattice cells, ensuring a sharp and well-defined interface.

It is important to note that several factors, including relaxation rates, density ratios, and interfacial forces, influence the smallest stable drop radius. The phase-field model employed in this study successfully simulated small drops with a radius as small as 5 lattice cells, demonstrating its robustness and adaptability for two-phase flow problems. Moreover, this framework effectively handles large density ratios, ensuring numerical stability and accuracy in scenarios with strong interfacial forces and significant density contrasts.

\input{Tables/dropTest}
\input{Figures/dropTest}

\subsection{Two-phase Poiseuille Flow}

Due to the inherent shear phenomena in Poiseuille flows, these problems serve as sensitive tests to compare numerical algorithms against analytical solutions \cite{Reis2007, Leclaire2012}. As illustrated in \figref{fig:poisueille_setup}, the computational domain consists of a two-dimensional channel with periodic boundary conditions along the horizontal direction and no-slip solid walls at the upper ($y=L$) and lower ($y=0$) boundaries.

The schematic in \figref{fig:poisueille_setup} demonstrates the stratification of fluids: the lighter fluid occupies the bottom half of the channel ($0 \leq y < L/2$) with density $\rho_L$ and viscosity $\mu_L$, while the heavier fluid resides in the top half ($L/2 \leq y \leq L$) with density $\rho_H$ and viscosity $\mu_H$. The flow is driven by a body force $F_b = \rho g \hat{x}$ in the horizontal direction, where $g$ represents the gravitational acceleration. In the absence of surface tension, the Navier-Stokes equation reduces to:
\begin{equation}
\frac{d}{dy}\left(\mu\frac{du_x}{dy}\right)+\rho g=0,
\label{eq:momentum_equation}
\end{equation}
where $u_x$ is the $x$-component of the velocity vector. The initial density and viscosity fields in the domain are given as:
\begin{subequations}\label{eq:initial_fields}
\begin{eqnarray}
&&\rho(y) = \frac{\rho_H + \rho_L}{2} - \frac{\rho_H - \rho_L}{2}\tanh\left(\frac{2y - L}{\xi}\right), \qquad
\label{eq:density_field}\\
&&\mu(y) = \frac{\mu_H + \mu_L}{2} - \frac{\mu_H - \mu_L}{2}\tanh\left(\frac{2y - L}{\xi}\right). \qquad
\label{eq:viscosity_field}
\end{eqnarray}
\end{subequations}

To validate the model, simulations were conducted on a computational domain with 64 vertical grid points. The fixed simulation parameters were $\xi=4\ lu$, $g=10^{-6}\ lu$, and $\tau_H=0.5\ lu$. Two density ratios ($\rho^\ast = 10$ and $\rho^\ast = 1000$) were tested, while the viscosity ratio was held constant at $\mu^\ast = 100$. \figref{fig:poiseuille_validation} presents the results, where the maximum velocity of the finite difference (FD) solution is used to normalize the velocity profiles. The results confirm excellent agreement between the LBM phase-field model and the second-order FD solutions for both density ratios.

To further evaluate the phase-field LBM model's accuracy, a grid convergence study was performed. The $L_2$-norm of the numerical error was calculated as:
\begin{equation}
\parallel\delta u\parallel_2 = \sqrt{\frac{\sum_{y}{(u_x - u_x^{FD})}^2}{\sum_{y}{(u_x^{FD})}^2}},
\label{eq:l2_norm_error}
\end{equation}
and plotted against the number of grid points in \figref{fig:poiseuille_accuracy} for $\rho^\ast = 1000$ and $\mu^\ast = 100$ ($\tau_L=0.5\ lu$). The convergence rate is observed to be approximately second order. 

The simulations employed a halfway bounce-back strategy \cite{Ladd1994} for no-slip boundary conditions. This approach does not require explicit solid nodes to represent wall boundaries, enabling the modeling of solid bodies with minimal lattice widths \cite{Kruger2017}, which enhances computational efficiency. Additionally, the halfway bounce-back strategy allows particles to return to the computational domain within one time step $\Delta t$, unlike the full bounce-back method, which requires $2\Delta t$. This not only reduces computational overhead but also improves numerical stability, making it particularly advantageous for large-scale simulations. \ref{sec:appendixB} provides additional details regarding this implementation.

\input{Figures/poisueilleSetup}
\input{Figures/poiseuille}

\subsection{Circular Interface in a Shear Flow}
A more complex problem involving shear flow is used to evaluate the accuracy of the model in simulating interfaces \cite{Rudman1997}. The simulation is performed in a periodic domain of size $L_0 \times L_0$, where $L_0 = 200$. The shear velocity is defined as follows:
\begin{subequations}
\begin{eqnarray}
&&u = U_0 \sin^2\left(\frac{\pi x}{L_0}\right)\sin\left(\frac{2\pi y}{L_0}\right)\cos\left(\frac{\pi t}{T}\right), \qquad
\label{eq:velocity_u}\\
&&v = -U_0 \sin^2\left(\frac{\pi y}{L_0}\right)\sin\left(\frac{2\pi x}{L_0}\right)\cos\left(\frac{\pi t}{T}\right). \qquad
\label{eq:velocity_v}
\end{eqnarray}
\end{subequations}

Initially, a circular interface with a radius $R = L_0 / 5$ is generated at the center of the domain at $(x, y) = (100, 60)$. The following parameters are fixed in lattice units: $U_0 = 0.04$, $M = 0.0002$, and $\xi = 3$. The Péclet number for this problem is calculated as:
\begin{equation}
Pe = \frac{U_0 \xi}{M},
\label{eq:peclet_number}
\end{equation}
The Péclet number is maintained constant at $Pe = 600$. The relative error between the analytical and numerical results is defined as:
\begin{equation}
\parallel \delta \phi \parallel_2 = \sqrt{\frac{\sum_{x,y} (\phi - \phi_0)^2}{\sum_{x,y} \phi_0^2}},
\label{eq:relative_error}
\end{equation}
where $\phi_0$ denotes the interface's initial position at $T = 0$. The time scale is expressed as:
\begin{equation}
T_f = L_0 / U_0.
\label{eq:time_scale}
\end{equation}

At $T = T_f$, the velocity field is reversed to restore the interface to its initial position. The simulation is continued until $T = 2T_f$. \tabref{tab:shear_flow_error} reports the relative error, and \figref{fig:shear_flow} shows the evolution of the interface shape over time. The excellent agreement with the theoretical solution demonstrates that the phase-field model accurately tracks the interface in complex flows.
\input{Tables/shearFlowError}

Furthermore, this problem evaluates the mass conservation property of the phase-field LB model. The system's total mass, normalized by the initial mass, is tracked over more than 25,000 iterations, and its evolution is plotted in \figref{fig:mass_evolution} as a function of dimensionless time. The results indicate that the current phase-field model exhibits excellent mass conservation properties.

\input{Figures/shearFlow}
\input{Figures/massCons}

\subsection{Rayleigh-Taylor Instability}
The Rayleigh-Taylor instability is utilized as a benchmark to evaluate the accuracy of the present model in solving complex flows and implementing body forces. This instability occurs when a heavier fluid lies above a lighter fluid in a gravitational field. Due to its significance in various natural and engineering phenomena, the problem has been extensively studied \cite{chandrasekhar1981hydrodynamic, Zu2013}. The computational domain is defined as $L_0 \times 4L_0$ for 2D cases and $L_0 \times L_0 \times 4L_0$ for 3D cases, with vertical solid walls and periodic boundaries in the horizontal directions. A perturbed interface is introduced based on the following equation:
\begin{equation}
x_0 = 0.1L \cos\left(\frac{2\pi x}{L}\right),
\label{eq:perturbation}
\end{equation}
where $x_0$ denotes the initial interface position.

The phase-field is initialized using the following expression:
\begin{equation}
\phi(x) = \phi_0 + \frac{\phi_H - \phi_L}{2} \tanh\left(\frac{|x - x_0|_\bot}{\xi/2}\right),
\label{eq:_initial_phase_field}
\end{equation}
where $|x - x_0|_\bot$ represents the perpendicular distance from a grid point in the computational domain to the perturbed interface $x_0$. 

The Rayleigh-Taylor instability is characterized by several dimensionless numbers, including the density ratio, viscosity ratio, Atwood number, and Reynolds number, which are defined as:
\begin{equation}
\begin{aligned}
\rho^\ast &= \frac{\rho_H}{\rho_L}, \quad
\mu^\ast = \frac{\mu_H}{\mu_L}, \quad
At = \frac{\rho_H - \rho_L}{\rho_H + \rho_L}, \quad
Re = \frac{\rho_H \sqrt{|\mathbf{g}L|}L}{\mu_H}.
\end{aligned}
\label{eq:dimensionless_numbers}
\end{equation}

where $g$ represents the gravitational acceleration. Additionally, the capillary and Péclet numbers are defined as:
\begin{gather}
Ca = \frac{\mu_H U_0}{\sigma},
\label{eq:capillary_number}\\
Pe = \frac{U_0 L}{M}.
\label{eq:peclet_number}
\end{gather}
where $U_0 = \sqrt{|\mathbf{g}L|}$. These parameters allow for comparison with other models in the literature. For consistency, the reference length is set to 256 lattice units (lu) in 2D simulations and 64 lu in 3D simulations. The dimensionless time is expressed as:
\begin{equation}
t^\ast = \frac{t}{t_0},
\label{eq:dimensionless_time}
\end{equation}
where $t_0 = \sqrt{\frac{L}{gAt}}$. Other simulation parameters are defined
in \tabref{tab:rayleigh_spec}.

\input{Tables/rayleighSpec}

The simulations use consistent parameters with the literature, allowing for validation of the results. For instance, surface tension is neglected in specific cases, such as case (c), to facilitate direct comparison with results from Ref. \cite{Zu2013}. Top and bottom boundaries are no-slip, while lateral boundaries are periodic. The defined computational setup allows for investigating the evolution of interface patterns, bubble growth, and spike formation in both 2D and 3D configurations.

The computational setup is illustrated in \figref{fig:rayleigh_bc}, showing the boundary conditions and the definitions of spike, saddle, and bubble points (\figref{fig:rayleigh_bc_3d}). For the low-density ratio case (a), \figref{fig:rayleigh_2d_low} demonstrates the interface evolution in the 2D computational domain, where the heavier fluid initially penetrates the lighter fluid symmetrically, forming a stable interface. Over time, counter-rotating vortices emerge, driven by Rayleigh-Taylor instability, and the interface develops complex patterns as small vortices are shed into the wake. In contrast, \figref{fig:rayleigh_2d_high} highlights the high-density ratio case (b), where the interface evolution is dominated by large spikes and bubbles due to the strong gravitational forces acting on the denser fluid. These results validate the model's capability to capture the dynamic behavior of the interface for both low- and high-density ratios.

\input{Figures/rayleighSetup}
\input{Figures/rayleigh2Dresults}

Similarly, the three-dimensional simulations are shown in \figref{fig:rayleigh_3d_low} and \figref{fig:rayleigh_3d_high}. For the low-density ratio case (c), the heavier fluid penetrates the lighter fluid under gravity, forming spikes and bubbles. The interface undergoes significant deformation, with roll-ups initially appearing near saddle points and evolving into mushroom-like structures along the spike flanks. Conversely, the high-density ratio case (d) demonstrates a different dynamic, where the shear force on the lighter fluid is insufficient to form mushroom-shaped spikes. Instead, the interface evolution is characterized by simpler bubble and spike formations, highlighting the distinct behaviors of low- and high-density ratio cases under Rayleigh-Taylor instability.

\input{Figures/rayleigh3Dresults}

Time histories of the dimensionless positions in the 2D and 3D computational domains are presented in \figref{fig:rayleigh_comparison}. The results align closely with previously published data, validating the model's accuracy in capturing the dynamics of Rayleigh-Taylor instability in both 2D and 3D simulations.

\input{Figures/rayleighComparison}

%% file: Figures/capillarySetup.tex
\begin{figure}[htbp]
    \centering
    \includegraphics[width=0.35\textwidth]{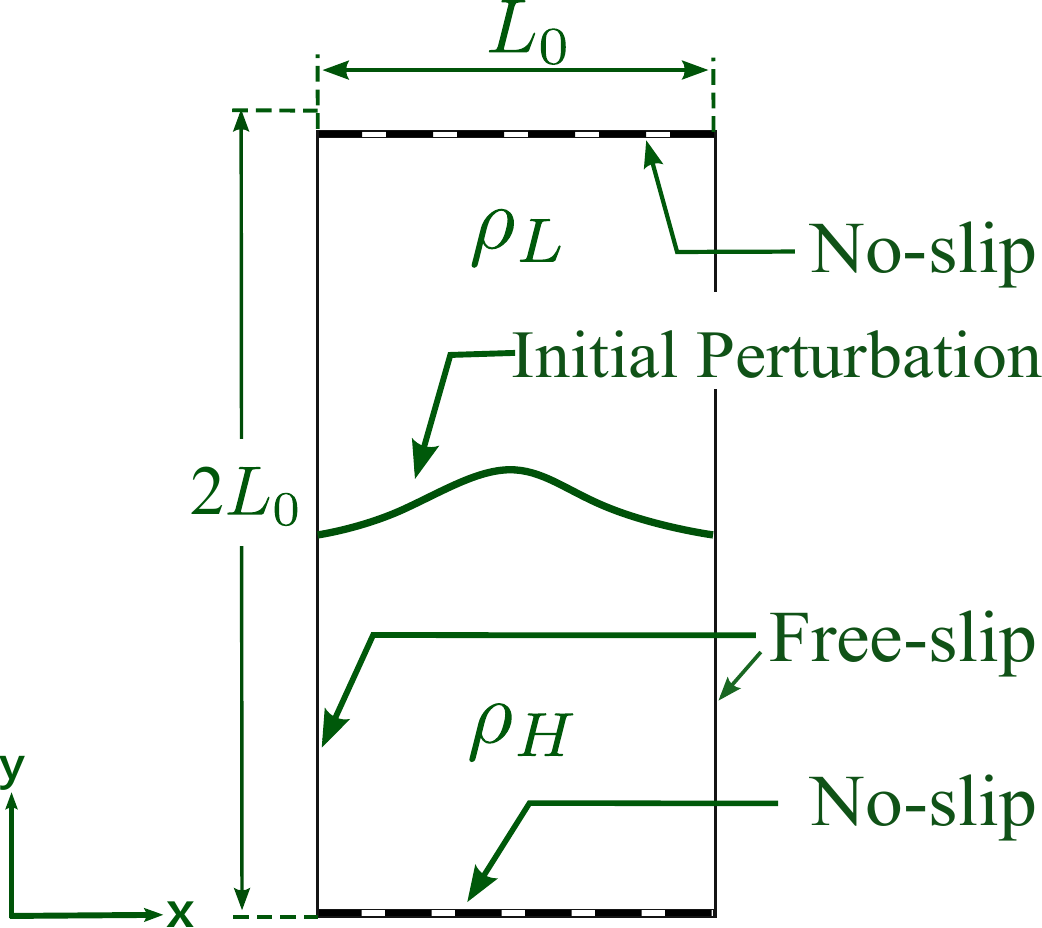} 
    \caption{Schematic representation of the simulation setup for capillary wave dynamics. The initial sinusoidal perturbation is applied at the interface between two immiscible fluids with densities $\rho_L$ (light fluid) and $\rho_H$ (heavy fluid). The top and bottom boundaries are subjected to no-slip conditions, while the vertical boundaries enforce free-slip conditions. The domain height is $2L_0$, and the domain width is $L_0$.}
    \label{fig:capillary_setup}
\end{figure}

%% file: Figures/capillary.tex
\begin{figure}[ht]
    \centering
    \begin{subfigure}[t]{0.48\textwidth}
        \centering
        \begin{tikzpicture}[scale=0.85]
        \begin{axis}[
            xlabel={$t^*$},
            ylabel={$\eta^*$},
            xmin=0, xmax=20,
            ymin=-1.4, ymax=1.4,
            ymajorgrids=true,
            grid style=dashed,
            legend style={at={(0.5,1.05)}, anchor=south, legend columns=2},
            legend cell align={left}
        ]
        
        \addplot[
            only marks,
            mark=*,
            mark options={fill=blue, draw=blue, opacity=0.5},
            mark size=2pt
        ]
        table[x=X,y=Y,col sep=comma] {Images/Capillary0005.csv};
        \addlegendentry{Current LBM}
        
        \addplot[
            color=red,
            solid,
            thick
        ]
        table[x=t,y=eta,col sep=comma] {Images/CapillaryAnalytical0005.csv};
        \addlegendentry{Prosperetti (1981)}
        
        \end{axis}
        \end{tikzpicture}
        \caption{Comparison of $\eta^*$ for viscosity $\nu = 0.0005$. The LBM results (blue circles) closely follow the analytical solution (red line).}
        \label{fig:capillary_a}
    \end{subfigure}
    \hfill
    \begin{subfigure}[t]{0.48\textwidth}
        \centering
        \begin{tikzpicture}[scale=0.85]
        \begin{axis}[
            xlabel={$t^*$},
            xmin=0, xmax=20,
            ymin=-1.4, ymax=1.4,
            ymajorgrids=true,
            grid style=dashed,
            legend style={at={(0.5,1.05)}, anchor=south, legend columns=2},
            legend cell align={left}
        ]
        
        \addplot[
            only marks,
            mark=*,
            mark options={fill=blue, draw=blue, opacity=0.5},
            mark size=2pt
        ]
        table[x=X,y=Y,col sep=comma] {Images/Capillary005.csv};
        \addlegendentry{Current LBM}
        
        \addplot[
            color=red,
            solid,
            thick
        ]
        table[x=t,y=eta,col sep=comma] {Images/CapillaryAnalytical005.csv};
        \addlegendentry{Prosperetti (1981)}
        
        \end{axis}
        \end{tikzpicture}
        \caption{Comparison of $\eta^*$ for viscosity $\nu = 0.005$. The simulation results show excellent agreement with the analytical solution.}
        \label{fig:capillary_b}
    \end{subfigure}
    \caption{Comparison of the normalized interface amplitude $\eta^*$ as a function of normalized time $t^*$ between the current LBM simulation and the analytical solution by Prosperetti (1981). Subfigure (a) corresponds to $\nu = 0.0005$, while subfigure (b) corresponds to $\nu = 0.005$.}
    \label{fig:capillary}
\end{figure}
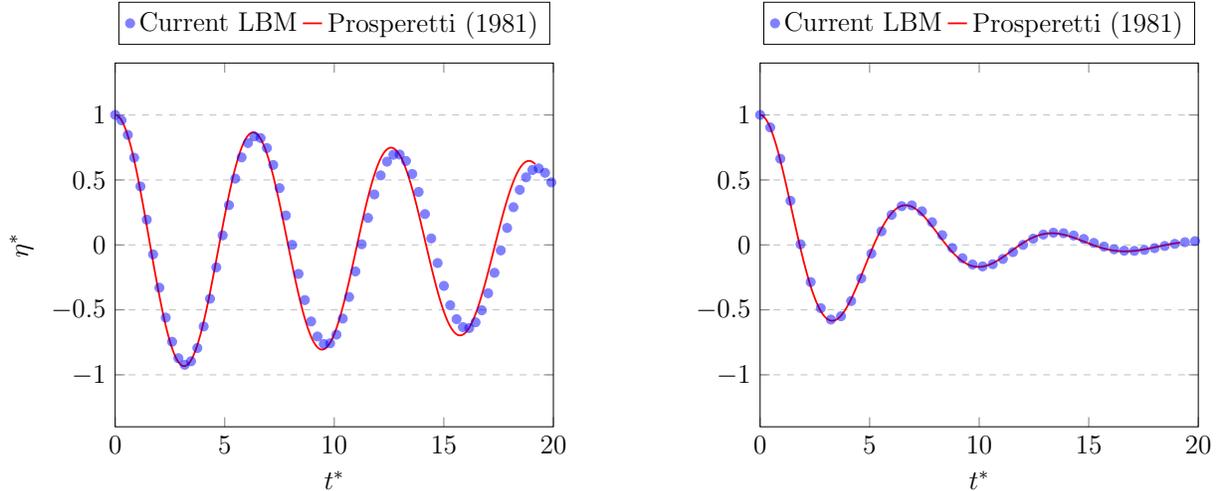

%% file: Tables/dropTest.tex
\begin{table}[ht]
\centering
\scriptsize
\renewcommand{\arraystretch}{1.5} 
\setlength{\tabcolsep}{10pt} 
\caption{Three-dimensional stationary drop test parameters and related errors. The table highlights error percentages for different surface tension values ($\sigma$).}
\label{tab:laplace_errors}
\begin{tabular}{l c c c}
\hline
\multirow{2}{*}{\textbf{R}} & \multicolumn{3}{c}{\textbf{Error Percentage ($E\%$)}} \\ 
\cmidrule(lr){2-4}
                            & $\sigma=0.001$ & $\sigma=0.005$ & $\sigma=0.010$ \\ 
\hline
15                          & $0.91$        & $1.01$         & $1.04$         \\
20                          & $0.73$        & $0.81$         & $0.82$         \\
25                          & $0.62$        & $0.63$         & $0.65$         \\
30                          & $0.40$        & $0.46$         & $0.46$         \\
35                          & $0.31$        & $0.33$         & $0.33$         \\
\hline
\end{tabular}
\end{table}

%% file: Figures/dropTest.tex
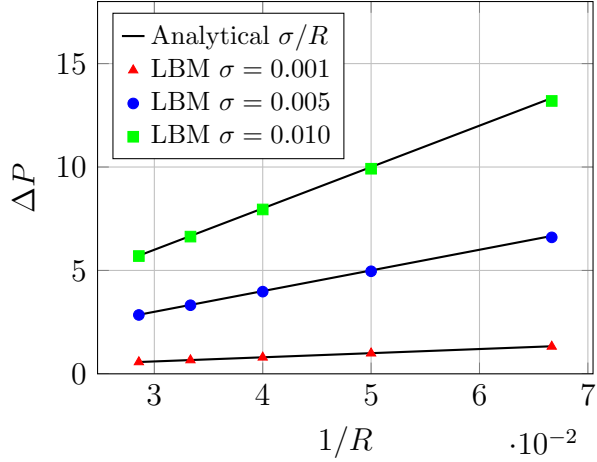
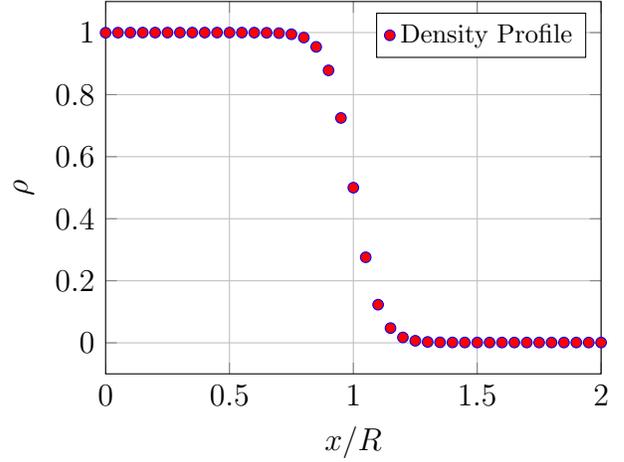
\begin{figure}[htbp]
    \centering
    \begin{subfigure}[t]{0.48\textwidth}
        \centering
        \begin{tikzpicture}
        \begin{axis}[
            xlabel={$1/R$},
            ylabel={$\Delta P$},
            legend style={at={(0.03,0.97)}, anchor=north west, font=\footnotesize},
            legend cell align={left},
            ymin=0, ymax=18,
            grid=major,
            width=\textwidth,
            height=0.8\textwidth,
            yticklabel style={/pgf/number format/fixed,/pgf/number format/precision=4},
        ]
            \addplot[black, solid, thick] table[x=R,y=Analytic_0.001,col sep=comma] {Data/laplace.txt};

            \addlegendentry{Analytical $\sigma/R$}
            
            \addplot[red, mark=triangle*, mark options={fill=red}, only marks] table[x=R,y=LBM_0.001,col sep=comma] {Data/laplace.txt};
            \addlegendentry{LBM $\sigma = 0.001$}
            
            \addplot[blue, mark=*, mark options={fill=blue}, only marks] table[x=R,y=LBM_0.005,col sep=comma] {Data/laplace.txt};
            \addlegendentry{LBM $\sigma = 0.005$}
            
            \addplot[green, mark=square*, mark options={fill=green}, only marks] table[x=R,y=LBM_0.010,col sep=comma] {Data/laplace.txt};
            \addlegendentry{LBM $\sigma = 0.010$}

            \addplot[black, solid, thick] table[x=R,y=Analytic_0.005,col sep=comma] {Data/laplace.txt};
            \addplot[black, solid, thick] table[x=R,y=Analytic_0.010,col sep=comma] {Data/laplace.txt};

        \end{axis}
        \end{tikzpicture}
        \caption{Laplace law tests for three different surface tension values ($\tau_d=\tau_g=0.5,\ \ \xi=4$), comparing LBM results and analytical solutions.}
        \label{fig:laplace_pressure}
    \end{subfigure}
    \hfill
    \begin{subfigure}[t]{0.48\textwidth}
        \centering
        \begin{tikzpicture}
        \begin{axis}[
            xlabel={$x/R$},
            ylabel={$\rho$},
            legend style={at={(0.97,0.97)}, anchor=north east, font=\footnotesize},
            legend cell align={left},
            xmin=0, xmax=2,
            ymin=-0.1, ymax=1.1,
            grid=major,
            width=\textwidth,
            height=0.8\textwidth,
        ]
            \addplot[blue, mark=*, mark options={fill=red}, only marks] table[x=x,y=rho,col sep=comma] {Data/dropProfile.txt};
            \addlegendentry{Density Profile}
        \end{axis}
        \end{tikzpicture}
        \caption{Density profile along the horizontal centerline for a drop with $R = 20$ after 50,000 iterations.}
        \label{fig:density_profile}
    \end{subfigure}
    \caption{Validation of the LBM model using Laplace law tests and density profile analysis. (a) Laplace law tests for three surface tension coefficients ($\sigma = 0.001$, $\sigma = 0.005$, and $\sigma = 0.010$) show agreement between LBM results and theoretical predictions. (b) The density profile confirms accurate representation of the interface for $\rho_d/\rho_g = 1000$.}
    \label{fig:laplace_test_density_profile}
\end{figure}

%% file: Figures/poisueilleSetup.tex
\begin{figure}[htbp]
    \centering
    \includegraphics[width=0.45\textwidth]{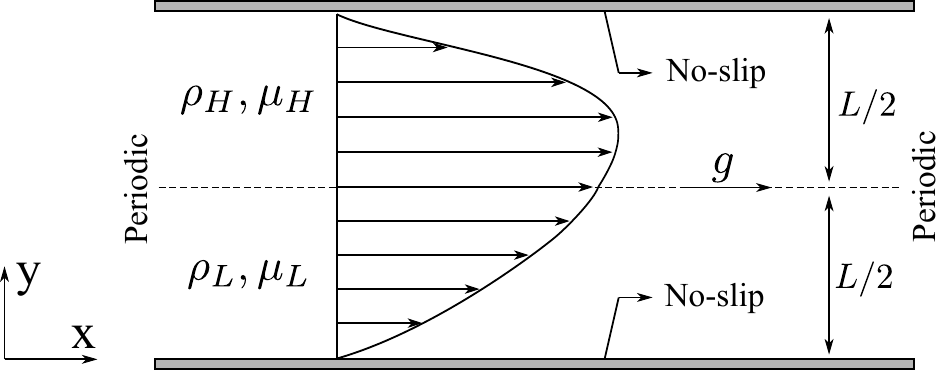} 
\caption{Schematic of the two-phase Poiseuille flow configuration: A periodic computational domain with density and viscosity stratification ($\rho_H, \mu_H$ in the upper half and $\rho_L, \mu_L$ in the lower half). The flow is driven by a body force $g$, with no-slip boundary conditions applied at the top and bottom walls. The domain height is $L$, split equally into two layers of $L/2$.}
    \label{fig:poisueille_setup}
\end{figure}

%% file: Figures/poiseuille.tex
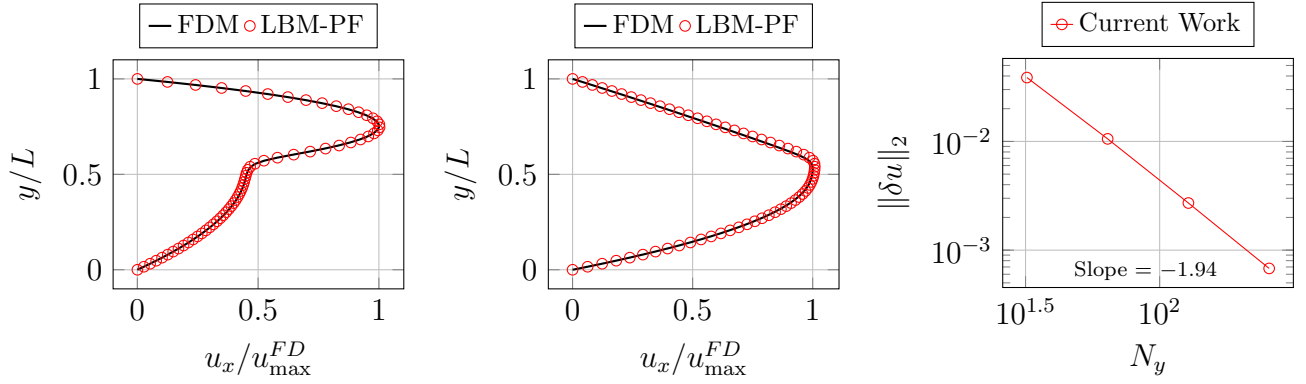
\begin{figure}[ht]
    \centering
    \begin{subfigure}[t]{0.32\textwidth}
        \centering
        \begin{tikzpicture}
        \begin{axis}[
            xlabel={$u_x / u_{\text{max}}^{FD}$},
            ylabel={$y / L$},
            width=\textwidth,
            height=0.85\textwidth,
            legend style={at={(0.5,1.05)}, anchor=south, legend columns=2,  font=\footnotesize},
            legend cell align={left},
            grid=major
        ]
            \addplot[black, solid, thick] table[x=xFDM,y=FDM,col sep=comma] {Data/p10.csv};
            \addlegendentry{FDM}
            \addplot[red, mark=o, mark options={solid}, only marks] table[x=xLBM,y=LBM,col sep=comma] {Data/p10.csv};
            \addlegendentry{LBM-PF}
        \end{axis}
        \end{tikzpicture}
        \caption{Normalized velocity profile for $\rho^* = 10$ and $\mu^* = 100$.}
    \end{subfigure}
    \hfill
    \begin{subfigure}[t]{0.32\textwidth}
        \centering
        \begin{tikzpicture}
        \begin{axis}[
            xlabel={$u_x / u_{\text{max}}^{FD}$},
            ylabel={$y / L$},
            width=\textwidth,
            height=0.85\textwidth,
            legend style={at={(0.5,1.05)}, anchor=south, legend columns=2,  font=\footnotesize},
            legend cell align={left},
            grid=major
        ]
            \addplot[black, solid, thick] table[x=xFDM,y=FDM,col sep=comma] {Data/P1000.csv};
            \addlegendentry{FDM}
            \addplot[red, mark=o, mark options={solid}, only marks] table[x=xLBM,y=LBM,col sep=comma] {Data/P1000.csv};
            \addlegendentry{LBM-PF}
        \end{axis}
        \end{tikzpicture}
        \caption{Normalized velocity profile for $\rho^* = 1000$ and $\mu^* = 100$.}
    \end{subfigure}
    \hfill
    \begin{subfigure}[t]{0.32\textwidth}
        \centering
        \begin{tikzpicture}
        \begin{loglogaxis}[
            xlabel={$N_y$},
            ylabel={$\| \delta u \|_2$},
            width=\textwidth,
            height=0.85\textwidth,
            legend style={at={(0.5,1.05)}, anchor=south, legend columns=2,  font=\footnotesize},
            legend cell align={left},
            grid=major
        ]
            \addplot[red, mark=o, mark options={solid}] table[x=x,y=ourLBM,col sep=comma] {Data/accuracy.txt};
            \addlegendentry{Current Work}
            \node[anchor=north east] at (axis cs:180,1e-3) {\scriptsize Slope = $-1.94$};
        \end{loglogaxis}
        \end{tikzpicture}
        \caption{Convergence study of two-phase Poiseuille flow at $\rho^* = 1000$ and $\mu^* = 100$ ($\tau_L = 0.5$).}
        \label{fig:poiseuille_accuracy}
    \end{subfigure}
    
    \caption{(a) Comparison between FDM solution (solid lines) and LBM-PF solution (symbols) for the normalized longitudinal velocity profile in two-phase Poiseuille flow with $\rho^* = 10$ and $\mu^* = 100$. (b) Same as (a) but for $\rho^* = 1000$ and $\mu^* = 100$. (c) Convergence study of two-phase Poiseuille flow at $\rho^* = 1000$ and $\mu^* = 100$ ($\tau_L = 0.5$) using a log-log plot.}
    \label{fig:poiseuille_validation}
\end{figure}

%% file: Tables/shearFlowError.tex
\begin{table}[ht]
\centering
\scriptsize
\renewcommand{\arraystretch}{1.5} 
\setlength{\tabcolsep}{10pt} 
\caption{Relative error for a circular interface in a shear flow based on Eq.~\eqref{eq:relative_error}. The table summarizes the parameters and the resulting error.}
\label{tab:shear_flow_error}
\begin{tabular}{l c c c c c}
\hline
\textbf{$L_0$} & \textbf{$U_0$} & \textbf{$M$} & \textbf{$\xi$} & \textbf{$Pe$} & \textbf{Error} \\ 
\hline
$200$ & $0.04$ & $0.0002$ & $3$ & $600$ & $8.58\times{10}^{-3}$ \\ 
\hline
\end{tabular}
\end{table}

%% file: Figures/shearFlow.tex
\begin{figure}[ht]
    \centering
    \includegraphics[width=\textwidth]{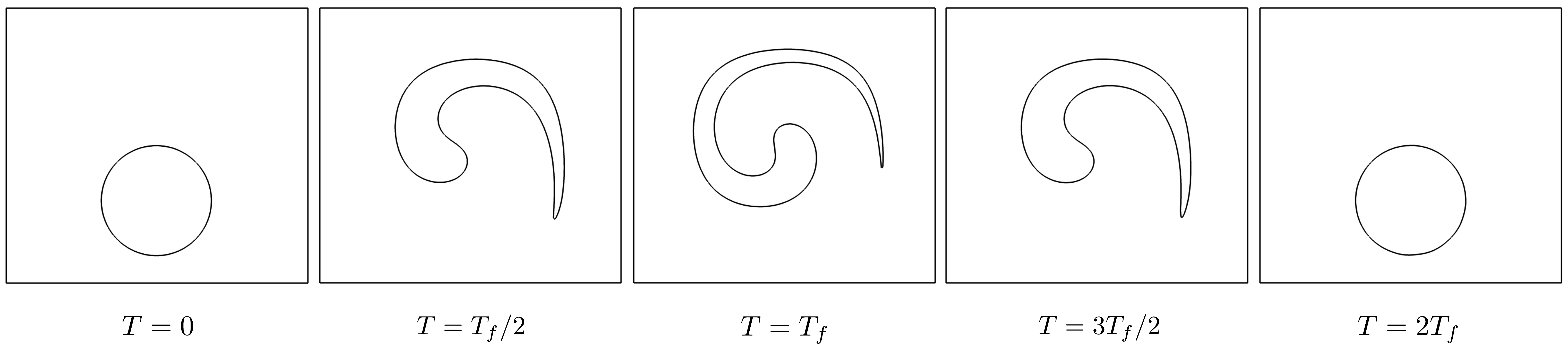} 
\caption{Evolution of a circular interface in a shear flow at $\text{Pe} = 600$. The interface ($\phi = 0$) is shown at five distinct time steps: $T = 0$, $T = T_f/2$, $T = T_f$, $T = 3T_f/2$, and $T = 2T_f$, illustrating the deformation and recovery process of the interface under shear flow.}
    \label{fig:shear_flow}
\end{figure}

%% file: Figures/massCons.tex
\begin{figure}[htbp]
    \centering
    \begin{tikzpicture}[scale=0.9]
    \begin{axis}[
        xlabel={$t^*$},
        ylabel={$\frac{\sum_{x, y} \rho}{\sum_{x, y} \rho_0}$},
        xmin=1, xmax=5,
        ymin=0.999998, ymax=1.000002,
        ytick={0.999998, 0.999999, 1.000000, 1.000001, 1.000002},
        yticklabel style={/pgf/number format/fixed,/pgf/number format/precision=6},
        grid=major,
        width=0.5\textwidth,
        height=0.35\textwidth,
        legend style={at={(0.97,0.03)}, anchor=south east, font=\footnotesize},
        legend cell align={left}
    ]
        \addplot[red, thick] table[x=t, y=phi, col sep=comma] {Data/mass.txt};
        \addlegendentry{Current Work}
    \end{axis}
    \end{tikzpicture}
    \caption{Time evolution of the normalized mass of the system for circular interface in a shear flow.}
    \label{fig:mass_evolution}
\end{figure}
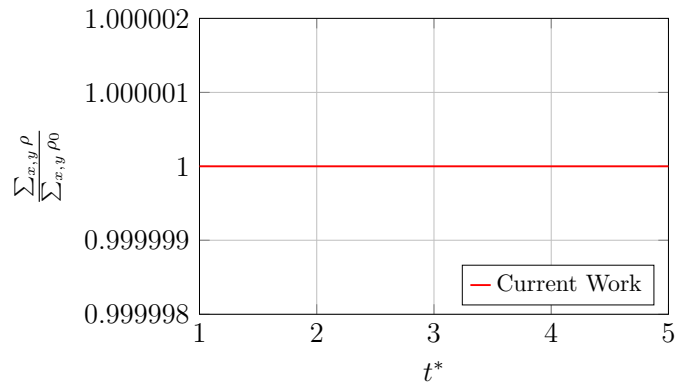

%% file: Tables/rayleighSpec.tex
\begin{table}[b]
\centering
\scriptsize
\renewcommand{\arraystretch}{1.5} 
\setlength{\tabcolsep}{8pt} 
\caption{Conditions of the Rayleigh-Taylor instability simulations. Key dimensionless numbers are included for each case.}
\label{tab:rayleigh_spec}
\begin{tabular}{l c c c c c c c}
\hline
\textbf{Case} & $\boldsymbol{\rho^\ast}$ & $\boldsymbol{\mu^\ast}$ & \textbf{Re} & \textbf{At} & \textbf{Pe} & \textbf{Ca} & $\boldsymbol{\xi}$ \\ 
\hline
a (2D) & $3$     & $1$   & $3000$  & $0.500$ & $1000$ & $0.26$ & $4$ \\
b (2D) & $1000$  & $100$ & $3000$  & $0.998$ & $1000$ & $0.44$ & $4$ \\
c (3D) & $3$     & $3$   & $128$   & $0.500$ & $744$  & $-$    & $5$ \\
d (3D) & $1000$  & $100$ & $3000$  & $0.998$ & $200$  & $8.7$  & $5$ \\
\hline
\end{tabular}
\end{table}

%% file: Figures/rayleighSetup.tex
\begin{figure}[htbp]
    \centering
    \begin{subfigure}[t]{0.47\textwidth} 
        \centering
        \includegraphics[width=0.5\textwidth, trim=0cm -2.5cm 0cm 0cm, clip]{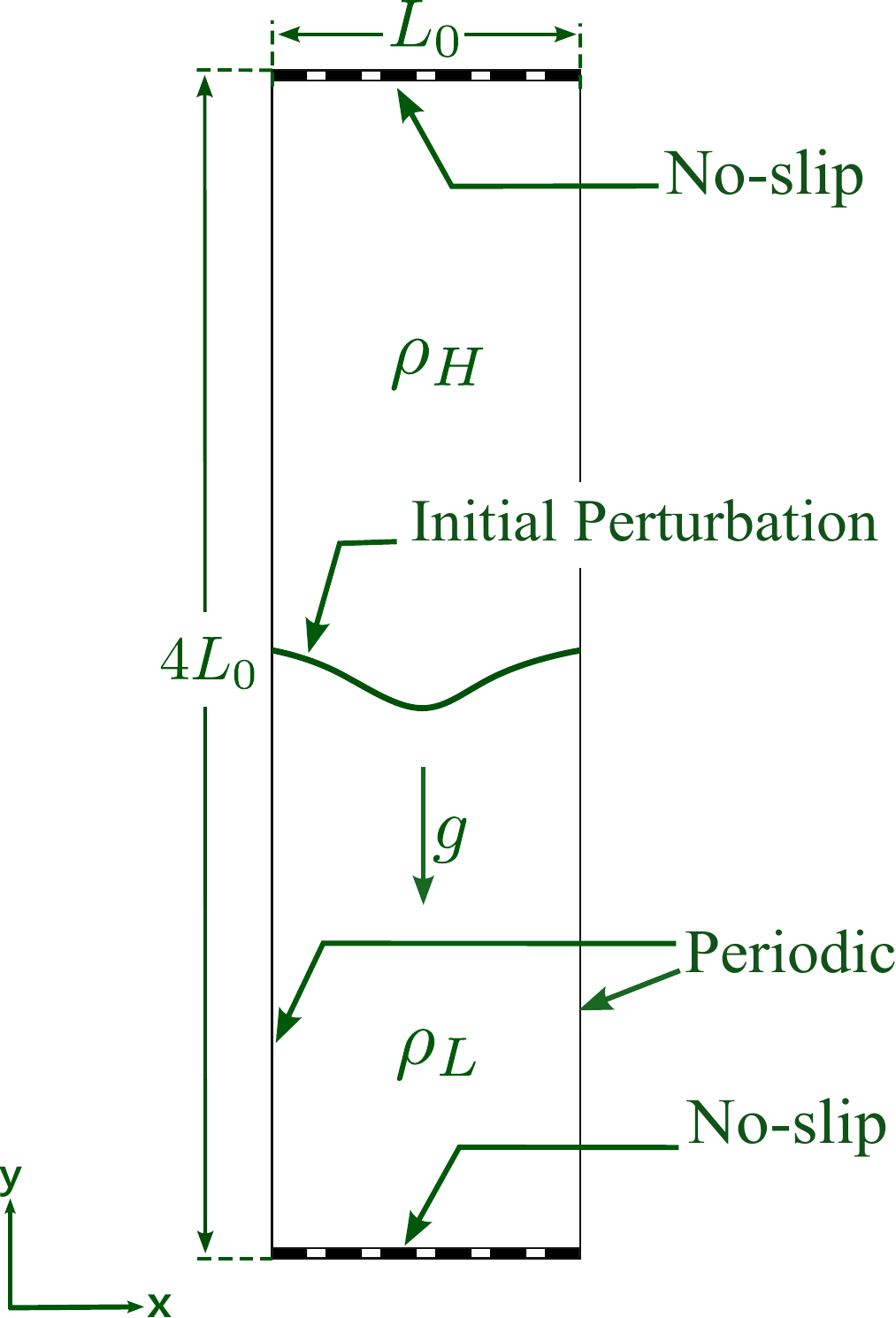} 
        \caption{Schematic of the 2D computational domain for the Rayleigh-Taylor instability. The domain is divided into two fluid layers, with the heavy fluid ($\rho_H$) on top and the light fluid ($\rho_L$) at the bottom. No-slip boundaries are applied vertically, while periodic boundaries are applied horizontally.}
        \label{fig:rayleigh_bc_2d}
    \end{subfigure}
    \hfill
    \begin{subfigure}[t]{0.47\textwidth} 
        \centering
        \includegraphics[width=0.5\textwidth]{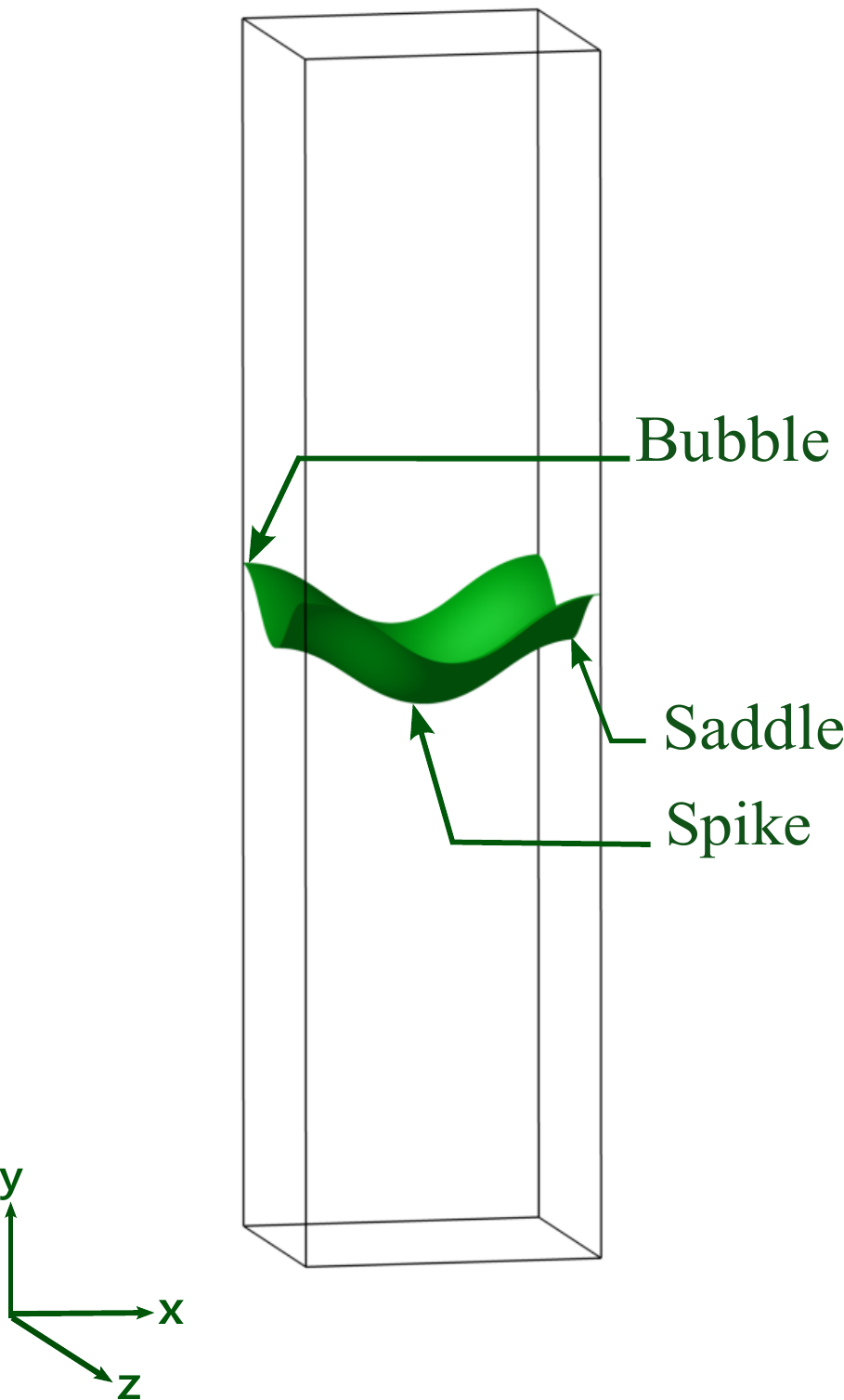} 
        \caption{Schematic of the 3D computational domain for the Rayleigh-Taylor instability. The interface evolution highlights the formation of bubbles, saddles, and spikes under gravitational forces.}
        \label{fig:rayleigh_bc_3d}
    \end{subfigure}
    \caption{Boundary conditions and initial setup for the Rayleigh-Taylor instability simulations in 2D and 3D. (a) shows the 2D setup with a solid wall at the top and bottom, and periodic boundaries horizontally. (b) illustrates the 3D domain with similar boundary conditions and highlights the interface dynamics.}
    \label{fig:rayleigh_bc}
\end{figure}

%% file: Figures/rayleigh2Dresults.tex
\begin{figure}[htbp]
    \centering
    \begin{subfigure}[t]{0.47\textwidth} 
        \centering
        \includegraphics[width=\textwidth, trim=0cm 0cm 0cm 0cm, clip]{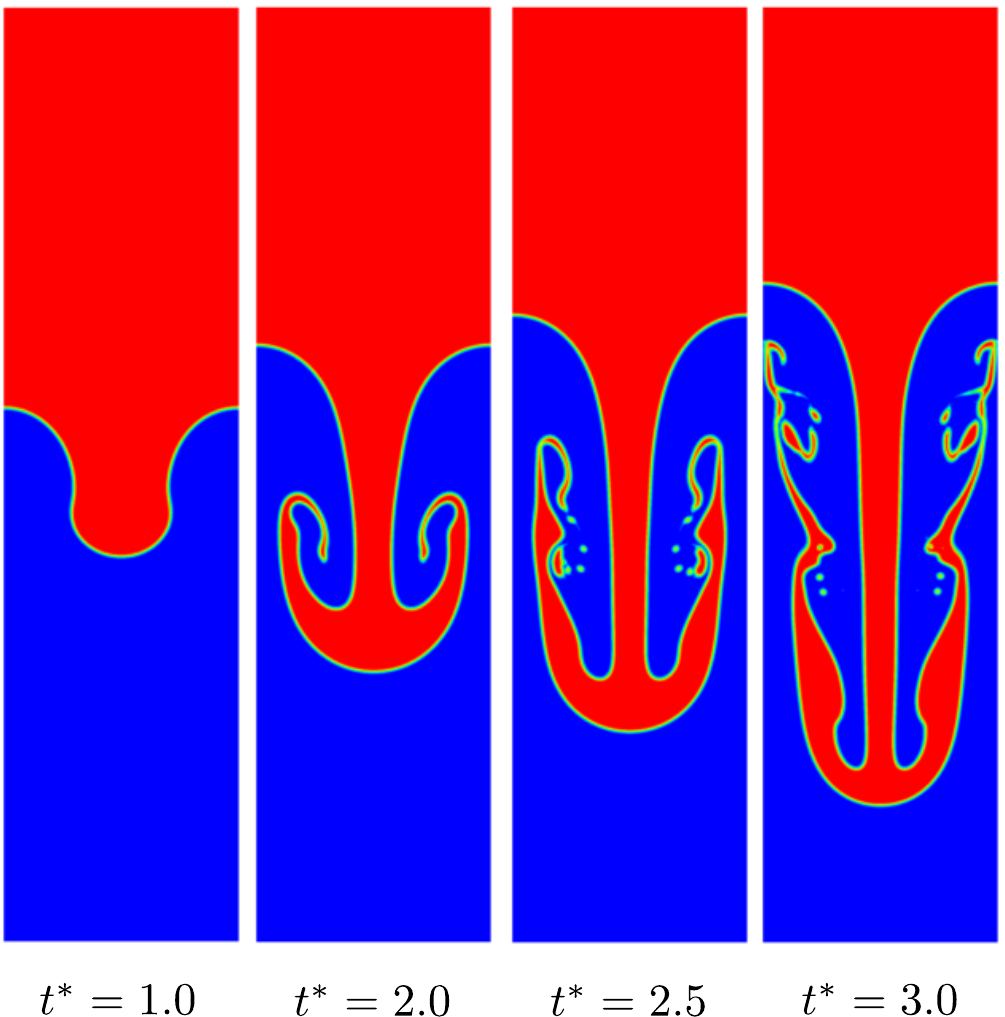} 
        \caption{Snapshots of the 2D Rayleigh-Taylor instability simulation for a low-density ratio case ($\rho^\ast = 3$, $\mu^\ast = 1$, $Re = 3000$, $At = 0.500$, $Pe = 1000$, $Ca = 0.26$, $\xi = 4$). The heavy fluid descends into the lighter fluid, forming counter-rotating vortices and complex interface patterns.}
        \label{fig:rayleigh_2d_low}
    \end{subfigure}
    \hfill
    \begin{subfigure}[t]{0.47\textwidth} 
        \centering
        \includegraphics[width=\textwidth]{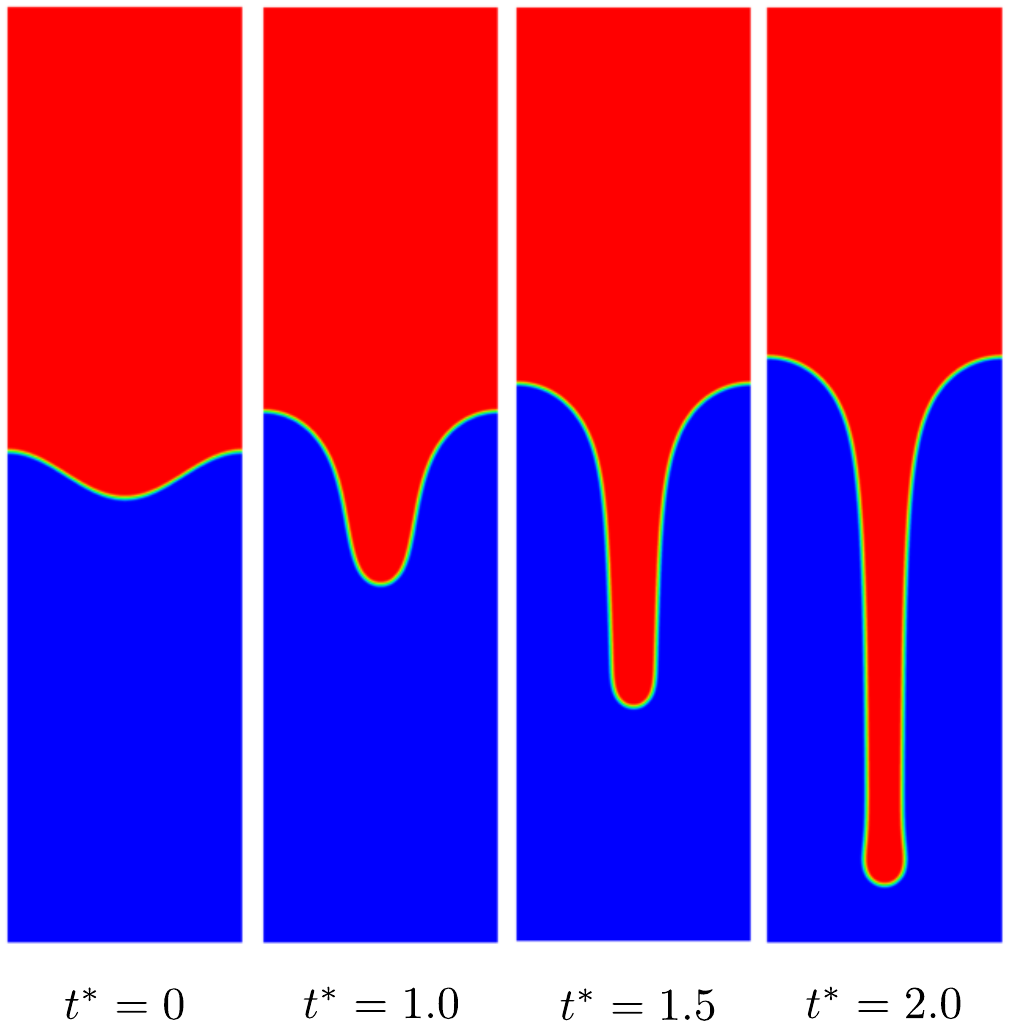} 
        \caption{Snapshots of the 2D Rayleigh-Taylor instability simulation for a high-density ratio case ($\rho^\ast = 1000$, $\mu^\ast = 100$, $Re = 3000$, $At = 0.998$, $Pe = 1000$, $Ca = 0.44$, $\xi = 4$). The interface evolution shows the formation of large bubbles and spikes, characteristic of high Atwood number flows.}
        \label{fig:rayleigh_2d_high}
    \end{subfigure}
    \caption{Comparison of interface evolution in 2D Rayleigh-Taylor instability simulations for low- and high-density ratios.}
    \label{fig:rayleigh_2d_interface}
\end{figure}

%% file: Figures/rayleigh3Dresults.tex
\begin{figure}[htbp]
    \centering
    \begin{subfigure}[t]{0.47\textwidth} 
        \centering
        \includegraphics[width=\textwidth, trim=0cm 0cm 0cm 0cm, clip]{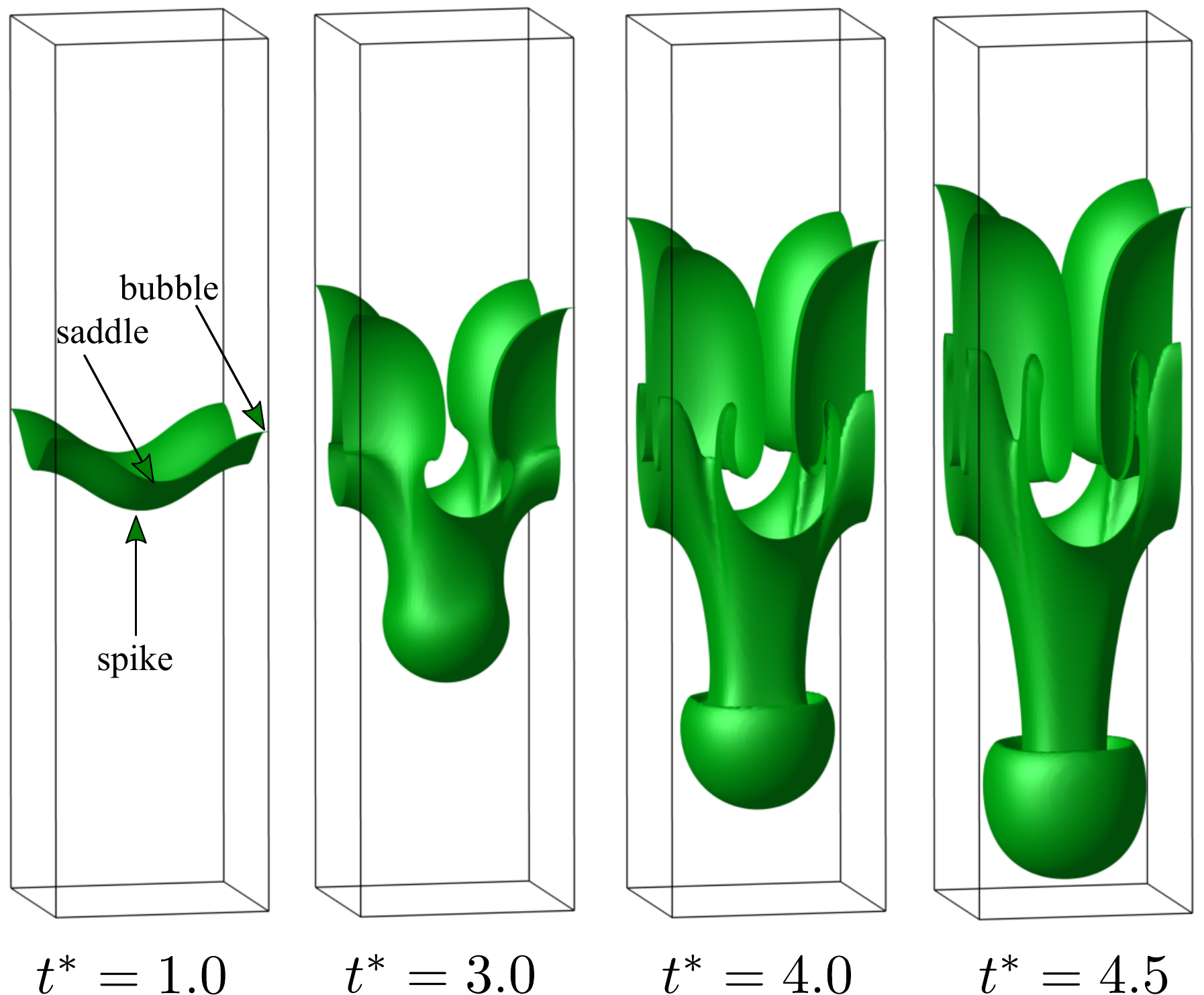} 
        \caption{Snapshots of the 3D Rayleigh-Taylor instability simulation for a low-density ratio case ($\rho^\ast = 3$, $\mu^\ast = 1$, $Re = 128$, $At = 0.500$, $Pe = 744$, $\xi = 5$). The interface evolution highlights the formation of bubbles, saddles, and spikes under gravitational forces over dimensionless times $t^* = 1.0, 3.0, 4.0$, and $4.5$.}
        \label{fig:rayleigh_3d_low}
    \end{subfigure}
    \hfill
    \begin{subfigure}[t]{0.47\textwidth} 
        \centering
        \includegraphics[width=\textwidth]{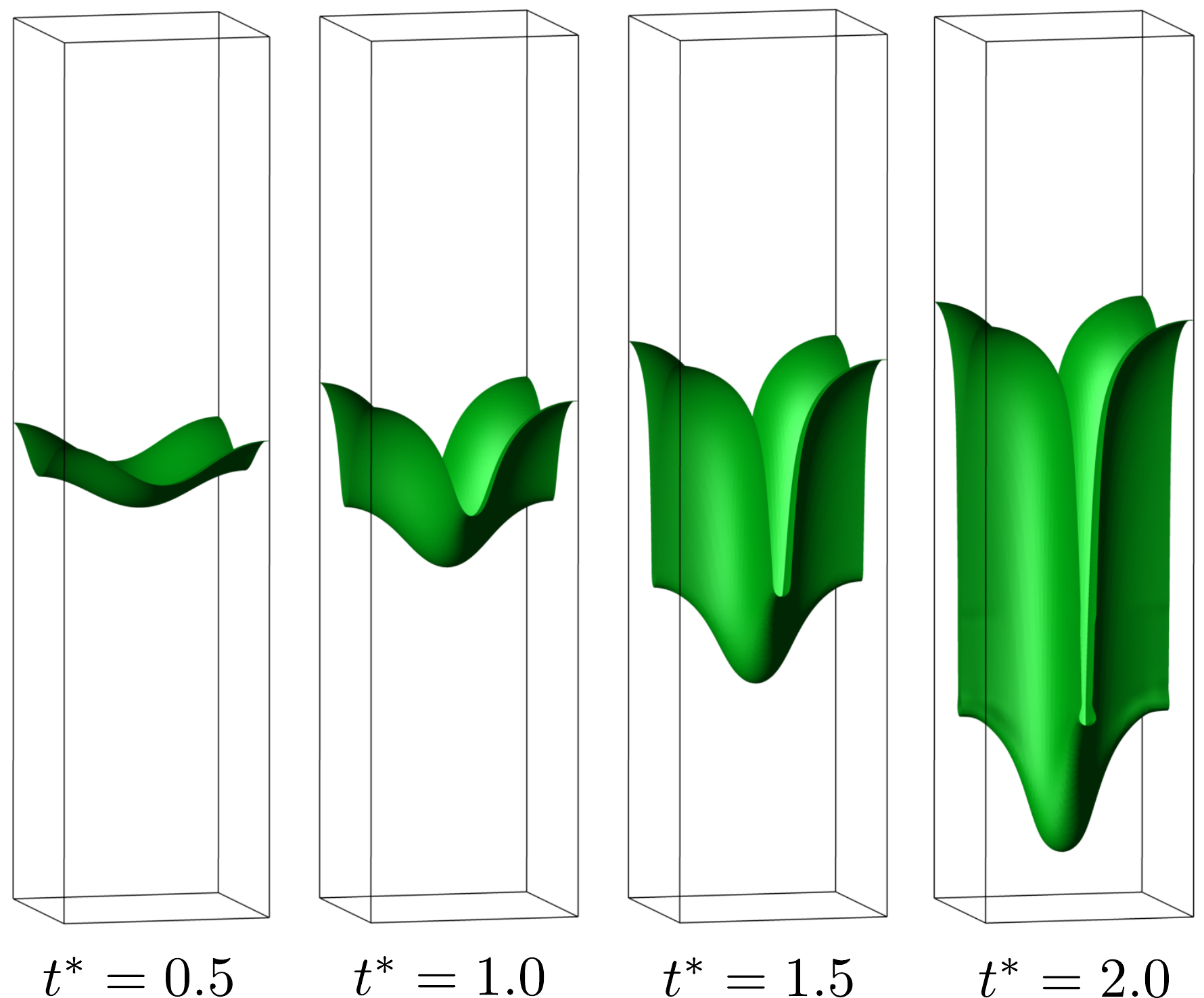} 
        \caption{Snapshots of the 3D Rayleigh-Taylor instability simulation for a high-density ratio case ($\rho^\ast = 1000$, $\mu^\ast = 100$, $Re = 3000$, $At = 0.998$, $Pe = 200$, $Ca = 8.7$, $\xi = 5$). The heavy fluid penetrates the light fluid, forming spikes and bubbles, with snapshots shown at dimensionless times $t^* = 0.5, 1.0, 1.5$, and $2.0$.}
        \label{fig:rayleigh_3d_high}
    \end{subfigure}
    \caption{Evolution of the interface pattern in the 3D Rayleigh-Taylor instability simulations.}
    \label{fig:rayleigh_3d_interface}
\end{figure}

%% file: Figures/rayleighComparison.tex
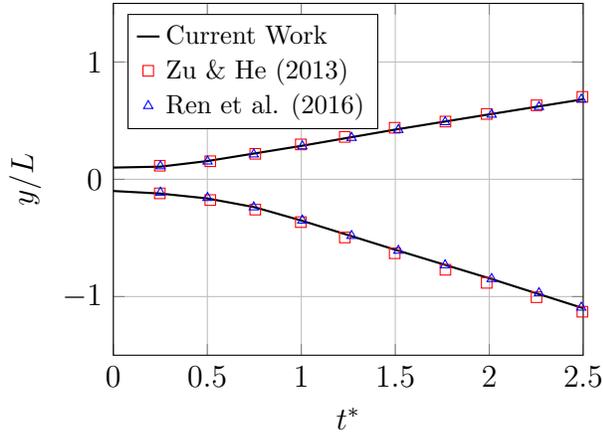
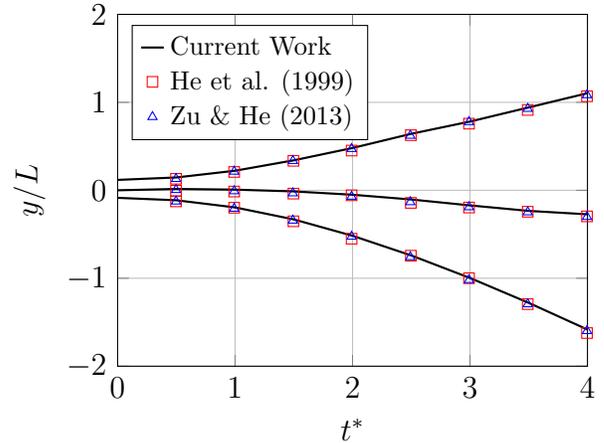
\begin{figure}[htbp]
    \centering
    \begin{subfigure}[t]{0.46\textwidth}
        \centering
        \begin{tikzpicture}
        \begin{axis}[
            xlabel={$t^*$},
            ylabel={$y/L$},
            legend style={at={(0.03,0.97)}, anchor=north west, font=\footnotesize},
            legend cell align={left},
            ymin=-1.5, ymax=1.5,
            xmin=0, xmax=2.5,
            grid=major,
            width=\textwidth,
            height=0.8\textwidth,
            yticklabel style={/pgf/number format/fixed,/pgf/number format/precision=4},
        ]
            \addplot[black, solid, thick] table[x=x,y=bubbleLBM,col sep=comma] {Data/rayleigh2D.csv};
            \addlegendentry{Current Work}

            \addplot[red, mark=square, only marks] table[x=xZu,y=bubbleZu,col sep=comma] {Data/rayleigh2D.csv};
            \addlegendentry{Zu \& He (2013)}

            \addplot[blue, mark=triangle, only marks] table[x=xRen,y=bubbleRen,col sep=comma] {Data/rayleigh2D.csv};
            \addlegendentry{Ren et al. (2016)}

            \addplot[black, solid, thick] table[x=x,y=spikeLBM,col sep=comma] {Data/rayleigh2D.csv};

            \addplot[red, mark=square, only marks] table[x=xZu,y=spikeZu,col sep=comma] {Data/rayleigh2D.csv};

            \addplot[blue, mark=triangle, only marks] table[x=xRen,y=spikeRen,col sep=comma] {Data/rayleigh2D.csv};

        \end{axis}
        \end{tikzpicture}
        \caption{Time evolution of the bubble front and spike tip for 2D Rayleigh-Taylor instability, comparing current work with results from \cite{Zu2013, Ren2016}.}
    \label{fig:rayleigh_comparison_2d}
    \end{subfigure}
    \hfill
    \begin{subfigure}[t]{0.46\textwidth}
        \centering
        \begin{tikzpicture}
        \begin{axis}[
            xlabel={$t^*$},
            ylabel={$y/L$},
            legend style={at={(0.03,0.97)}, anchor=north west, font=\footnotesize},
            legend cell align={left},
            ymin=-2, ymax=2,
            xmin=0, xmax=4.0,
            grid=major,
            width=\textwidth,
            height=0.8\textwidth,
        ]
            \addplot[black, solid, thick] table[x=x,y=bubbleLBM,col sep=comma] {Data/rayleigh3D.csv};
            \addlegendentry{Current Work}

            \addplot[red, mark=square, only marks] table[x=xHe,y=bubbleHe,col sep=comma] {Data/rayleigh3D.csv};
            \addlegendentry{He et al. (1999)}

            \addplot[blue, mark=triangle, only marks] table[x=xZu,y=bubbleZu,col sep=comma] {Data/rayleigh3D.csv};
            \addlegendentry{Zu \& He (2013)}

            \addplot[black, solid, thick] table[x=x,y=saddleLBM,col sep=comma] {Data/rayleigh3D.csv};

            \addplot[red, mark=square, only marks] table[x=xHe,y=saddleHe,col sep=comma] {Data/rayleigh3D.csv};

            \addplot[blue, mark=triangle, only marks] table[x=xZu,y=saddleZu,col sep=comma] {Data/rayleigh3D.csv};

            \addplot[black, solid, thick] table[x=x,y=spikeLBM,col sep=comma] {Data/rayleigh3D.csv};

            \addplot[red, mark=square, only marks] table[x=xHe,y=spikeHe,col sep=comma] {Data/rayleigh3D.csv};

            \addplot[blue, mark=triangle, only marks] table[x=xZu,y=spikeZu,col sep=comma] {Data/rayleigh3D.csv};
            
        \end{axis}
        \end{tikzpicture}
        \caption{Time evolution of the bubble front, saddle point, and spike tip for 3D Rayleigh-Taylor instability, comparing current work with results from \cite{He1999_Zhang, Zu2013}.}
    \label{fig:rayleigh_comparison_3d}
    \end{subfigure}
    \caption{Time evolution of the bubble front, saddle, and spike tip for  Rayleigh-Taylor instability.}
    \label{fig:rayleigh_comparison}
\end{figure}

%% file: Sections/Results.tex
\section{Liquid Jet Breakup}
\label{sec:results}
Understanding the breakup dynamics of liquid jets is essential for various engineering and scientific applications, including fuel injection, inkjet printing, and biomedical sprays. The breakup process is governed by a complex interplay of inertia, viscosity, surface tension, and external forces, requiring accurate numerical models to capture these interactions. This section presents a GPU-accelerated phase-field lattice Boltzmann framework designed to efficiently simulate liquid jet breakup across a wide range of density ratios and flow conditions.

\subsection{Setup}
The boundary conditions and computational domain play a crucial role in determining the accuracy and reliability of numerical studies for liquid jet breakup. The setup for this investigation is illustrated in \figref{fig:jet_setup}, highlighting the essential boundary conditions, domain layout, and forces affecting jet breakup.

The computational domain, shown in \figref{fig:jet_domain}, is defined as $6D_j \times 6D_j \times 20D_j$, where $D_j$ is the jet diameter. As depicted in \figref{fig:jet_boundary_conditions}, the inlet boundary at the top introduces the liquid jet, while the side walls are modeled as free-slip boundaries. The convective boundary at the bottom is designed to allow the fluid to exit without reflecting back into the computational domain. Further details on the treatment of boundary conditions can be found in \ref{sec:appendixB}. The inflow region at the inlet boundary is defined as ${(x-x_c)}^2 + {(z-z_c)}^2 < {(D_j/2)}^2$, where $x_c$ and $z_c$ denote the center of the circular inlet on the $x-z$ plane. The velocity profile at the inlet is uniform, ensuring no artificial disturbances. Gravity acts in the vertical direction with an acceleration of $g = (0, g, 0)$, influencing the dispersed phase of the liquid jet.

In addition to boundary conditions, the forces governing jet breakup are illustrated in \figref{fig:jet_forces}. These include inertial forces, gravitational forces, capillary forces (surface tension), and aerodynamic forces (drag). The interaction of these forces determines the breakup length ($L_b$) and the size of the resulting droplets ($d_e$). The gravitational body force acting on the fluid is expressed as:
\begin{equation}
F_g = \left(\rho(x,t) - \rho_L\right)g,
\label{eq:gravity_force}
\end{equation}
where $\rho(x,t)$ is the local density, and $\rho_L$ is the density of the lighter fluid.

This numerical setup effectively captures the dynamics of liquid jet breakup, providing a comprehensive framework for analyzing the interplay of forces and boundary conditions in jet simulations.

\input{Figures/jetSetup}

\subsection{Comparison with experiments}
We investigated the breakup of a liquid jet numerically using the phase-field lattice Boltzmann (LB) model. The primary objective was to characterize the breakup length, drop size, and the transition between dripping and jetting regimes. The simulations were conducted using real water and air properties, with parameters chosen to match the experimental study of \cite{Sunol2015}. In these simulations, the D3Q19 lattice structure and the WMRT collision operator were employed to enhance numerical stability under high-density conditions. The inlet velocity was set to $u_j = 0.05$, with the dispersed phase density $\rho_j = 1.0$. The Reynolds, Weber, Froude, and Ohnesorge numbers, as well as the dimensionless time, were used to describe the flow and are defined as follows:
\begin{gather}
Re = \frac{\rho_j u_j D_j}{\mu_j}, \quad 
We = \frac{\rho_j u_j^2 D_j}{\sigma}, \quad 
Fr = \frac{u_j}{\sqrt{g D_j}}, \quad 
t^\ast = \frac{t}{D_j/u_j}, \quad 
Oh = \frac{\sqrt{We}}{Re}.
\label{eq:non_dimensional_params}
\end{gather}
In all simulations, the Ohnesorge number was fixed at $Oh = 4.4 \times 10^{-3}$, consistent with \cite{Sunol2015}, and the cases are summarized in \tabref{tab:jet_spec}.

\input{Tables/jetSpec}

\figref{fig:jet_dripping} shows the interface evolution for the dripping regime at $We = 0.447$ and $We = 1.79$. In this regime, surface tension and gravitational forces dominate, leading to the formation of a pendant drop at the nozzle tip. As the drop grows under gravity, the liquid column connecting it to the nozzle thins due to capillary forces. Eventually, the neck becomes unstable and pinches off, resulting in the detachment of an axisymmetric droplet. This process repeats cyclically, producing discrete drops without significant perturbations along the liquid column. The observed behavior aligns with classical dripping phenomena, where drop formation and detachment are entirely controlled by the interplay between gravity and surface tension, with no significant inertial effects.

In contrast, \figref{fig:jet_jetting} illustrates the jetting regime at $We = 3.52$ and $We = 4.05$. In this regime, inertial forces dominate over surface tension, allowing the liquid jet to extend further before breaking up. As the jet exits the nozzle, small perturbations, known as Rayleigh waves, develop along the liquid column due to capillary instabilities. These waves grow in amplitude as the jet moves downstream, eventually reaching a critical point where the wave amplitude equals the jet radius. At this stage, surface tension amplifies the disturbance, leading to the periodic breakup of the liquid column into droplets. This process, governed by the classical Rayleigh-Plateau instability, results in a well-defined drop formation pattern.

The breakup length, $L_b$, is one of the most critical parameters in characterizing jet breakup. \figref{fig:jet_breakup_length} shows that $L_b / D_j$ increases with the Weber number in both regimes, with a linear trend observed in the jetting regime. Experimental results from \cite{Sunol2015} confirm this trend, and our numerical results are in excellent agreement with their findings. The phase-field model effectively captures the distinct breakup mechanisms in each regime, demonstrating its reliability for studying liquid jet breakup.

\figref{fig:jet_droplet_size} presents the dimensionless drop size, $d_e / D_j$, as a function of the Weber number. In the dripping regime, surface tension dominates, resulting in larger drops. As the Weber number increases, inertial forces become more significant, reducing the drop size. This transition reflects the shift from surface-tension-driven breakup in the dripping regime to inertia-dominated breakup in the jetting regime. The results also show frequent coalescence events in the jetting regime, which further influence the mean drop size. Our numerical results align well with experimental observations, highlighting the phase-field LB model's ability to predict the complex dynamics of liquid jet breakup with high accuracy.

\input{Figures/dripping}

\input{Figures/jetting}

\input{Figures/dropSize}

\subsection{Transition of breakup regimes}
Jet breakup regimes in liquid-gas systems are classified into five categories, as shown in \figref{fig:jet_breakup_overview}. The regimes are as follows: I corresponds to dripping, II to Rayleigh breakup, III to the first wind-induced breakup, IV to the second wind-induced breakup, and V to atomization. The transition between dripping and jetting (Rayleigh regime) was experimentally observed in Ref. \cite{Sunol2015} at $2 < We < 3$. Our results align with this range, but we conducted additional simulations within $2 < We < 3$ to determine the critical Weber number (${We}_{cr}$) more precisely.

To illustrate the transition between the dripping and jetting regimes, two simulations are shown in \figref{fig:jet_transition}. At $We=2.17$, \figref{fig:jet_transition_a} demonstrates a dripping regime where gravity and surface tension dominate, leading to periodic drop formation. As $We$ increases to 2.27, \figref{fig:jet_transition_b} reveals the onset of jetting, where inertia enables the liquid to extend further before breaking up due to Rayleigh instabilities. These results indicate that the critical Weber number for the transition is approximately ${We}_{cr} \approx 2.2$. The lattice Boltzmann simulation effectively captures the interplay between hydrodynamic forces, as schematically shown in \figref{fig:jet_forces}. The stability of the phase-field LB model enables simulations with large density and viscosity contrasts, making it well-suited for characterizing interfacial instabilities across a wide range of Weber numbers.

\input{Figures/transition}

In this study, the liquid jet simulations span two distinct regimes: dripping and jetting. As illustrated in \figref{fig:Oh_Re}, the Reynolds number ($Re$) increases while the Ohnesorge number ($Oh$) remains constant across test cases. For instance, a dripping regime is observed at $We=2.17$ and $Re=332$, where surface tension prevails. However, with a slight increase in $Re$, the liquid jet transitions into the jetting regime. It is worth noting that the boundary between the dripping and jetting regimes, as depicted in \figref{fig:Oh_Re}, may not be sharply defined and could vary depending on specific conditions.

\input{Figures/OhRe}

%% file: Figures/jetSetup.tex
\begin{figure}[htbp]
    \centering
    \begin{subfigure}[t]{0.31\textwidth} 
        \centering
        \includegraphics[width=0.8\textwidth, trim=0cm 0cm 0cm 0cm, clip]{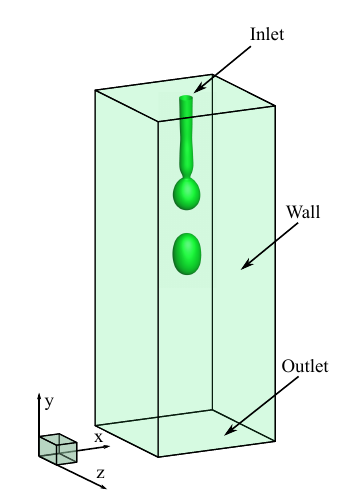} 
        \caption{Visualization of the computational domain for the liquid jet setup. The domain size is $6D_j\times6D_j\times20D_j$, where $D_j$ represents the jet diameter. The inlet boundary at the top introduces the liquid jet.}
        \label{fig:jet_domain}
    \end{subfigure}
    \hfill
    \begin{subfigure}[t]{0.31\textwidth} 
        \centering
        \includegraphics[width=0.8\textwidth, trim=0cm -0.8cm 0cm 0cm, clip]{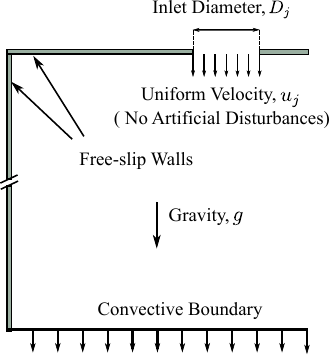} 
        \caption{Boundary conditions for the liquid jet simulation. A circular inlet with a diameter of $D_j$ imposes a uniform velocity $u_j$. Free-slip walls are applied to the sides, while a convective boundary condition is used at the outlet. Gravity acts along the vertical direction.}
        \label{fig:jet_boundary_conditions}
    \end{subfigure}
    \hfill
    \begin{subfigure}[t]{0.31\textwidth} 
        \centering
        \includegraphics[width=0.8\textwidth, trim=0cm -0.5cm 0cm 0cm, clip]{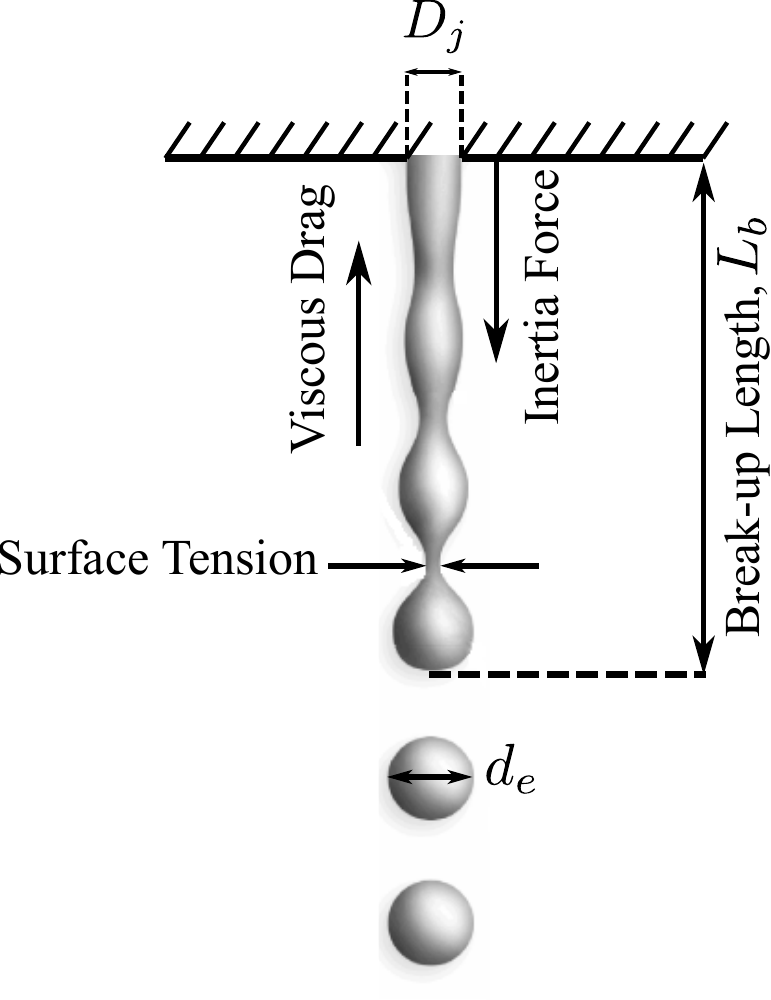} 
        \caption{Schematic representation of jet breakup forces. The interplay of surface tension, viscous drag, and inertial forces leads to the formation of droplets. The breakup length ($L_b$) and equilibrium droplet diameter ($d_e$) are highlighted.}
        \label{fig:jet_forces}
    \end{subfigure}
    \caption{Liquid jet setup for numerical simulations. (a) The computational domain layout. (b) Boundary conditions, including a uniform velocity inlet, free-slip walls, and a convective outflow boundary. (c) Forces affecting jet breakup, including viscous drag, inertia, and surface tension, with key parameters like breakup length and droplet diameter.}
    \label{fig:jet_setup}
\end{figure}

%% file: Tables/jetSpec.tex
\begin{table}[htbp]
\centering
\scriptsize
\renewcommand{\arraystretch}{1.5} 
\setlength{\tabcolsep}{8pt} 
\caption{Conditions of the liquid jet simulations. [*] {These conditions are chosen as the same as the target experimental study.}}
\label{tab:jet_spec}
\begin{tabular}{lcccccc}
\hline
\textbf{Case} & \boldmath{$\rho^\ast$} & \boldmath{$\mu^\ast$} & \boldmath{$Re$} & \boldmath{$We$} & \boldmath{$Fr$} \\ 
\hline
a$^*$ & 814 & 55 & 151 & 0.44 & 2.60 \\ 
b & 814 & 55 & 211 & 0.88 & 3.65 \\ 
c$^*$ & 814 & 55 & 228 & 1.03 & 3.94 \\ 
d$^*$ & 814 & 55 & 302 & 1.79 & 5.22 \\ 
e & 814 & 55 & 332 & 2.17 & 5.74 \\ 
f & 814 & 55 & 340 & 2.27 & 5.87 \\ 
g & 814 & 55 & 347 & 2.37 & 6.01 \\ 
h & 814 & 55 & 362 & 2.58 & 6.26 \\ 
i$^*$ & 814 & 55 & 397 & 3.10 & 6.86 \\ 
j$^*$ & 814 & 55 & 408 & 3.27 & 7.05 \\ 
k & 814 & 55 & 423 & 3.52 & 7.31 \\ 
l$^*$ & 814 & 55 & 454 & 4.05 & 7.84 \\ 
\hline
\end{tabular}
\end{table}

%% file: Figures/dripping.tex
\begin{figure}[htbp]
    \centering
    \begin{subfigure}[t]{0.47\textwidth} 
        \centering
        \includegraphics[width=\textwidth, trim=0cm -1cm 0cm 0cm, clip]{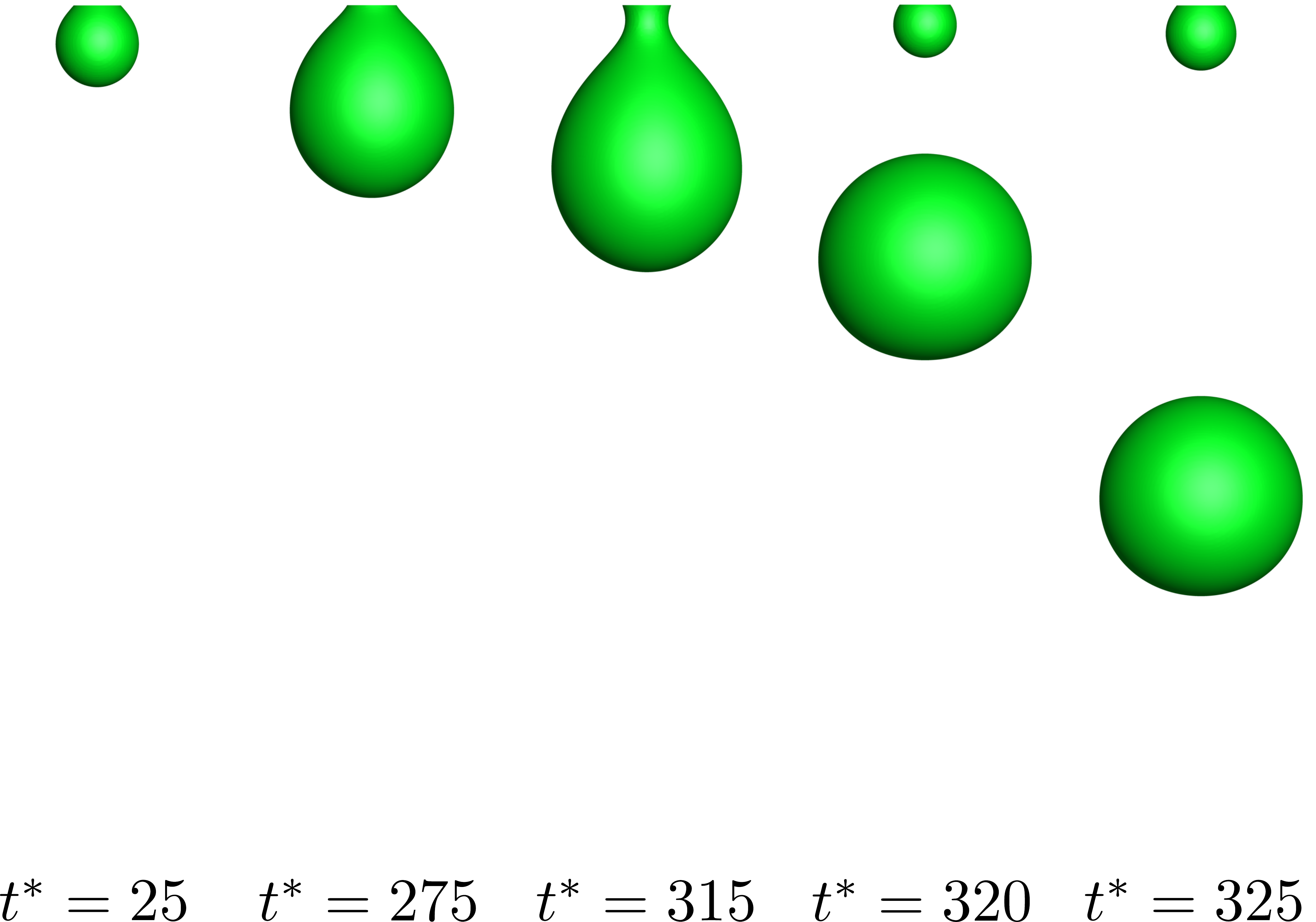} 
        \caption{Interface evolution of liquid jet breakup in the dripping regime. (a) $We = 0.44$, $Re = 151$, $Fr = 2.60$. The snapshots capture drop formation at different time steps, where the breakup occurs predominantly at the tip of the nozzle.}
        \label{fig:jet_dripping_a}
    \end{subfigure}
    \hfill
    \begin{subfigure}[t]{0.47\textwidth} 
        \centering
        \includegraphics[width=\textwidth, trim=0cm -1cm 0cm 0cm, clip]{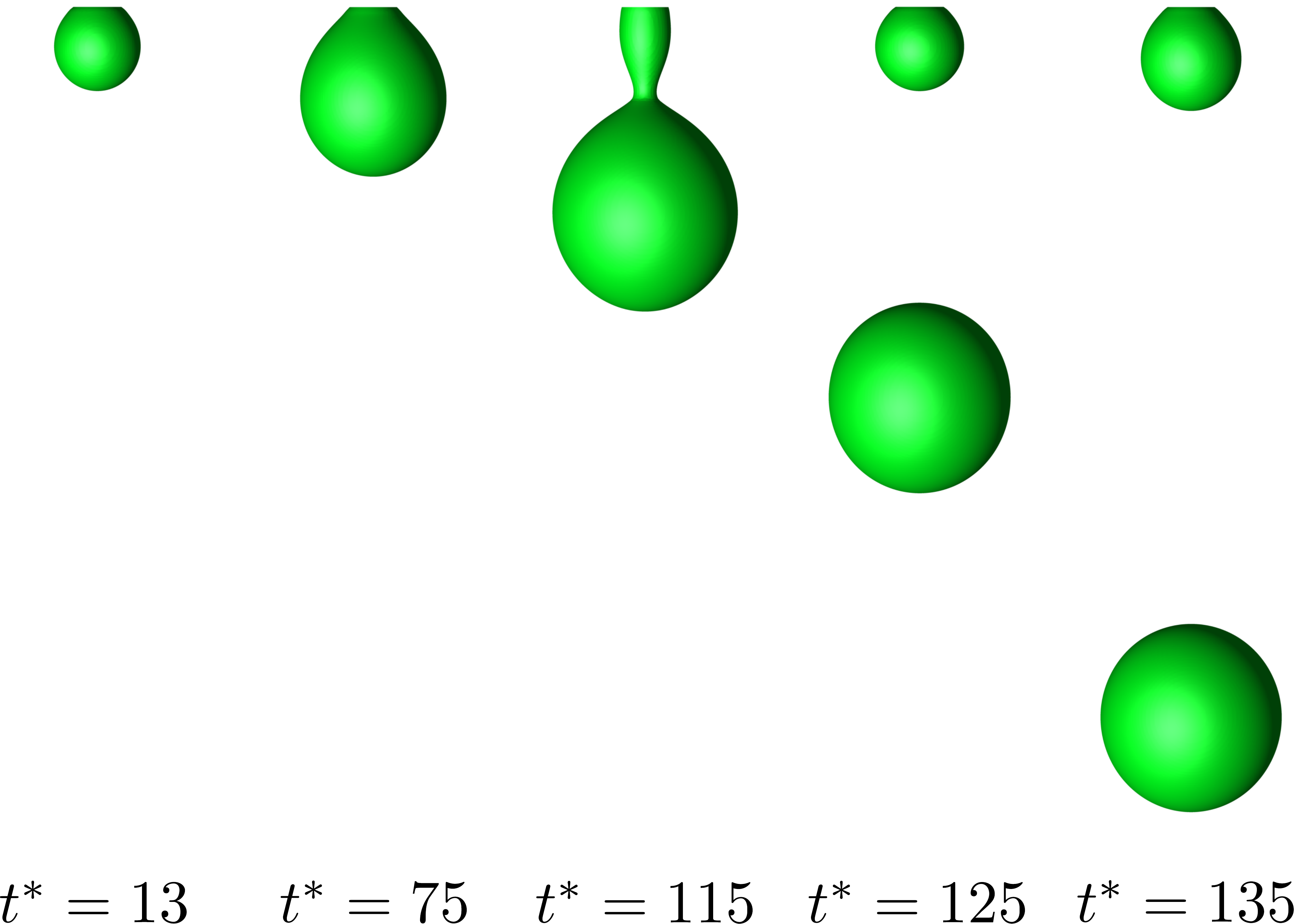} 
        \caption{Interface evolution of liquid jet breakup in the dripping regime. (b) $We = 1.79$, $Re = 302$, $Fr = 5.22$. The snapshots show the generation of larger droplets and longer breakup lengths compared to (a) due to increased inertial effects.}
        \label{fig:jet_dripping_b}
    \end{subfigure}
    \caption{Interface evolution of liquid jet breakup in the dripping regime. (a) $We = 0.44$, $Re = 151$, $Fr = 2.60$. (b) $We = 1.79$, $Re = 302$, $Fr = 5.22$. The computational domain is set to $120 \times 120 \times 300$. In the dripping regime, drop formation occurs mainly at the tip of the nozzle, with increased Weber and Reynolds numbers leading to larger droplets and longer breakup lengths.}
    \label{fig:jet_dripping}
\end{figure}

%% file: Figures/jetting.tex
\begin{figure}[htbp]
    \centering
    \begin{subfigure}[t]{0.47\textwidth} 
        \centering
        \includegraphics[width=\textwidth, trim=0cm -1cm 0cm 0cm, clip]{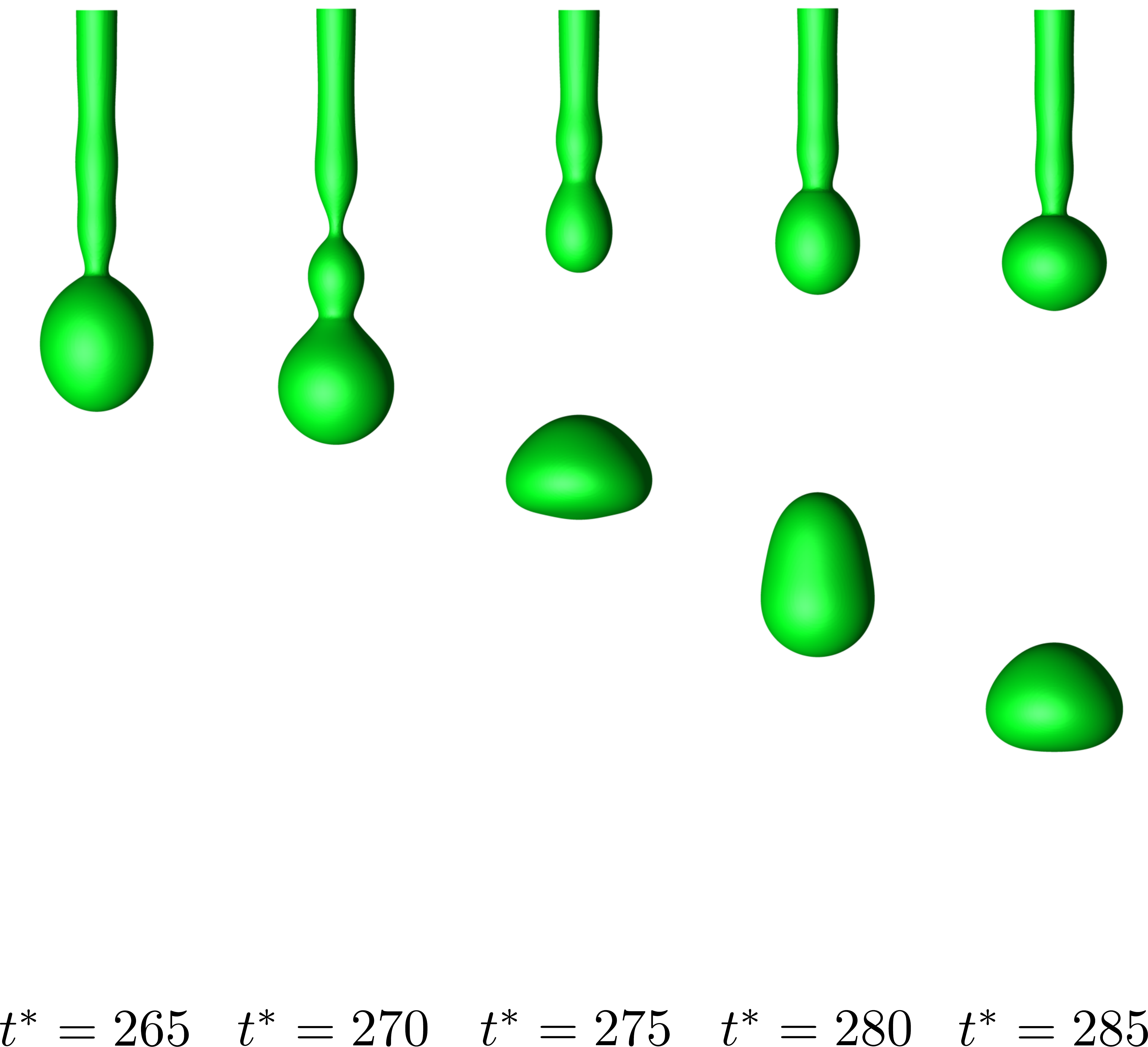} 
        \caption{Interface evolution of liquid jet breakup in the jetting regime. (a) $We = 3.52$, $Re = 423$, $Fr = 7.31$. The Rayleigh waves grow on the axisymmetric jet, leading to the formation of drops through capillary instability.}
        \label{fig:jet_jetting_a}
    \end{subfigure}
    \hfill
    \begin{subfigure}[t]{0.47\textwidth} 
        \centering
        \includegraphics[width=\textwidth, trim=0cm -1cm 0cm 0cm, clip]{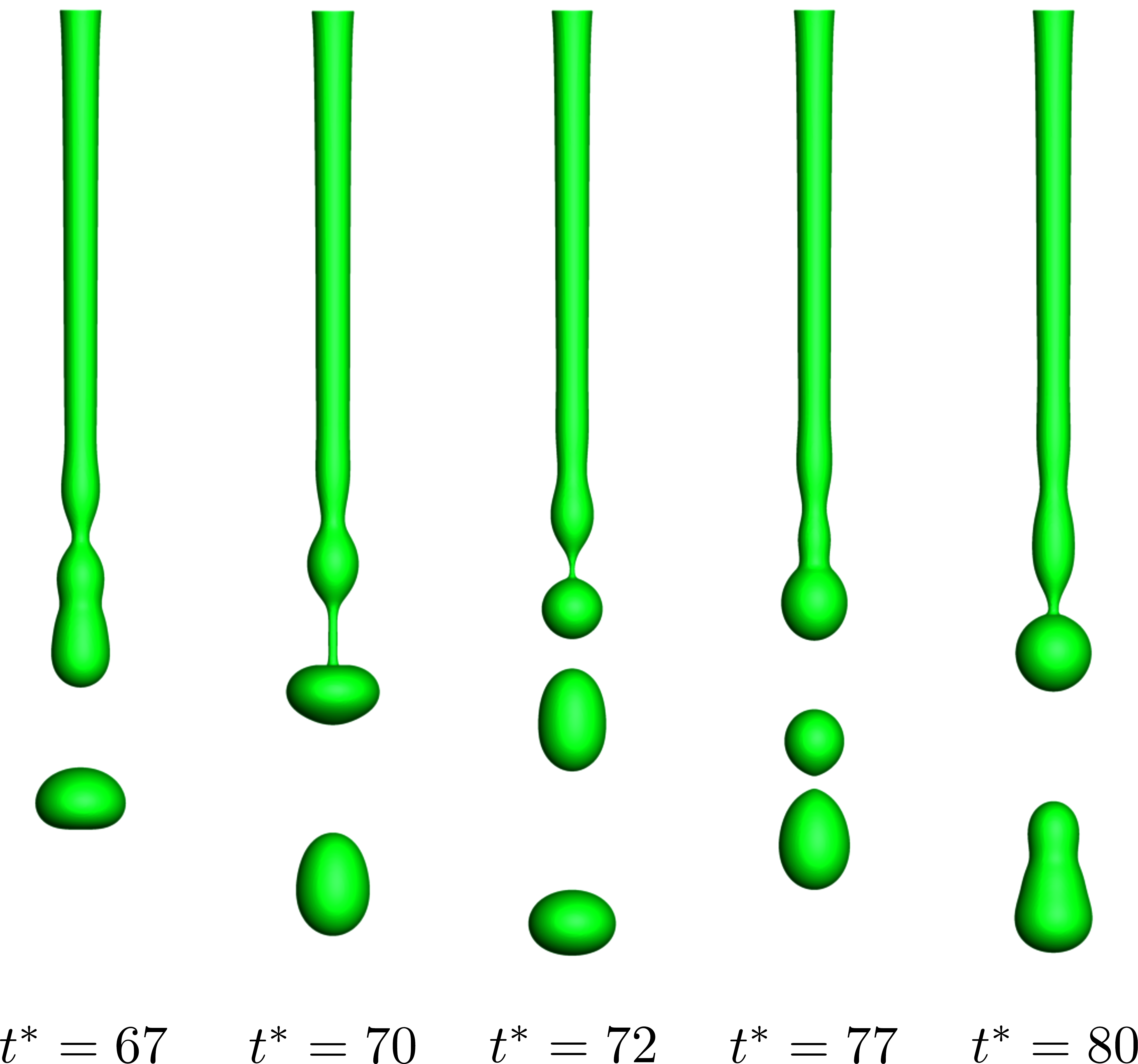} 
        \caption{Interface evolution of liquid jet breakup in the jetting regime. (b) $We = 4.05$, $Re = 454$, $Fr = 7.84$. The breakup occurs further downstream, producing a series of droplets along the column of the liquid jet.}
        \label{fig:jet_jetting_b}
    \end{subfigure}
    \caption{Interface evolution of liquid jet breakup in the jetting regime. (a) $We = 3.52$, $Re = 423$, $Fr = 7.31$. (b) $We = 4.05$, $Re = 454$, $Fr = 7.84$. The computational domain is set to $120 \times 120 \times 400$. In the jetting regime, Rayleigh waves dominate, causing the axisymmetric liquid jet to form droplets as it progresses.}
    \label{fig:jet_jetting}
\end{figure}

%% file: Figures/dropSize.tex
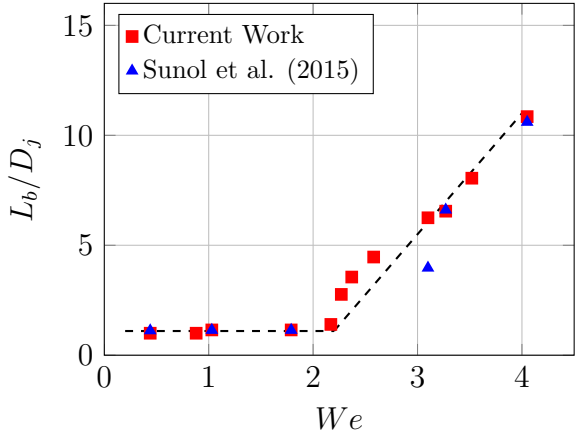
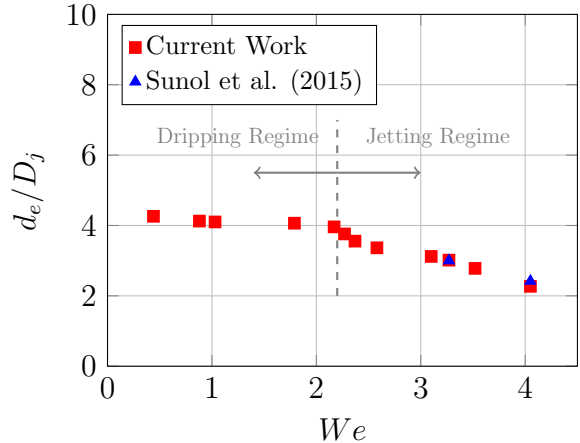
\begin{figure}[htbp]
    \centering
    \begin{subfigure}[t]{0.46\textwidth}
        \centering
        \begin{tikzpicture}
        \begin{axis}[
            xlabel={$We$},
            ylabel={$L_b / D_j$},
            legend style={at={(0.03,0.97)}, anchor=north west, font=\footnotesize},
            legend cell align={left},
            ymin=0, ymax=16,
            xmin=0, xmax=4.5,
            grid=major,
            width=\textwidth,
            height=0.8\textwidth,
        ]
            \addplot[red, mark=square*, only marks, thick] table[x=weLBM,y=lLBM,col sep=comma] {Data/jetBreakupLength.csv};
            \addlegendentry{Current Work}

            \addplot[blue, mark=triangle*, only marks, thick] table[x=weSunol,y=lSunol,col sep=comma] {Data/jetBreakupLength.csv};
            \addlegendentry{Sunol et al. (2015)}

            \addplot[dashed, thick] coordinates {
                (0.2, 1.1)
                (2.2, 1.1)
            };

            \addplot[dashed, thick] coordinates {
                (2.2, 1.1)
                (4, 11)
            };
        \end{axis}
        \end{tikzpicture}
        \caption{Breakup length ($L_b / D_j$) as a function of the Weber number ($We$) for liquid jet breakup. Current numerical results are compared with experimental data from \cite{Sunol2015}.}
        \label{fig:jet_breakup_length}
    \end{subfigure}
    \hfill
    \begin{subfigure}[t]{0.46\textwidth}
        \centering
        \begin{tikzpicture}
        \begin{axis}[
            xlabel={$We$},
            ylabel={$d_e / D_j$},
            legend style={at={(0.03,0.97)}, anchor=north west, font=\footnotesize},
            legend cell align={left},
            ymin=0, ymax=10,
            xmin=0, xmax=4.5,
            grid=major,
            width=\textwidth,
            height=0.8\textwidth,
        ]
            \addplot[red, mark=square*, only marks, thick] table[x=weLBM,y=dLBM,col sep=comma] {Data/dropMeanSize.csv};
            \addlegendentry{Current Work}

            \addplot[blue, mark=triangle*, only marks, thick] table[x=weSunol,y=dSunol,col sep=comma] {Data/dropMeanSize.csv};
            \addlegendentry{Sunol et al. (2015)}

            \addplot[dashed, thick, gray] coordinates {
                (2.2, 2)
                (2.2, 7)
            };

            \addplot[->, thick, gray] coordinates {
                (2.2, 5.5)
                (3, 5.5)
            };

            \addplot[->, thick, gray] coordinates {
                (2.2, 5.5)
                (1.4, 5.5)
            };

            \node[anchor=south west, gray, scale=0.7] at (axis cs:0.4, 6.0) {Dripping Regime};
            \node[anchor=south west, gray, scale=0.7] at (axis cs:2.4, 6.0) {Jetting Regime};
        \end{axis}
        \end{tikzpicture}
        \caption{Droplet size ($d_e / D_j$) as a function of the Weber number ($We$). The transition from dripping to jetting regimes is highlighted, showing agreement with experimental results from \cite{Sunol2015}.}
        \label{fig:jet_droplet_size}
    \end{subfigure}
    \caption{Comparison of numerical and experimental results for liquid jet breakup. (a) Breakup length ($L_b / D_j$) as a function of $We$, showing increasing length with higher Weber numbers. (b) Droplet size ($d_e / D_j$) decreases with increasing $We$, highlighting the transition between dripping and jetting regimes. Current work (LBM-PF) aligns well with experimental findings from \cite{Sunol2015}.}
    \label{fig:jet_breakup_analysis}
\end{figure}

%% file: Figures/transition.tex
\begin{figure}[htbp]
    \centering
    \begin{subfigure}[t]{0.47\textwidth} 
        \centering
        \includegraphics[width=\textwidth, trim=0cm -1cm 0cm 0cm, clip]{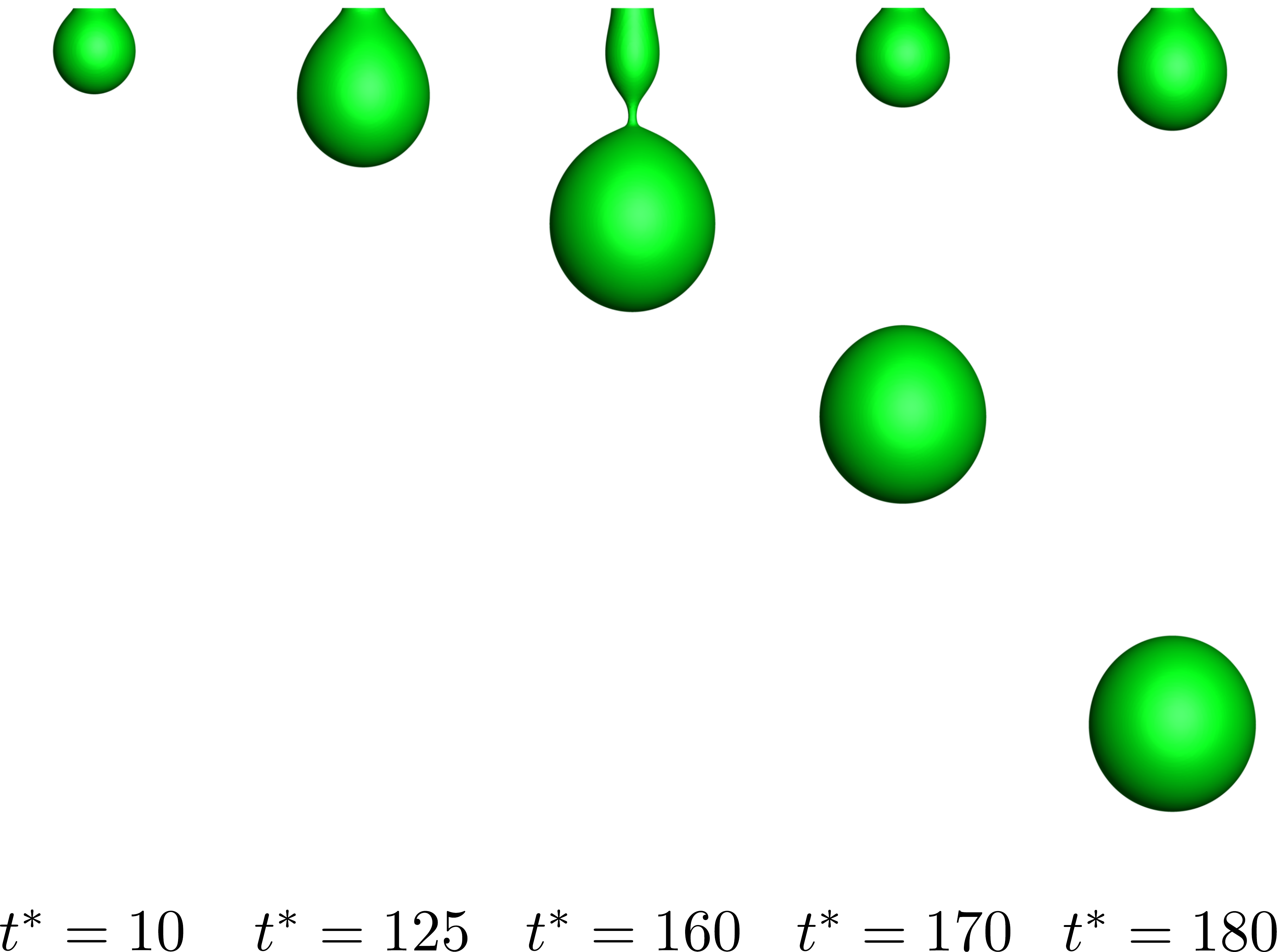} 
        \caption{Interface evolution of liquid jet breakup in the transition regime. (a) $We = 2.17$, $Re = 332$, $Fr = 5.74$. The Rayleigh waves begin to grow on the liquid jet surface, leading to longer breakup lengths and an increased transition toward jetting behavior.}
        \label{fig:jet_transition_a}
    \end{subfigure}
    \hfill
    \begin{subfigure}[t]{0.47\textwidth} 
        \centering
        \includegraphics[width=\textwidth, trim=0cm -1cm 0cm 0cm, clip]{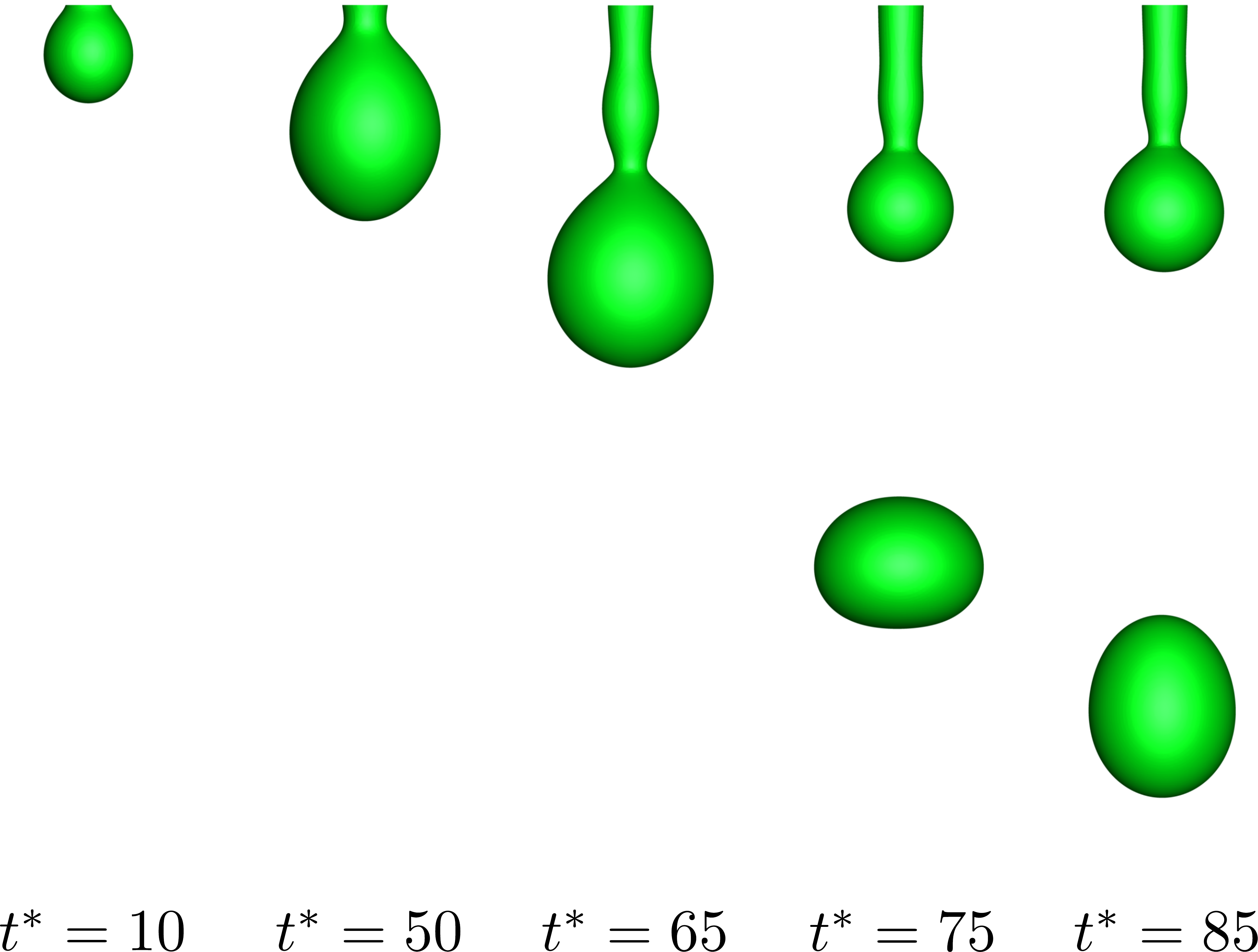} 
        \caption{Interface evolution of liquid jet breakup in the transition regime. (b) $We = 2.27$, $Re = 340$, $Fr = 5.87$. The perturbations on the liquid jet surface grow faster, and the breakup length increases, transitioning further toward the jetting regime.}
        \label{fig:jet_transition_b}
    \end{subfigure}
    \caption{Interface evolution of liquid jet breakup in the transition regime. (a) $We = 2.17$, $Re = 332$, $Fr = 5.74$. (b) $We = 2.27$, $Re = 340$, $Fr = 5.87$. The computational domain is set to $120 \times 120 \times 300$. Increasing the Weber number leads to growing perturbations on the liquid jet surface, extending the breakup length and moving toward jetting behavior.}
    \label{fig:jet_transition}
\end{figure}

%% file: Figures/OhRe.tex
\begin{figure}[ht]
    \centering
    \begin{tikzpicture}[scale=0.9]
\begin{loglogaxis}[
    xlabel={$Re$},
    ylabel={$Oh$},
    xmin=100, xmax=620,
    ymin=0.001, ymax=1,
    xtick={100,200,300,400,500,600},
    xticklabels={100, 200, 300, 400, 500, 600}, 
    grid=both, 
    minor grid style={dotted}, 
    major grid style={solid}, 
    width=0.75\textwidth,
    height=0.6\textwidth,
    legend style={at={(0.03,0.97)}, anchor=north west, font=\footnotesize},
    legend cell align={left},
    log basis x={10}, 
    log basis y={10}, 
    tick label style={font=\normalsize}, 
    label style={font=\normalsize}, 
]

        \addplot[red, solid, thick, mark=*, only marks, mark size=3.5] table[x=ReDripping, y=OhDripping, col sep=comma] {Data/ReOh.csv};
        \addlegendentry{Dripping Regime}

        \addplot[black, solid, thick, mark=triangle*, only marks, mark size=3.5] table[x=ReJetting, y=OhJetting, col sep=comma] {Data/ReOh.csv};
        \addlegendentry{Jetting Regime}

        \node[anchor=north] at (axis cs:500, 0.5) {
            \includegraphics[width=0.5cm]{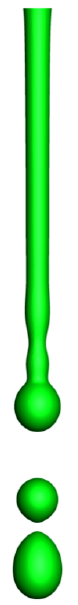} 
        };

        \node[anchor=north] at (axis cs:350, 0.3) {
            \includegraphics[width=0.5cm]{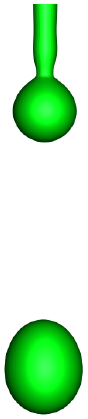} 
        };

        \node[anchor=north] at (axis cs:200, 0.3) {
            \includegraphics[width=0.5cm]{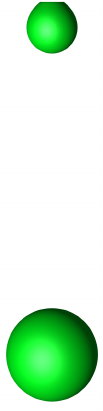} 
        };

        \node[dashed, anchor=north, draw=black, line width=0.2pt] at (axis cs:500, 0.5) {
    \includegraphics[width=0.5cm]{Images/OhRePic1.pdf}
};

\node[dashed,anchor=north, draw=black, line width=0.2pt] at (axis cs:350, 0.3) {
    \includegraphics[width=0.5cm]{Images/OhRePic2.pdf}
};

\node[dashed,anchor=north, draw=black, line width=0.2pt] at (axis cs:200, 0.3) {
    \includegraphics[width=0.5cm]{Images/OhRePic3.pdf}
};

        \addplot [
            fill=blue, 
            fill opacity=0.1, 
            draw=none 
        ] coordinates {
            (100, 0.1) 
            (620, 0.0015) 
            (620, 1)  
            (100, 1)  
        };

                \addplot [
            fill=red, 
            fill opacity=0.1, 
            draw=none 
        ] coordinates {
            (100, 0.001) 
            (620, 0.001) 
            (620, 0.0015)  
            (100, 0.1)  
        };

        \addplot[solid, thick, black] coordinates {
        (100, 0.1) 
        (620, 0.0015) 
    };

    \addplot[->, thick, black] coordinates {
        (450, 0.005) 
        (480, 0.01) 
    };

    \addplot[->, thick, black] coordinates {
        (340, 0.005) 
        (350, 0.015) 
    };
    
        \addplot[->, thick, black] coordinates {
        (210, 0.005) 
        (200, 0.015) 
    };

\node[anchor=center, black, scale=0.9] at (axis cs:250, 0.5) {Jetting Regime};
\node[anchor=center, black, scale=0.9] at (axis cs:150, 0.002) {Dripping Regime};

    \end{loglogaxis}
    \end{tikzpicture}
\caption{Phase diagram of liquid jet breakup regimes as a function of Reynolds number ($Re$) and Ohnesorge number ($Oh$). The diagram highlights two distinct regimes: the dripping regime (red circles) and the jetting regime (black triangles). The boundary between these regimes is defined by the critical $Oh$ as a function of $Re$ \cite{Ohnesorge1936}. Insets show representative snapshots of the liquid jet morphology for selected cases, illustrating the transition from axisymmetric drops in the dripping regime to the formation of Rayleigh waves and breakup patterns in the jetting regime.}

    \label{fig:Oh_Re}
\end{figure}

%% file: Sections/Conclusion.tex
\section{Conclusion}
\label{sec:conclusion}


This study developed an efficient GPU-accelerated phase-field lattice Boltzmann (LB) framework for simulating immiscible two-phase flows in three-dimensional domains. Leveraging the D3Q19 lattice arrangement and the weighted MRT collision operator (WMRT), the model ensures high accuracy and numerical stability while effectively capturing interfacial dynamics. The integration of the Allen-Cahn equation further enhances interface tracking in liquid-gas systems. By implementing the framework in CUDA, we achieved significant computational acceleration, enabling large-scale, high-fidelity simulations on personal computers with optimized memory usage and parallel processing capabilities. The open-source implementation of this framework is available at \href{https://github.com/mshadkhah/CLIP}{\textcolor{green!40!black}{GitHub}} page.

The proposed model was rigorously validated through various benchmark problems, including the stationary drop test, two-phase Poiseuille flow, capillary wave dynamics, a circular interface in shear flow, and Rayleigh-Taylor instability. The stationary drop test demonstrated excellent agreement with Laplace's analytical solution, achieving a maximum error of $1.04\%$. The two-phase Poiseuille flow test confirmed nearly second-order accuracy and stability under body forces. The capillary wave simulations accurately captured the temporal evolution of interfacial oscillations, showing close agreement with analytical predictions and demonstrating the model's ability to handle surface tension-driven dynamics. The Rayleigh-Taylor instability simulations, conducted with properties matching prior studies \cite{He1999_Zhang, Zu2013}, showed excellent consistency in predicting the bubble front, spike tip, and saddle point positions.

To further evaluate the model, liquid jet breakup simulations were performed based on experimental data from Ref. \cite{Sunol2015}. The model successfully handled high-density contrast and complex interface dynamics, maintaining numerical stability. Quantitative comparisons of breakup length and drop sizes with experimental results highlighted the model's reliability in predicting jet breakup characteristics. The GPU-accelerated implementation enabled these high-resolution simulations, significantly reducing computational costs. In contrast, performing these simulations using a sequential or CPU-parallelized approach would be computationally expensive and impractical for large-scale three-dimensional cases. 

Simulations conducted across the Ohnesorge regime map for liquid-gas systems provided valuable insights into jet breakup regimes. A critical Weber number (${We}_{cr} \approx 2.2$) was identified as the transition point between dripping and jetting regimes. For $We < 2.2$, the dripping regime was dominated by surface tension forces, resulting in larger drops forming at the nozzle outlet. In contrast, for $We > 2.2$, the jetting regime was characterized by rapid surface perturbations and Rayleigh instability, causing drops to form before significant growth.

%% file: Sections/Appendix.tex
\appendix
\section{Transformation Matrices}
\label{sec:appendixA}

The transformation matrix $M$ for D2Q9 model is given by:

{\scriptsize
\begin{equation}
M =\begin{pmatrix}
1 & 1 & 1 & 1 & 1 & 1 & 1 & 1 & 1  \\
-4 & -1 & -1 & -1 & -1 & 2 & 2 & 2 & 2  \\
4 & -2 & -2 & -2 & -2 & 1 & 1 & 1 & 1  \\
0 & 1 & 0 & -1 & 0 & 1 & -1 & -1 & 1  \\
0 & -2 & 0 & 2 & 0 & 1 & -1 & -1 & 1  \\
0 & 0 & 1 & 0 & -1 & 1 & 1 & -1 & -1  \\
0 & 0 & -2 & 0 & 2 & 1 & 1 & -1 & -1  \\
0 & 1 & -1 & 1 & -1 & 0 & 0 & 0 & 0  \\
0 & 0 & 0 & 0 & 0 & 1 & -1 & 1 & -1  
\end{pmatrix}.
\label{eq:A1}
\end{equation}
}

The weighted transformation matrix $M$ for D3Q19 model is given by:
\setcounter{MaxMatrixCols}{20}

{\scriptsize
\begin{equation}
M =\begin{pmatrix}
1 & 1 & 1 & 1 & 1 & 1 & 1 & 1 & 1 & 1 & 1 & 1 & 1 & 1 & 1 & 1 & 1 & 1 & 1 \\
0 & 1 & -1 & 0 & 0 & 0 & 0 & 1 & -1 & 1 & -1 & 1 & -1 & 1 & -1 & 0 & 0 & 0 & 0 \\
0 & 0 & 0 & 1 & -1 & 0 & 0 & 1 & -1 & -1 & 1 & 0 & 0 & 0 & 0 & 1 & -1 & 1 & -1 \\
0 & 0 & 0 & 0 & 0 & 1 & -1 & 0 & 0 & 0 & 0 & 1 & -1 & -1 & 1 & 1 & -1 & -1 & 1 \\
0 & 0 & 0 & 0 & 0 & 0 & 0 & 1 & 1 & -1 & -1 & 0 & 0 & 0 & 0 & 0 & 0 & 0 & 0 \\
0 & 0 & 0 & 0 & 0 & 0 & 0 & 0 & 0 & 0 & 0 & 0 & 0 & 0 & 0 & 1 & 1 & -1 & -1 \\
0 & 0 & 0 & 0 & 0 & 0 & 0 & 0 & 0 & 0 & 0 & 1 & 1 & -1 & -1 & 0 & 0 & 0 & 0 \\
0 & 2 & 2 & -1 & -1 & -1 & -1 & 1 & 1 & 1 & 1 & 1 & 1 & 1 & 1 & -2 & -2 & -2 & -2 \\
0 & 0 & 0 & 1 & 1 & -1 & -1 & 1 & 1 & 1 & 1 & -1 & -1 & -1 & -1 & 0 & 0 & 0 & 0 \\
-1 & 0 & 0 & 0 & 0 & 0 & 0 & 1 & 1 & 1 & 1 & 1 & 1 & 1 & 1 & 1 & 1 & 1 & 1 \\
0 & -2 & 2 & 0 & 0 & 0 & 0 & 1 & -1 & 1 & -1 & 1 & -1 & 1 & -1 & 0 & 0 & 0 & 0 \\
0 & 0 & 0 & -2 & 2 & 0 & 0 & 1 & -1 & -1 & 1 & 0 & 0 & 0 & 0 & 1 & -1 & 1 & -1 \\
0 & 0 & 0 & 0 & 0 & -2 & 2 & 0 & 0 & 0 & 0 & 1 & -1 & -1 & 1 & 1 & -1 & -1 & 1 \\
0 & 0 & 0 & 0 & 0 & 0 & 0 & 1 & -1 & 1 & -1 & -1 & 1 & -1 & 1 & 0 & 0 & 0 & 0 \\
0 & 0 & 0 & 0 & 0 & 0 & 0 & -1 & 1 & 1 & -1 & 0 & 0 & 0 & 0 & 1 & -1 & 1 & -1 \\
0 & 0 & 0 & 0 & 0 & 0 & 0 & 0 & 0 & 0 & 0 & 1 & -1 & -1 & 1 & -1 & 1 & 1 & -1 \\
1 & -2 & -2 & -2 & -2 & -2 & -2 & 1 & 1 & 1 & 1 & 1 & 1 & 1 & 1 & 1 & 1 & 1 & 1 \\
0 & -2 & -2 & 1 & 1 & 1 & 1 & 1 & 1 & 1 & 1 & 1 & 1 & 1 & 1 & -2 & -2 & -2 & -2 \\
0 & 0 & 0 & -1 & -1 & 1 & 1 & 1 & 1 & 1 & 1 & -1 & -1 & -1 & -1 & 0 & 0 & 0 & 0

\end{pmatrix}.
\label{eq:A2}
\end{equation}
}
\section{Boundary Treatments}
\label{sec:appendixB}

Boundary conditions are critical in ensuring the accuracy and stability of the lattice Boltzmann method (LBM), significantly influencing the overall solution. This section presents the implementation of commonly used boundary conditions within the phase-field model framework. Whether applied globally across the domain or locally to specific boundaries, their effects propagate through the flow field, shaping the simulation outcomes. To aid in understanding boundary treatments, \figref{fig:bc_velocity} illustrates a representative 2D/3D channel flow, highlighting the application of boundary conditions and their role in maintaining solution continuity.

\input{Figures/bcVelocity}

\subsection{Periodic Boundary Conditions}
Periodic boundary conditions allow the fluid exiting one side of the domain to reenter through the opposite side, ensuring mass and momentum conservation while maintaining a periodic flow solution. These conditions are particularly effective for simulations with repeating or infinite domain properties.

The computation involves two main steps: collision and streaming.

\textbf{Collision step:}
\begin{align}
f_a^+\left(x,\ t\right) &= f_a\left(x,t\right) + \Omega_a(x,t) + F_a(x,t),
\label{eq:collision_f}\\ 
g_a^+\left(x,\ t\right) &= g_a\left(x,t\right) - \frac{g_a\left(x,t\right) - {\bar{g}}_a^{eq}\left(x,t\right)}{\tau_\phi+1/2} + F_a^\phi(x,t).
\label{eq:collision_g}
\end{align}

\textbf{Streaming step:}
\begin{align}
f_a\left(x,t+\delta t\right) &= f_a^+\left(x-e_a\delta t,\ t\right),
\label{eq:streaming_f}\\ 
g_a\left(x,t+\delta t\right) &= g_a^+\left(x-e_a\delta t,\ t\right).
\label{eq:streaming_g}
\end{align}

During the streaming step, the boundary nodes require information from adjacent imaginary nodes to complete the velocity vector connections. As shown in \figref{fig:bc_velocity}, imaginary boundary nodes are defined at $(x_0-n)$ and $(x_\pi-n)$, where $n$ is the unit normal vector pointing toward fluid nodes. The unknown distribution functions at these locations are calculated as follows:
\begin{align}
f_a^+\left(x_0-n\right) &= f_a^+\left(x_\pi\right),\quad f_a^+\left(x_\pi-n\right) = f_a^+\left(x_0\right),
\label{eq:periodic_f}\\ 
g_a^+\left(x_0-n\right) &= g_a^+\left(x_\pi\right),\quad g_a^+\left(x_\pi-n\right) = g_a^+\left(x_0\right),
\label{eq:periodic_g}\\
\varphi\left(x_0-n\right) &= \varphi\left(x_\pi\right),\quad \varphi\left(x_\pi-n\right) = \varphi\left(x_0\right).
\label{eq:periodic_phi}
\end{align}

Here, $\varphi$ represents macroscopic quantities such as density $\rho$, velocity $u$, or other phase-field variables, used to evaluate their gradients across boundaries. These periodic conditions maintain flow continuity and enable the simulation of domains with repetitive characteristics.
\input{Figures/bcWall}

\subsection{\label{app:subsec_free_slip}Free-slip Boundary}
The free-slip boundary condition allows the tangential component of the fluid velocity to remain unrestricted while enforcing zero normal velocity at solid nodes \cite{Kruger2017}. This condition is particularly useful for modeling frictionless walls. When implementing a free-slip boundary, populations leaving the boundary node $x_b$ at time $t$ encounter the free-slip wall surface at time $t + \Delta t/2$, as illustrated in \figref{fig:free_slip}.

The algorithm for free-slip conditions replaces the populations on wall nodes pointing into fluid nodes $e_a$ with the corresponding opposing populations $e_{\bar{a}}$ on the fluid node. At time $t + \Delta t$, the populations stream back to $x_b$ or its neighboring nodes. The standard streaming step is modified as follows to accommodate the free-slip boundary condition:
\begin{align} 
f_{\bar{a}}\left(x_b + e_{\bar{a}} \delta t, t + \delta t \right) &= f_a^+\left(x_b, t \right),
\label{eq:free_slip_f} \\
g_{\bar{a}}\left(x_b + e_{\bar{a}} \delta t, t + \delta t \right) &= g_a^+\left(x_b, t \right),
\label{eq:free_slip_g}
\end{align}
where $e_{a,t} = e_{\bar{a},t}$ represents populations in the tangential direction. By ensuring $e_a = e_{\bar{a}}$, the free-slip boundary condition enforces zero normal velocity while maintaining tangential flow.

\subsection{\label{app:subsec_no_slip}No-slip Boundary}
The no-slip boundary condition enforces zero velocity at solid walls by reversing the populations that interact with the boundary. In this study, the halfway bounce-back method is employed for solid boundaries, as described in \cite{Ladd2001}. When populations propagate toward a rigid barrier, the no-slip condition reflects them back to their origin. 

For a no-slip boundary, populations leaving the boundary node $x_b$ at time $t$ collide with the wall surface at time $t + \Delta t/2$, as shown in \figref{fig:no_slip}. The incoming populations at $x_b$ are replaced by the outgoing populations on the same node. The halfway bounce-back method introduces a time delay of $\Delta t$. The modified streaming step for the no-slip boundary is given by:
\begin{align} 
f_{\bar{a}}\left(x_b, t + \delta t \right) &= f_a^+\left(x_b, t \right),
\label{eq:no_slip_f} \\
g_{\bar{a}}\left(x_b, t + \delta t \right) &= g_a^+\left(x_b, t \right).
\label{eq:no_slip_g}
\end{align}

These boundary conditions ensure accurate and stable enforcement of velocity constraints at solid walls, preserving the physical behavior of the flow in simulations.
\subsection{Velocity Boundary}
The implementation of velocity boundary conditions is crucial for ensuring accurate flow simulations in lattice Boltzmann models. As shown in \figref{fig:bc_velocity}, the streaming step cannot be fully completed at the inlet boundary because several distribution function populations are unknown. Specifically, these are populations in the direction of $\alpha = 1, 5, 8$ for the D2Q9 model and $\alpha = 1, 7, 9, 11, 13$ for the D3Q19 model.

At the inlet boundary ($x_{\text{in}}$), the velocity distribution function $f_a^\prime = f_a + 1/2F_a$, defined in terms of Eqs.~\eqref{eq:force_term} and \eqref{eq:normalized_pressure}, must satisfy the following condition:
\begin{equation}
\sum_{a}{f_a^\prime(x_{\text{in}}) e_a} = \sum_{a}{f_a^{eq}(x_{\text{in}}) e_a} = u_{\text{in}},
\label{eq:velocity_boundary_u}
\end{equation}
where $u_{\text{in}}$ is the predefined inlet velocity. Considering the nonequilibrium component of $f_a^\prime$, the normal velocity component to the inlet boundary is given by:
\begin{equation}
f_a^\prime(x_{\text{in}}) - f_a^{eq}(x_{\text{in}}) = f_{\bar{a}}^\prime(x_{\text{in}}) - f_{\bar{a}}^{eq}(x_{\text{in}}),
\label{eq:velocity_boundary_normal}
\end{equation}
ensuring momentum conservation in the normal direction. Excess momentum in the tangential directions is expressed as:
\begin{equation}
M(x_{\text{in}}) = \sum_{a \in \Lambda} e_a \left[f_a^\prime(x_{\text{in}}) - f_a^{eq}(x_{\text{in}})\right],
\label{eq:velocity_boundary_tangential}
\end{equation}
where $\Lambda = \{a | e_a \cdot n = 0\}$. To conserve momentum, $M$ can be redistributed among the unknown populations as:
\begin{equation}
\begin{split}
f_a^\prime(x_{\text{in}}) &= f_{\bar{a}}^\prime(x_{\text{in}}) + \left[f_a^{eq}(x_{\text{in}}) - f_{\bar{a}}^{eq}(x_{\text{in}})\right] \\
& \quad - \frac{1}{N c^2} e_a \cdot M(x_{\text{in}}),
\end{split}
\label{eq:velocity_boundary_distribution}
\end{equation}
where $N$ is the total number of unknown populations with $e_a \cdot n \neq 0$, which equals $2$ for the D2Q9 model and $4$ for the D3Q19 model.

Mass conservation at the inlet boundary is ensured by the distribution function $g_a$, as required by:
\begin{equation}
\sum_{a}{g_a(x_{\text{in}})} = \sum_{a}{g_a^{eq}(x_{\text{in}})} = \varphi_{\text{in}},
\label{eq:mass_conservation}
\end{equation}
where $\varphi_{\text{in}}$ is the inlet phase-field variable. For the nonequilibrium component of $g_a$, the following relationship holds:
\begin{equation}
g_a(x_{\text{in}}) - g_a^{eq}(x_{\text{in}}) = -\left[g_{\bar{a}}(x_{\text{in}}) - g_{\bar{a}}^{eq}(x_{\text{in}})\right].
\label{eq:phase_field_nonequilibrium}
\end{equation}

The excess density at the inlet boundary is given by:
\begin{equation}
\vartheta(x_{\text{in}}) = \sum_{a \in \Lambda} g_a^{\text{neq}}(x_{\text{in}}),
\label{eq:excess_density}
\end{equation}
where $g_a^{\text{neq}} = g_a - g_a^{eq}$ represents the nonequilibrium component of the phase-field distribution function. The unknown distribution functions can then be calculated as:
\begin{align}
g_a(x_{\text{in}}) &= g_a^{eq}(x_{\text{in}}) - g_{\bar{a}}^{\text{neq}}, & \text{if } e_a \cdot n = 0, \label{eq:unknown_distribution_normal} \\
g_a(x_{\text{in}}) &= g_a^{eq}(x_{\text{in}}) - g_{\bar{a}}^{\text{neq}} - \frac{\vartheta(x_{\text{in}})}{N}, & \text{if } e_a \cdot n \neq 0. \label{eq:unknown_distribution_tangential}
\end{align}

First-order derivatives and Laplacians of macro variables at the inlet boundary can be computed using one-sided biased differences. However, because the inlet boundary typically consists of a single-phase fluid, it is unnecessary to evaluate derivatives of macro variables related to interfacial forces at the inlet.

\input{Figures/bcConvective}
\subsection{Convective Boundary}
For open boundaries, the convective boundary condition is one of the most effective methods for maintaining numerical stability and accuracy \cite{Lou2013, Yang2013}. At the outlet boundary, as shown in \figref{fig:bc_convective}, the convective boundary condition is applied to ensure proper outflow behavior.

The convective boundary condition for the distribution function is expressed as:
\begin{equation}
\frac{\partial \chi}{\partial t} + U \frac{\partial \chi}{\partial y} = 0, \qquad y = N,
\label{eq:convective_bc}
\end{equation}
where $U$ represents the characteristic velocity normal to the outlet boundary \cite{Orlanski1976}. The velocity $U$ can be chosen using one of the following schemes:
\begin{equation}
U =
\begin{cases}
U_{\text{max}}(t) \equiv \max \{ u(x, N-1, t) \}, \\
U_{\text{ave}}(t) \equiv \frac{1}{M+1} \sum_{x}{u(x, N-1, t)}, \\
U_{\text{local}}(x, N-1, t) \equiv u(x, N-1, t),
\end{cases}
\label{eq:velocity_options}
\end{equation}
where $0 \leq x \leq M$, and $M+1$ is the number of nodes in the $\left(N-1\right)$th layer along the $x$-direction. It is important to note that $U$ is time-dependent for all three schemes and is computed based on the distribution function from the previous time step at the outlet. In the local velocity scheme, $U_{\text{local}}$ also depends on the spatial location.

To discretize Eq.~\eqref{eq:convective_bc}, this study employs a first-order implicit scheme. The time and spatial derivatives are approximated as:
\begin{gather}
\frac{\partial \chi}{\partial t}(x, N) = \frac{\chi(x, N, t+\delta t) - \chi(x, N, t)}{\delta t}, 
\label{eq:time_derivative} \\
\frac{\partial \chi}{\partial y}(x, N) = \frac{\chi(x, N, t+\delta t) - \chi(x, N-1, t+\delta t)}{\delta y}.
\label{eq:spatial_derivative}
\end{gather}

Substituting these approximations into Eq.~\eqref{eq:convective_bc} gives:
\begin{equation}
\chi(x, N, t+\delta t) = \frac{\chi(x, N, t) + \lambda \chi(x, N-1, t+\delta t)}{1 + \lambda},
\label{eq:implicit_convective}
\end{equation}
where $\lambda = U(t+\delta t) \delta t / \delta y$, determined after the streaming step in the $\left(N-1\right)$th layer. Using this formulation, the distribution functions at the outlet boundary are updated as:
\begin{gather}
f_i(x, N, t+\delta t) = \frac{f_i(x, N, t) + \lambda f_i(x, N-1, t+\delta t)}{1 + \lambda},
\label{eq:convective_f} \\
g_i(x, N, t+\delta t) = \frac{g_i(x, N, t) + \lambda g_i(x, N-1, t+\delta t)}{1 + \lambda}.
\label{eq:convective_g}
\end{gather}

This approach ensures that mass and momentum conservation are preserved while maintaining numerical stability at the outlet boundary.

%% file: Figures/bcVelocity.tex
\begin{figure}[htbp]
    \centering
    \includegraphics[width=0.45\textwidth]{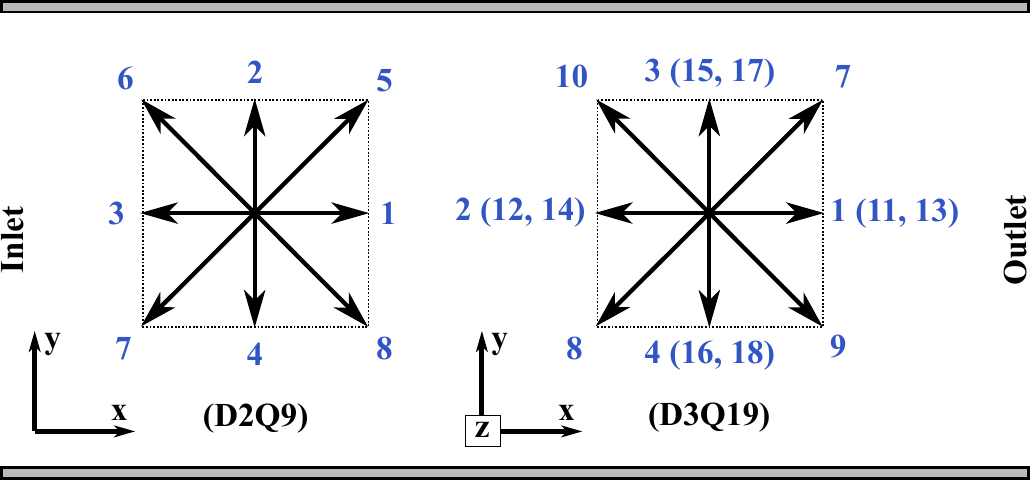} 
\caption{Discrete velocity vectors for the D2Q9 model and the projection of the discrete velocity vectors for the D3Q19 model in a channel. The z-axis is pointing into the paper. The arrows represent the discrete velocity directions, while the labels indicate the corresponding velocity indices. The inlet and outlet boundaries are shown to highlight the flow direction.}
    \label{fig:bc_velocity}
\end{figure}

%% file: Figures/bcWall.tex
\begin{figure}[htbp]
    \centering
    \begin{subfigure}[t]{0.47\textwidth} 
        \centering
        \includegraphics[width=0.9\textwidth, trim=0cm 0cm 0cm 0cm, clip]{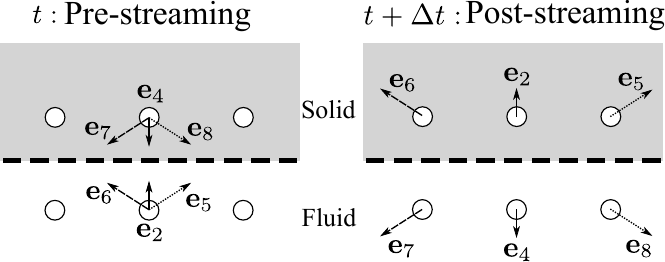} 
\caption{Schematic of populations streaming across the boundary between solid wall nodes and fluid nodes for free-slip boundary conditions. The arrows demonstrate the populations' directions, with tangential velocities remaining unrestricted and normal velocities set to zero at the solid wall surface. Individual populations maintain consistent arrow styles for clarity.}

        \label{fig:free_slip}
    \end{subfigure}
    \hfill
    \begin{subfigure}[t]{0.47\textwidth} 
        \centering
        \includegraphics[width=0.9\textwidth, trim=0cm 0.0cm 0cm 0cm, clip]{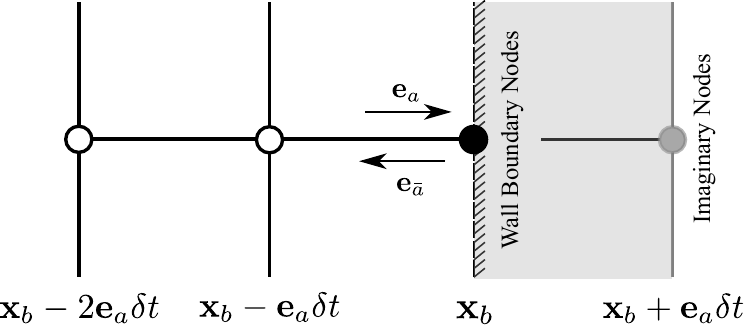} 
\caption{Schematic of wall boundary nodes surrounded by regular lattices for no-slip boundary conditions. The outgoing populations at the wall boundary node $x_b$ are reflected back to their origin, ensuring zero velocity at the solid wall. Imaginary nodes at $x_b + e_a\Delta t$ assist in implementing the bounce-back scheme effectively.}
        \label{fig:no_slip}
    \end{subfigure}
    \caption{The figure illustrates the mechanics of free-slip and no-slip boundary conditions, highlighting the role of pre-streaming, post-streaming, and the use of imaginary nodes to maintain boundary accuracy.}
    \label{fig:bc_wall}
\end{figure}

%% file: Figures/bcConvective.tex
\begin{figure}[htbp]
    \centering
    \includegraphics[width=0.40\textwidth]{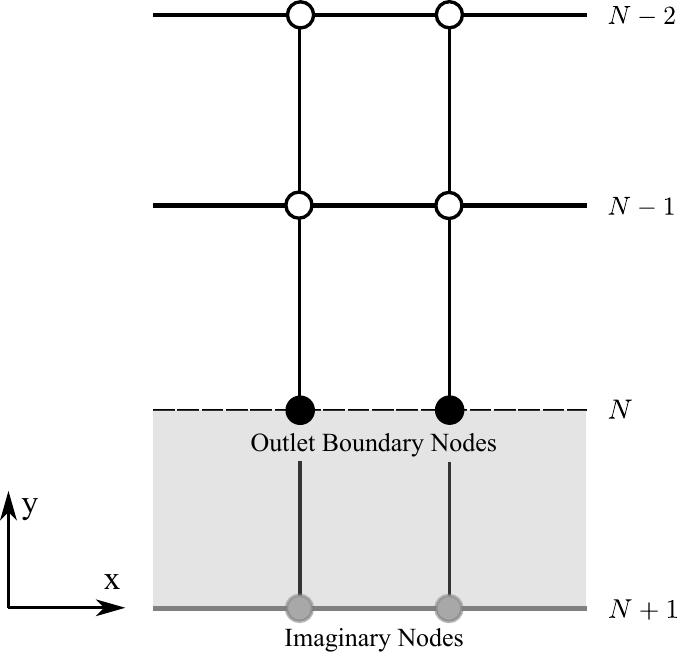} 
\caption{Schematic representation of the outlet boundary and imaginary nodes used in the implementation of convective boundary conditions. The gray shaded area highlights the imaginary nodes ($N+1$) required to compute the streaming step, while the outlet boundary nodes ($N$) interact with the fluid nodes at layer $N-1$.}

    \label{fig:bc_convective}
\end{figure}